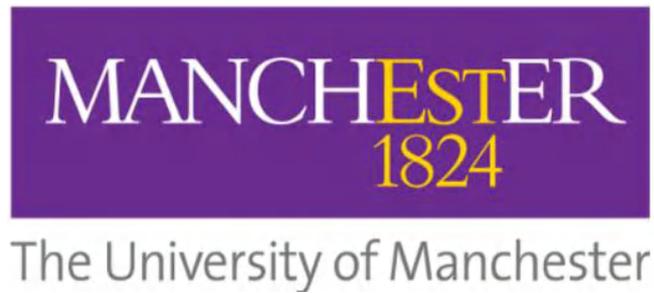

# Wind tunnel actuation movement system

A dissertation submitted to The University of Manchester for the degree of

Master of Science Mechanical Engineering design

in the Faculty of Science and Engineering

2020

Qiaoqiao Ren

(10474511)

School of engineering

Department of Mechanical, Aerospace and Civil Engineering

The University of Manchester

# List of Contents

# List of figures

# List of tables

Total words: 20793

# Abstract

In this dissertation project, an actuation system was designed for the supersonic wind tunnel in Mach 5 at the University of Manchester. The aim of this project is to build a remote-control actuation system which could adjust the angle of attack for the aerodynamic shape to save researchers' time and improve the experimental efficiency. This project involves the model-supporting system, a six-component wind tunnel balance, a control system design, a virtual angle of attack adjustment interface and LabVIEW programming implementation, the angle of attack adjustment range is $[-20°, +20°]$. The three-dimensional model of the mechanical part and its engineering drawing were finished in SolidWorks, and the control system including the sensors and rotary encoder control, the closed-loop control of the stepper motor and the wind tunnel balance feedback. The performance of the wind tunnel balance can be known in advance by finite element analysis. Finally, the virtual operating system was built based on the LabVIEW and Arduino interactive programs.

# Declaration

This dissertation is a presentation of my original design and research work. Efforts are made to clearly indicate wherever contributions of others are involved, with due reference to the literature, and acknowledgement of research and discussions. The work was done under the guidance of Dr Mark Quinn, at the University of Manchester, United Kingdom.

# Copyrights

i. The author of the dissertation (including any appendices and /or schedules to this dissertation) owns certain copyright or related rights in it (the "Copyright") and s/he has given The University of Manchester certain rights to use such Copyright, including for administrative purposes.

ii. Copies of this dissertation, either in full or in extracts and whether in hard or electronic copy, may be made only in accordance with the Copyright, Designs and Patents Act 1988 (as amended) and regulations issued under it or, where appropriate, in accordance with licensing agreements which the University of Manchester has entered into. This page must form part of any such copied made.

iii. The ownership of certain Copyright, patents, designs, trademarks and other intellectual property (the "Intellectual Property") and any reproductions of copyright works in the dissertation, for example graphs and tables ("Reproductions") which may be described in this dissertation, may not be owned by the author and may be owned by third parties. Such Intellectual Property and Reproductions cannot and must not be made available for use without the prior written permission of the owner(s) of the relevant Intellectual Property and/or Reproductions

iv. iv. Further information on the conditions under which disclosure, publication and commercialization of this dissertation, the Copyright and any Intellectual Property and/or Reproductions described in it may take place is available in the University IP Policy (see http://documents.manchester.ac.uk/display.aspx?DocID=487 ), in any relevant Dissertation restriction declarations deposited in the University Library, and The University Library's regulations. (see http://www.manchester.ac.uk/library/aboutus/regulations )

# Acknowledgement

First of all, I would like to express my gratitude to my supervisor Dr Mark Quinn, who gave me a lot of support and encouragement both in study and life. Because of the coronavirus, he had a massive workload every day, but he still insisted on having a weekly meeting and giving me a lot of feedback and guidance, which let me learned a lot from this project. He also put forward many new ideas for this project which not only helped me learn lots of new technology in this filed but also inspired me to continue my study.

Secondly, I would like to thank my family and friends for giving me a lot of help and support during my graduation.

Finally, I would like to thank the staff of the University of Manchester, who have been answering our questions and providing support online during this challenging time.

# 1 Introduction

## 1.1 Background

This study aimed to improve the progress of wind tunnel experiments and improve efficiency by implementing a model actuation system to help researchers change the angle of attack for the model automatedly.

When the wind tunnel experiments were conducted in the nineteenth century, they were used to study aircraft. The development of wind tunnels accompanied the development of the aeroplane and now wind tunnel testing is wildly used in various fields, which generate airflow and are controlled artificially to simulate the flow of air around an aircraft or entity, such as wind tunnel testing for cars and sports equipment. A critical device in wind tunnel experiments is the actuation system. However, small wind tunnels usually do not have one because of the height and area limitation, so the researchers have no choice but to adjust the model manually, which wastes time.

Because there are many types of wind tunnels, there are lots of studies and controversies about various wind tunnel actuation systems. To implement a robotic model motion system capable of moving aerodynamic shapes to multiple angles of attack without human intervention in the University of Manchester hypersonic wind tunnel would help researchers save time and simplify the experimental procedures.

This study, which will help further the knowledge on actuation systems, is essential to the engineering field, particularly aerospace engineering because it simplifies the process of the experiment and avoids the repeated work of adjusting the angle of the model.

## 1.2 Actuation system

Model actuation systems are devices for model positioning in the wind tunnel test section,

which makes it possible for the aerodynamic shape to adjust the angle of attack, the measurement of the model pose can be fed back to the position and attitude of the control system in real-time to realise the real-time feedback during the adjustment of the angle of attack.

Aerodynamic shape (typically a scale model) must be connected to a supporting system in the wind tunnel test segment which makes it possible to adjust its position, according to different instructions sent by researchers. This device must be rigid and change the angle of attack precisely. What's more, remote control angle of attack adjustment could reduce the influence of the flow disturbing in the test section.

## 1.3 Actuation system design specification

### 1.3.1 Overview

The wind tunnel in this project is the supersonic wind tunnel in Manchester, the basic size and model are shown in Figure 1 Size of the wind tunnel. In the present stage, the angle of attack of the model can only be manually adjusted. Each change in the angle of attack requires manual removal of the fixture, which is a waste of time. It is difficult to ensure the accuracy of manual adjustment, so this project will establish an actuation movement system for remote adjustment of the angle of attack and combine with real-time feedback wind tunnel balance to make the testing process more convenient and accurate.

### 1.3.2 Size of the wind tunnel

The size of the hypersonic wind tunnel is shown in Figure 1, which is measured in the hypersonic wind tunnel laboratory. The size of the actuation system will be determined by the limitations posed by the size of the wind tunnel and the position of the nozzle. Figure 1 shows that the diameter of the nozzle is 160mm, the distance between the top of the nozzle to the floor of the wind tunnel is 255mm, and the range from the bottom of the nozzle to the floor of the wind tunnel is 95mm. What's more, it is 280mm from the outlet of the nozzle to the air intake. The length of the nozzle $L_n$ is 154 mm, the length of the hypersonic wind tunnel $L_t$ is 575mm and the width $W_t$ of the wind tunnel is 348mm.

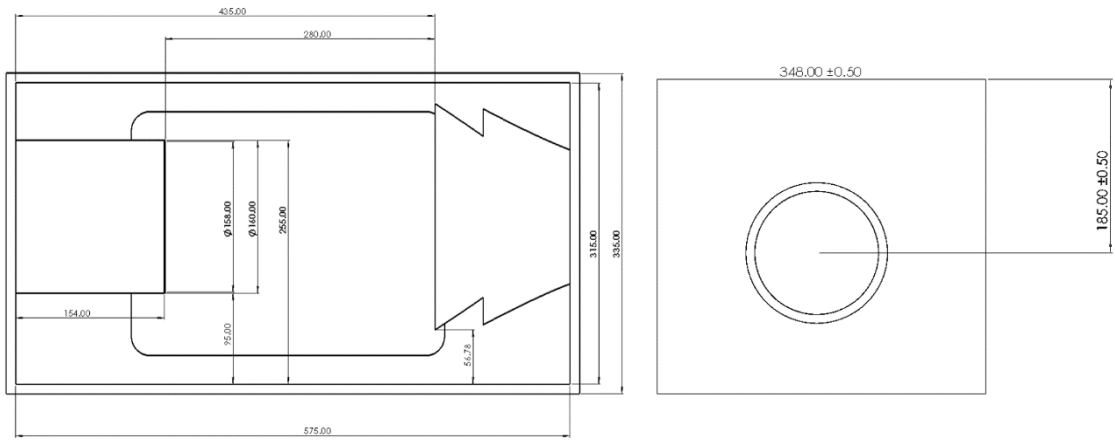

*Figure 1 Size of the wind tunnel*

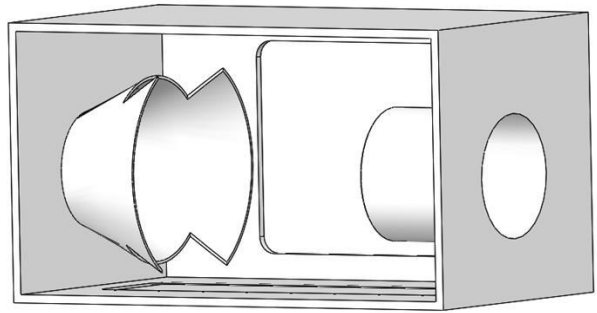

*Figure 2 Three-dimensional model of wind tunnel*

### 1.3.3 Research aim and objectives

This study aimed to design a remote control angle adjustment system to operate inside the wind tunnel during the experiment in order to save researchers' time.

Objectives:

1. Develop a reasonable mechanical concept design for the attack angle adjustment model in the wind tunnel

2. Design a six-component wind tunnel balance to obtain the deflection angle under the load

3. Design a control system which can control the model to achieve the adjustment of attack angle and build a closed-loop control system.

4. Build a virtual angle of attack operation interface based on LabVIEW for the researcher.

5. Verify the feasibility of the program with Arduino and other hardware devices.

6. Test the stability of the model interface in ANSYS.

### 1.3.4 System requirements

1. Size: Must be able to be placed in the supersonic wind tunnel at the University of Manchester and with a suitable blockage ratio;

2. Loading: The maximum force in the wind tunnel is 200N, while 50N forces are applied in most cases. It is necessary to ensure its structural stability and run smoothly under both a static and dynamic load;

3. Cost: The maximum cost is £500;

4. Accurate attack angle adjustment: Able to adjust the pitch angle with an error of 5%. Because of the boundary effect and the blockage factor, the results are affected and need to be minimised;

5. Product lifetime: Guaranteed to last for five years, so regular maintenance is required;

6. Convenience: The device has a remote control so that researchers don't have to adjust it manually;

7. Power supply: During the test, the actuation system can be powered by a power supply module and a 9V1A adapter;

8. Cabling: Wiring including USB cable, F-M wires and M-M wire will not interfere with the operation of the test;

9. Non-interference: The supporting system cannot interference the final image of the model by using the Schlieren system;

10. Driving system: The height of the driving system should be lower than the top of the model;

11. The workspace of the model should always be limited in the area of the nozzle section.

*Table 1 Program design specification*

| Program | Wind tunnel actuation movement system |
|---|---|
| Variable drag force | 0–200N |
| Power supply | Power supply module and 9V1A adapter |
| The size limitation of the whole system | Smaller than $280mm \times 348mm \times 255mm$ |
| Driving system | Smaller than $280mm \times 348mm \times 95mm$ |
| Model height | $95mm - 255mm$ |
| Cost | £500 |
| Weight | Smaller than 5kg |
| Product lifetime | Five years |
| Cabling | USB cable, F-M wires and M-M wire |

### 1.3.5 Methodology

The model supporting system can be designed according to the literature and existing design solutions and then using SolidWorks to get the three-dimensional model. The wind tunnel balance could feedback the aerodynamic forces and moment to the supporting system to correct the deflection, the performance of the wind tunnel balance could be predicted in the ANSYS. Finally, remotely control could be achieved by Arduino and LabVIEW interactive programming, and implement a virtual angle of attack model interface for researchers based on the LabVIEW.

### 1.3.6 Measuring impact

In this project, the projected attack angle of the model will be compared to the actual angle of attack to ensure the accuracy of the experiment. Additionally, a research survey could be conducted on the researchers who will subsequently be using the machine to evaluate their feedback and identify problems with the system.

### 1.3.7 Security

Failure of the actuation movement system may cause damage to the test section. At the same time, if the structural stability of the drive system is not enough and the parts lose their fixation, the damage will be caused to the wind tunnel.

# 2 Literature review

This literature review will discuss actuation model systems in the wind tunnel, including wind tunnel experiments, the mechanical structure which can be used to achieve the model pitch movement in Manchester supersonic wind tunnel, control system which can control the model actuation system accurately, structure strength analysis, convenient virtual angle of attack operation interface for researchers and wind tunnel balance.

## 2.1 Wind tunnel experiment theories

There are various kinds of wind tunnels with different uses and aims, depending on the different classification, such as the speed and way of the circuit.

From NASA, wind tunnels are usually of two types, open-loop tunnel or closed-loop tunnel, according to the way of circuits. For the open return tunnel, the wind tunnel is placed in a room and the air in the wind tunnel test area is flowing. The atmosphere can be circulated through the room where it is located. Figure 3 shows the wind tunnel and the airflow around the wind tunnel. Another type of wind tunnel is the closed-loop tunnel. Air is mainly generated by a series of fans and rotating blades. Air is circulated in the tunnel but it does not exchange air with the outside.

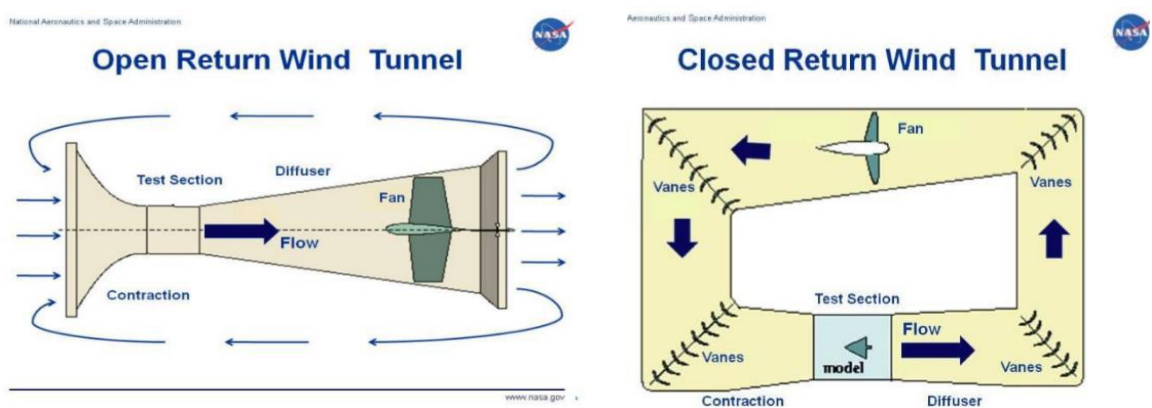

*Figure 3 Open-circuit and closed-circuit wind tunnel (NASA)*

The model actuation system in this project was designed for the supersonic wind tunnel laboratory at the University of Manchester, and the size of the enclosed jet test section is limited. Figure 4 showed the layout of the wind tunnel.

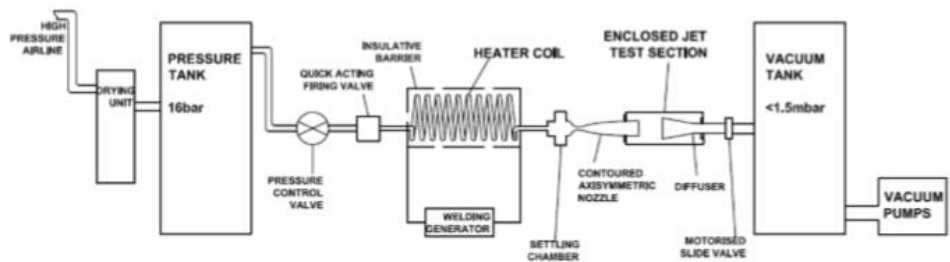

*Figure 4 University of Manchester HSST schematic layout(Erdem, 2011)*

Size is one of the limiting factors of wind tunnels is that it has a power requirement; hence, the test section is kept small and only scaled models of bodies are kept for testing. In order to reduce the effects of wall interference, due to shock reflection from wind tunnels, the length of the test section needs to be small, so that the reflected shocks go outside the test section length and do not interfere with the flow in the test section.

In supersonic flow over typical launch vehicles with a spherical cap at the nose, the detached shock at the nose will extend to the tunnel wall. These shock waves on hitting the tunnel wall will reflect back toward the model which can again reflect back to the tunnel wall and back to the model again and multiple shock reflection phenomena can be observed if the models' lengths are very large. These shocks hitting the model can dramatically change the aerodynamic loads on the model from those that would be expected in free flight. Hence it is crucial to find out the optimum length of the launch vehicle model to be tested in order to get accurate results without these wall interference effects. It is challenging to know the size of the model to overcome the above-mentioned tunnel wall interference effects prior to experiments, although some estimates can be predicted with previous experimental studies.

Depending on the kind of wind tunnel, the requirements for the actuation system of the wind tunnel are also different. Compared with low-speed wind tunnels, the structural strength of the drive system is very high. Also, between large and small wind tunnels, large wind tunnels have lower requirements for the size of the drive system and are even better controlled.

## 2.2 Blockage ratio

As (Eltayesh et al., 2019)says, the blockage is usually defined as the supporting model's projected area in the flow direction to the domain's cross-sectional area around the model. This has a significant influence on the experimental result when doing the research in supersonic wind tunnels; in wind tunnel testing of sub-scaled models for subsonic, transonic and supersonic flows, there are two types of blockage effects: body blockage effects and wall interference-caused blockage effects. The results obtained from wind tunnel testing need to be corrected for blockage effects, buoyancy, wall interference and STI (strut, tare and interference) effects. The body blockage effects include solid blockage, due to model being inserted into the wind tunnel and wake blockage, due to a higher velocity as boundary layers grow around the model, putting the model on a pressure gradient and results need to be corrected. The wall interference effects are dependent on Reynolds number and caused due to shock/boundary layer interaction.

Choi and Kwon (1998) think the blockage effect is a very crucial factor which affects the test results. They also talked about the aerodynamic features of two test models based on different wind tunnel tests with varying angles of attack and different blockage effects (Choi and Kwon, 1998). And they found that the test model's aerodynamic characteristics are almost identical for models with up to 10% blockage ratios, while for the models with over 10% (Choi and Kwon, 1998), the feature showed some discrepancies, (Ryi et al., 2015) mentioned that the blockage effect is due to the limited nozzle size and test section length of the wind tunnel. In wind tunnel test studies, in general, to obtain accurate aerodynamic measurements, the wind tunnel blockage ratio of the model should not exceed 5%.

One of the significant constraints in the wind tunnel testing is the size of the model to be tested as wind tunnels are small in size compared to the vehicles in question to be investigated. The size restriction includes all the dimensions of length, breadth and height. Normally the breadth and height dimension will give rise to the blockage effect in the wind tunnel testing and the length dimension will give rise to the tunnel wall interference effect. Therefore, it is essential to know the blockage ratio before testing in order to get reliable results. Here is the function of the blockage ratio. (Eltayesh et al., 2019)

$$b_r = \frac{d_m}{\sqrt{a-(\delta_{bl}.P_{test})}} \qquad Equation\ 2\text{-}1$$

Where A= $a - (\delta_{bl}.P_{test})$, so $b_r = \frac{d_m}{\sqrt{A)}}$

$d_m$ is the maximum model diameter

$a$ is physical test section cross-sectional area

$\delta_{bl}$ is boundary layer displacement thickness

$P_{test}$ is the test section perimeter

## 2.3 Mach number

In fluid mechanics, an important dimensionless parameter representing the compressibility of fluid is expressed as Ma, which is defined as the ratio of the velocity v at a point in the flow field to the local sound velocity c at that point, i.e. Ma=v/c (Pope and Gollin, 1965).

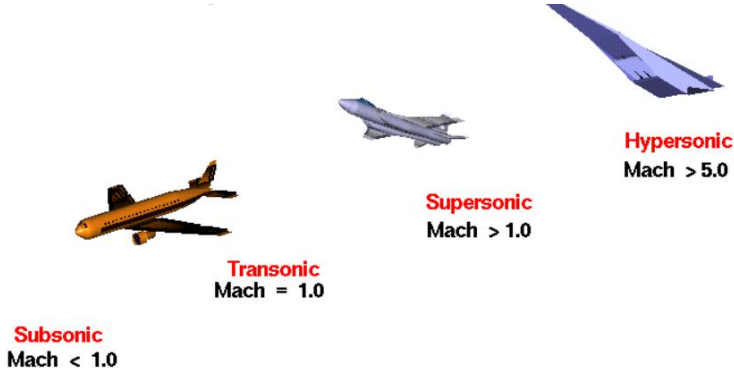

*Figure 5 Different wind tunnel(NASA)*

*Table 2 Different types of wind tunnel classified by Mach number*

| Mach number | Wind tunnels |
|---|---|
| M<1 | Subsonic wind tunnel |
| M=1, or near 1 | Transonic wind tunnel |
| 1<M<3 | Supersonic wind tunnel |
| 3<M<5 | High supersonic wind tunnel |

| M>5 | Hypersonic wind tunnel |
| M>>5 | High Hypersonic wind tunnel |

The geometry of the wind tunnel nozzle determines the Mach number, as shown in Table 1. Subsonic wind tunnels operate at a very low Mach number(M<1). When M=1, the wind tunnel can be described as a transonic wind tunnel. A supersonic wind tunnel's Mach number range is from 1 to 3. If the Mach number is greater than 3, but not greater than 5, it is called a hypersonic wind tunnel. At this speed, the overall temperature of the airflow is high, which will overheat the nozzle throat, so cooling measures should be taken Mach numbers exceeding five are called hypersonic speeds.

$$Ratio = \frac{object\ speed}{speed\ of\ sound} = Mach\ number \qquad Equation\ 2\text{-}2$$

Mach number is a crucial parameter in aerodynamics, and the wind tunnel in the University of Manchester is a supersonic wind tunnel, the Mach number is 5. Thus we can also infer the size of the support structure and the projection area of restriction which can be influenced by Mach number, through Mach number is easy to calculate fluid velocity, and combine with the temperature of the wind tunnel, and it is concluded that in test section inside the support structure of the force, thus the structural strength is calculated.

## 2.4 Different angle adjustment system in the wind tunnel

### 2.4.1 The wire-driven parallel suspension system

Xiao Yangwen.et al (2010) developed a new wire-driven parallel suspension system based on the active suspension for wind tunnel tests project conducted by the French National Aerospace Research Centre, which is shown in Figure 6. The eight-wire-driven system was built to achieve the six degrees of freedom motion control of the aerodynamic shape, besides, through the measurement system for wire tension and the data acquisition system, the aerodynamic parameter of the model can be obtained.

This new type of parallel manipulator's structure is reconfigurable and straightforward so that the workspace is large, therefore, the experimental results won't experience interference and the blockage ratio is small. What's more, it has a high load capacity and weight ratio. In addition, compared to other complex structure, the eight wires system is easy to assemble and costs little, however, when it is used in the supersonic system, this wire-driven system is not stable and hard to stop at the desired angle, due to the high-speed wind, and the error is large when adjusting the angle of attack of the aerodynamic shape.

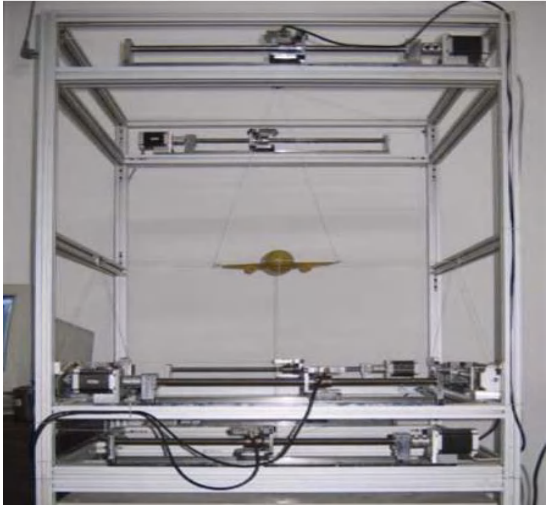

*Figure 6 Prototype of WDPSS-8(Xiao Yangwen .et al, 2010)*

### 2.4.2 Bracing system

En-xia (2001) introduced a system of hanging braces of a large angle of attack. Based on the traditional bracing system, the interference of the bracket can be reduced. The model is suspended between the upper and lower walls of the test section of the wind tunnel with a tension line. There is a balance in the model and a turntable on the upper and lower walls of the wind tunnel. The elevation angle of the model can be changed by changing the rotation of the rotary table. The rotation angle of the model is changed by the up-and-down movement of the end tensioning line with the disk.

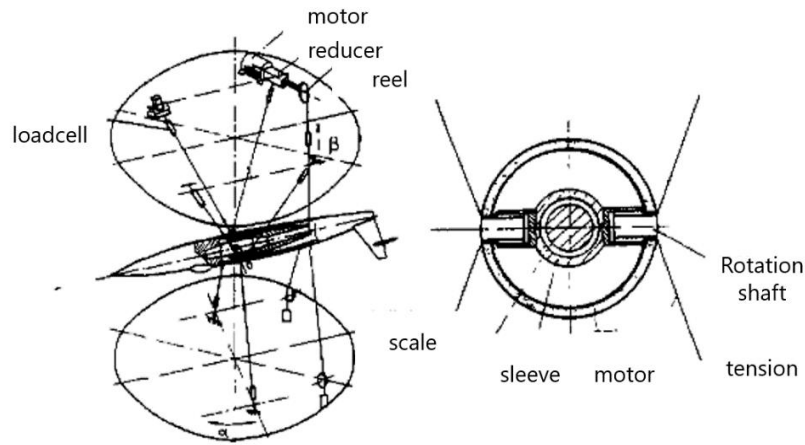

*Figure 7 The basic structure of the bracing system*

All the institutions in this hanging bracing system are installed outside of the wind tunnel test section, which minimises the interference flow field for the experiment. What's more, this bracing system achieves both a pitching angle and sideslip angle adjustment. At the same time, the two angles of attack and sideslip are independent of each other and controlled separately. Both of them can be continuously changed.

In this research, the research object is a supersonic wind tunnel; the structural strength of the model needs to be considered. The bracing system is in good condition when working in the low-speed wind tunnel. However, when using in the supersonic wind tunnel, the structural strength is not enough and the tested model cannot be very stable, so the accuracy of the adjustment angle is low in the high-speed wind tunnel. Besides, all the institutions are installed outside the wind tunnel in En-xia (2001)'s design. In this research, the support system should fix in the wind tunnel but outside the airflow range, so the machine cannot be too complicated and the size is also limited.

### 2.4.3  Crescent sting support system

The crescent supporting system is usually used for sting mounted models, which can achieve pitch and yaw angle variation, and the rotation centre of pitch and yaw are on the centre line and fixed and, through the rotation of a turntable, can achieve the variation of yaw angle.

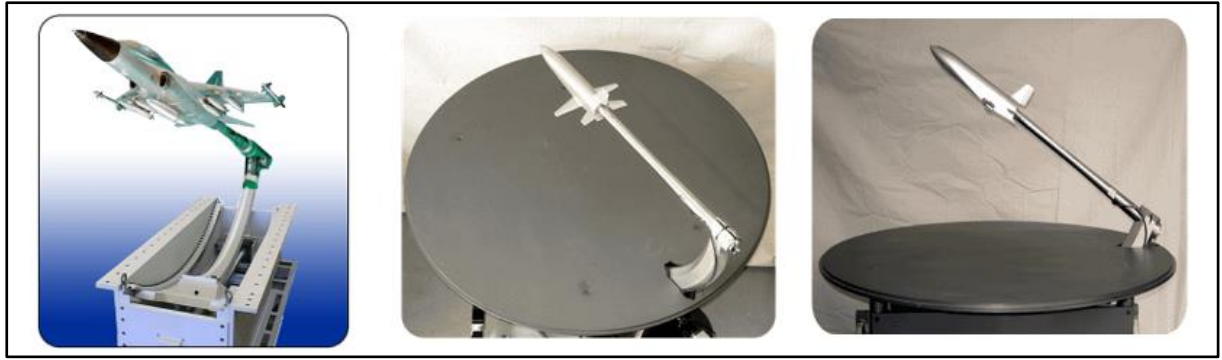

*Figure 8 Pitch, yaw, roll movement for the crescent sting support system*

Changhai (2010) designed a crescent sting support system and conducted a finite element analysis. On the premise of ensuring the design index, the mass of the structure was significantly reduced and the blockage effect was reduced under the condition that the strength and stiffness were satisfied.

For the automotive crescent sting supporting system, the actuation system usually mounts outside of the wind tunnel, which can not meet my design requirement, and there are limitations of size and height in the University of Manchester high supersonic wind tunnel, which will place restrictions on angle adjustment.

### 2.4.4 Magnetic support system

Takagi et al. (2016), who designed a magnetic suspension and balanced system for 85 suction-type supersonic wind tunnels, mentioned that the magnetic supporting system, which supports the model through magnetic force, is an excellent method of moving a model in a wind tunnel without a real supporting system. The main working principle is the interaction of the magnet inside the model and the external coil located in the test area. For this system, there is no support interference and the force is evaluated by magnetic function.

In the test section of the wind tunnel, the exposure of the support structure will interfere with the experimental results. Magnetic levitation technology solves this problem very well. Without a mechanical support structure, it will not cause interference to the test.

As for its disadvantages, the technology is not particularly mature now and it is challenging to accomplish in HSST. Because the model is challenging to make levitate under high aerodynamic load, it is challenging to realise maglev technology in the high-speed wind tunnel.

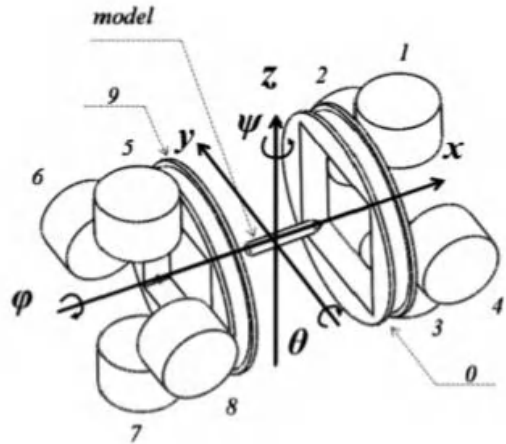

Figure 9 Configuration of electromagnets for MSBS(Takagi et al., 2016)

## 2.4.5 The supporting system based on a lead screw

Chu Weihua (2012) did research on the kinematic relationship between the big angle of attack, the sideslip angle and multi-axis control. Chu Weihua (2012) produced a design to achieve a big attack angle adjustment, which uses a triangular slider linkage. Still, it is hard to realise linear velocity control by using the control hypothesised in this Chu Weihua (2012)'s research, however, it is not necessary to achieve linear velocity control in this project, the relationship between displacement and angle of attack could be given through the transmission function and be implemented in LabVIEW.

Chu Weihua (2012) used a motor to drive the system, connected a coupling and a screw rod for lateral displacement. Now many support systems of models are driven by motor screw rods to convert rotary motion into linear motion, and then use mechanical structures to achieve straight lines motion and angle transformation. The best method to reduce the interference in the wind tunnel is to place the mechanical structure outside the test section. At the same time, the size of the model is controllable and easy to design. The rigidity of the mechanical structure will also be relatively stable, which is a better choice in this project.

## 2.5 Actuation system

### 2.5.1 Hydraulic

The hydraulic transmission is relatively small in size, compact in structure, light in weight and large in driving torque, which can meet the requirements of stepless speed regulation. (Manring and Fales, 2019), which causes the hydraulic drive to have higher external temperature requirements, high energy loss and low efficiency. Xuanyin (1999) introduced a hydraulic wind tunnel tail brace system, which is a Servo hydraulic actuation way by using the high-precision hydraulic cylinder to ensure the performance and improve the service life.

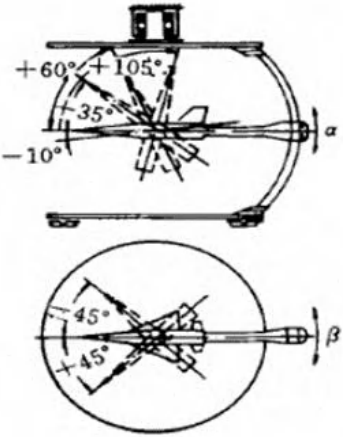

*Figure 10 Principle of hydraulic wind tunnel tail brace system(Xuanyin, 1999)*

The hydraulic system is widely used and has many advantages, but also has some disadvantages. As Figure 11 shown, the structure of the hydraulic system is complex, and it is not convenient to install in the wind tunnel. Meanwhile, it needs to be reset during the wind tunnel test. Finally, the hydraulic system needs to be cleaned and the pump and filter need to be maintained regularly.

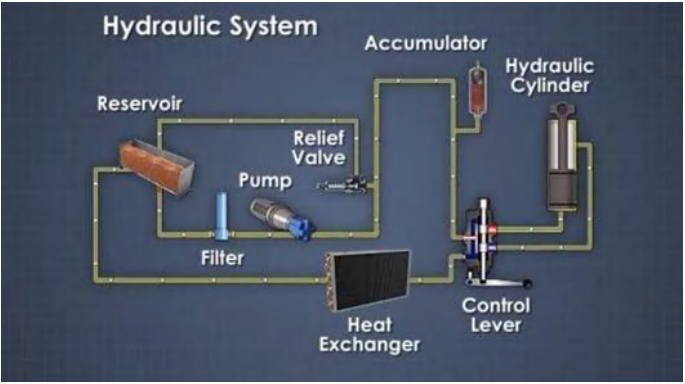

*Figure 11 Components for hydraulic system（vector solution）*

The operating temperature range of the University of Manchester supersonic wind tunnel is 300–950K, which can reach higher temperatures, which may cause leakage. At the same time, there are many components in the hydraulic system, which require more space and are more troublesome to install.

### 2.5.2 Pneumatic

Hao Yufei (2016) created a four-fingered pneumatic actuated elastomeric robot, which only requires some basic control for pneumatic actuation. There are only two conditions in this system, which are open the grippers claw to approach the objectives and contact it, so Hao Yufei (2016) chose the pneumatic system cause he didn't need to achieve the continuous control of the system. The pneumatic uses the medium (air) as the working medium, which is easy to obtain, convenient to drive land non-polluting to the environment, is low cost and is safe. It responds quickly to the hydraulic drive system with pneumatic actuation, simple maintenance, clean working medium and there is no medium deterioration, replenishment or other issues. Toshiro Noritsugu (2008) developed a novel three-degree-freedom system by using the pneumatic actuation system, which is suitable for the large wind tunnel actuation.

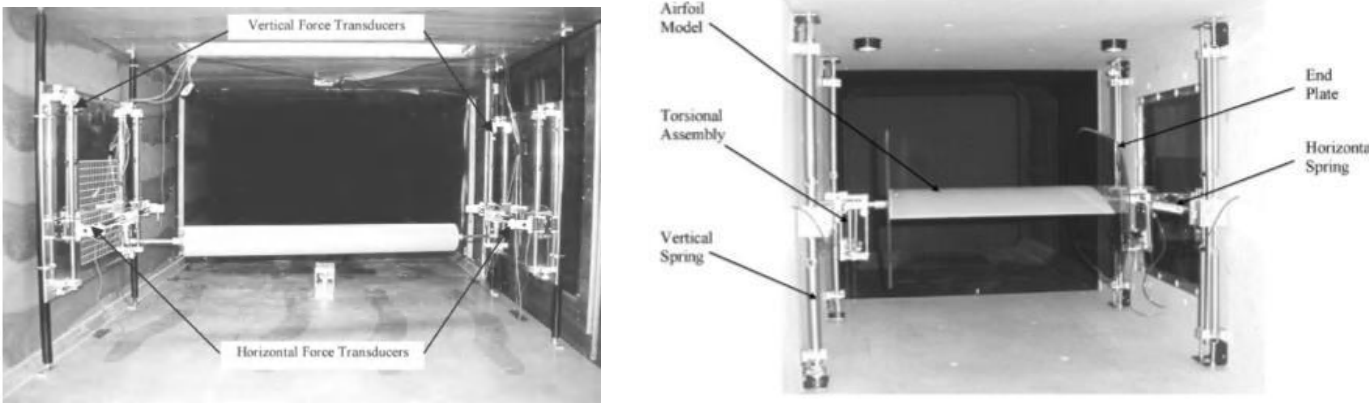

*Figure 12 Four-bar support system(Hao Yufei, 2016)*

The air compressor is used pneumatically. The volume is coupled with noise. The sealing requirements are high and the thrust is too small to achieve precise intermediate position adjustment. Usually, it is used in two extreme positions. And it also suffers from compression loads which are depending on the aerodynamic forces, closed-loop control can be a must. Therefore, in this project, the angle of attack cannot be accurately adjusted without closed-loop feedback.

### 2.5.3 Electric

The electric actuation system has high accuracy and convenient speed adjustment, but the cost is high at high thrust. At the same time, in the supersonic wind tunnel, the overall structure is subjected to relatively large forces, so a motor with higher torque is required to support the movement. The increased size and weight of the motor affect the flexibility and portability of the entire device. But, due to the smaller size of the University of Manchester sonic wind tunnel, the engine used is lower and the torque is relatively small. Under high sound speed conditions, the maximum wind force is nearly 200N, and the stepper motor is easy to get stuck under the aerodynamic force, therefore the motor should be placed outside the test section. At the same time, the torque required by this project is not high. Because of the size limitations and experiment safety of the wind tunnel, electric actuation is the better choice (Zhang and Cai, 2014).

## 2.6 Control system

In this project, in order to achieve the precise adjustment of the angle, closed-loop control of the system is required. Rodríguez-Sevillano et al., (2014) said that LabVIEW is a visual programming language that can easily see the logical relationships of the programme. It is also a virtual work platform that can effectively handle different test tasks. (National Instruments 2019). In addition, National Instruments also provides a variety of I/O cards that can be installed with plug-ins which can efficiently study programming structure without installing hardware (Rodríguez-Sevillano et al., 2014). In this project, LabVIEW act as an upper machine, which can design a visual user interface and is easy for researchers to operate.

In this context, the Arduino which is a powerful tool for automatic control and monitoring, has been used worldwide. Due to its lower price and relatively small size, it can be used in a range of application. An Arduino board could be used as an automatic system in the system, because it includes lots of I/O pins and an AT mega microcontroller and because the platform has an integrated development environment (Espinoza et al., 2015). Arduino can control the motor's drive and forward and backward rotation, which can receive signals from sensors and has many other applications in different fields (Molle et al., 2011). In this project, it is enough to achieve control by using Arduino and LabVIEW.

Closed-loop control feedbacks the input system through the information of system output changes to achieve precise control of the angle of attack, and avoid the result from deviating too much from the system's predetermined target, mainly by returning the input to the output, comparing the feedback information and then controlling the comparison result.

LabVIEW can achieve closed-loop control easily, such as PID control. PID control is widely used in supersonic and transonic tunnels (Espinoza et al., 2015). It can be used to control the Mach number and pressure. In addition, compressed air can run the wind tunnel. PID control can also quickly set the fluid speed. In a subsonic wind tunnel, a direct current motor and three-phase rectifier used by (Dinca and Corcau, 2014) and developing a PI control algorithm in LabVIEW achieved the closed-loop control. Similarly, (Xuan et al., 2010), in order to improve the network's robustness, implemented a proportional feedback control system in union with a diffuse controller by adding to a neural network. In order to achieve a closed-loop control; in this project, the deflection angles must be fed back to the supporting system to achieve real-time feedback. What's more, the actuators also need closed-loop control. Therefore, there are two closed-loop controls in this system.

## 2.7 Position and measurement system feedback

In the wind tunnel experiment, the measurement of the model pose is essential. The pose of the aerodynamic shape can be known through the Schlieren system. Therefore, the model support system cannot interfere with the experiment results of the aerodynamic shape. At the same time, the wind tunnel balance can be used to measure the deflection, forces and moments of the aerodynamic shape under load when the wind tunnel is working normally. This can compare the desired angle of attack of the aerodynamic shape with the current angle of the aerodynamic shape in real-time, so as to give feedback to the supporting system.

## 2.7.1 Wind tunnel balances

A wind tunnel balance is a multi-component force sensor that can measure the aerodynamic load of the aerodynamic shape during wind tunnel testing (NASA). The balance is designed to measure some or all of the three forces and three moments of the aerodynamic shape; these forces are axial force, normal force and lateral force, while the moments are pitch moment, roll moment and yaw moment.

In industry, these systems have strain gauge balances that feed their force data into simple models of the support (to calculate deflection) and they use this to move the system back to the requested angle of attack (as it deflects under load) (NASA). The aerodynamic shape is attached to the wind tunnel balance and is supported by the model interface, the model support system for changing the angle of attack.

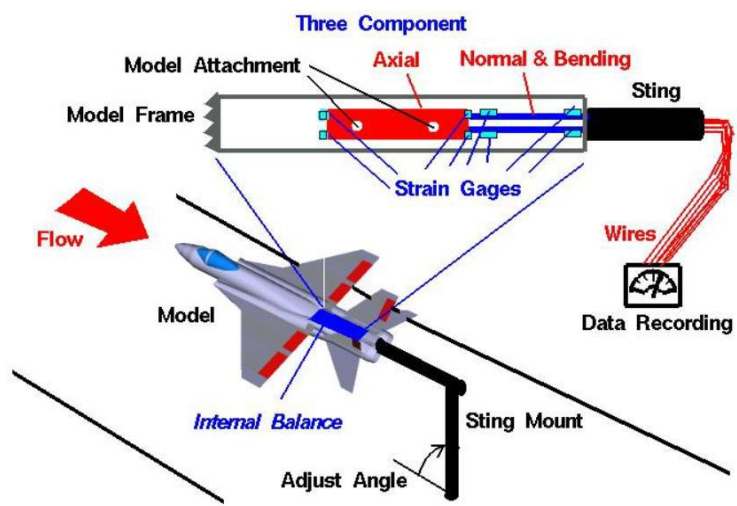

*Figure 13 Wind tunnel balance（NASA）*

## 2.7.2 Schlieren system

In the supersonic wind tunnel test at the University of Manchester, the Schlieren system was used to provide diagnostic information about the flow around the model. (NASA) The principle of Schlieren photography is similar to shadow map technology. Since light will bend when encountering fluid density changes, the Schlieren system can be used to visually indicate that the airflow leaves the surface of the object. As shown in Figure 14, the Schlieren system can be used for wind tunnel testing. There are two concave mirrors on the side. The light from the light source passes through a slit and the reflected lights are parallel rays. Then the reflected light is collected by the other mirror. The light is concentrated on the point of the blade and, then, the light continues to shine onto the camera is recorded.

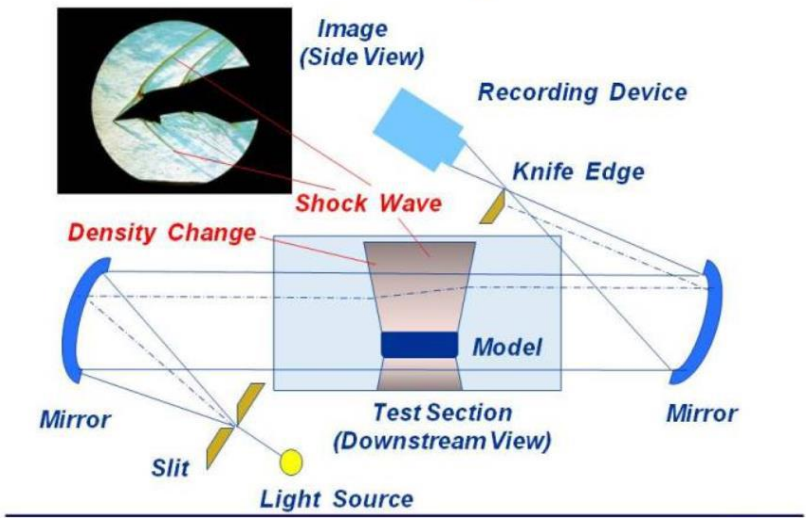

*Figure 14 Principal of Schlieren system(NASA)*

As shown in Figure 14, the parallel light encounters a knife-edge in the test area and the resulting image recorded by the camera shows dark lines where there is a density gradient. From this, it can be concluded that the aerodynamic shape completely blocked the passage of light in the experiment and the black image of the model can be seen. Therefore, there should be no large obstructions on both sides of the model. Otherwise, the imaging of the obstructions will become a black image, which will interfere with the imaging of the model and affect the observation of researchers. Therefore, the model support system mustn't interfere with the aerodynamic shape image.

## 2.8 Conclusion

In conclusion, this project involves lots of aspects. For the mechanical design, the lead method screw is the best supporting structure that can achieve the angle of attack adjustment. The bracing system and strip system reduces the interference of the supporting system but they usually suit large and low-speed wind tunnels. For the supersonic wind tunnel, the structural strength is not enough. And the crescent sting support system could achieve adjustment for the angle of yaw and angle of attack, but the supporting system is usually mounted outside of the wind tunnel and, because of the limitation of the wind tunnel size, the range of angle adjustment will be limited. The magnetic levitation support system is a good technology, because it doesn't have the support interference problems and can achieve six-axis control, but it is hard to achieve in HSST in Manchester because of the heavy load and the limited size of the wind tunnel. Table 3 shows the comparison of different aspects of supporting structures:

A= System based on the lead screw

B= Bracing supporting system

C= Wire-driven parallel suspension system

D= Crescent Sting support system

E= Magnetic supporting system

*Table 3 Comparison of supporting structure*

| Metrics | Weight (1-5) | Rating | | | | | Weight x Rating | | | | |
|---|---|---|---|---|---|---|---|---|---|---|---|
| | | A | B | C | D | E | A | B | C | D | E |
| Robustness | 5 | 4 | 2 | 3 | 4 | 4 | 20 | 10 | 15 | 20 | 20 |
| Size | 5 | 4 | 4 | 5 | 3 | 5 | 20 | 20 | 25 | 20 | 15 |
| Precision of angle adjustment | 5 | 4 | 3 | 2 | 4 | 5 | 20 | 15 | 10 | 10 | 25 |
| Cost | 3 | 5 | 3 | 2 | 3 | 1 | 15 | 9 | 6 | 9 | 3 |
| Difficulty of assemblies | 2 | 4 | 2 | 3 | 3 | 1 | 8 | 4 | 6 | 6 | 2 |
| Scoring: Total weights | | | | | | | 83 | 58 | 62 | 65 | 65 |
| Rank | | | | | | | 1 | 5 | 4 | 2 | 2 |
| Continue? | | | | | | | yes | no | no | no | no |

There are three types of the actuation system. The best one is the electric system. Firstly, compared with the hydraulic system, the electric system is much cleaner and does not require additional equipment. However, for the hydraulic system, the air storage tank is outside of the wind tunnel, the structure of the wind tunnel needs to be changed and, because of the limited space in the wind tunnel, the electric system is the better choice. The pneumatic is point-controlled so that the angle cannot be continuously adjusted. Generally speaking, these are the two limitations of pneumatic control, even though the aerodynamic risk is relatively low and it is relatively clean. In this project, the electric system is much better.

*Table 4 Comparison of Actuation systems*

| Metrics | Weight(1-5) | Rating | | | Weight x Rating | | |
|---|---|---|---|---|---|---|---|
| | | electric | hydraulic | Pneumatic | electric | hydraulic | Pneumatic |
| Size | 5 | 5 | 3 | 3 | 25 | 15 | 15 |
| Controllability | 5 | 5 | 4 | 4 | 25 | 20 | 20 |
| Risk | 5 | 4 | 2 | 3 | 20 | 15 | 10 |
| Speed control | 3 | 5 | 4 | 3 | 15 | 12 | 9 |
| Cost | 3 | 3 | 3 | 3 | 9 | 9 | 9 |
| Difficulty of | 2 | 4 | 3 | 3 | 8 | 6 | 6 |
| Environmental | 2 | 4 | 3 | 5 | 8 | 6 | 6 |
| Scoring: Total weights | | | | | 110 | 83 | 75 |
| Rank | | | | | 1 | 2 | 3 |
| Continue? | | | | | yes | no | no |

Last but not least, some coefficients, such as the Mach number of the Manchester wind tunnel, were used to calculate the blockage coefficient in order to obtain the projected area limit and height limit of the support system. And for the control system, LabVIEW and Arduino were used for joint programming and closed-loop control (A. A. Rodnguez-Sevillano), the virtual angle of attack system operation system could be designed based on LabVIEW.

# 3 Fluid mechanics

Fluid mechanics analysis is crucial in wind tunnel tests, since they can help us know the force condition of the model supporting system under the load and some limitations in the wind tunnel, such as the blockage ratio, in advance. Based on the above conditions and limitations, a support system that meets the requirements can be given.

## 3.1 Blockage ratio

The importance has already been given in the literature review. For the model support system to work normally, the blockage ratio should be designed within a normal range. As Figure 15 is shown, the reasonable blockage ratio range under different Mach numbers are given. The two curves in Figure 15 represent the theoretical and actual curves respectively. In this project, the Mach number is 5. It can be seen from Figure 15 that the blockage ratio is 0.42, and the maximum allowable projection area can be obtained according to the given blockage ratio and the equation given below. (Pope and Gollin, 1965).

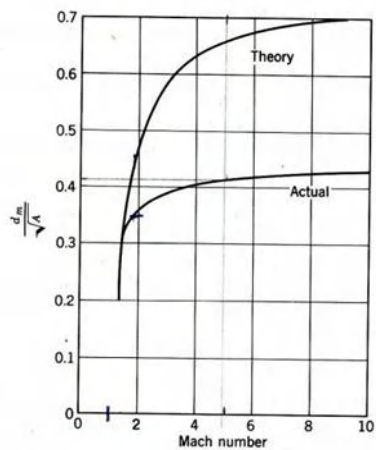

*Figure 15 Blockage ratio for aerodynamic shapes(Pope and Gollin, 1965)*

$$Blockage\ ratio = \frac{D_m}{\sqrt{A_{nozzle} - (\delta_{bl} * P_{nozzle})}} \qquad Equation\ 3\text{-}1$$

Where $a$ means the nozzle area of the supersonic wind tunnel which is 76 mm, while $P_{test}$ is the nozzle exit perimeter which is $152\pi$ and $d_m$ represents the maximum allowable diameter of the model, and $\delta_{bl}$ is the thickness of the boundary layer, and the value is 11 mm.

The maximum model diameter can be derived through Equation 3-2

$$d_m = Blockage\ ratio \times \sqrt{a - (\delta_{bl} * P_{test})} \qquad Equation\ 3\text{-}2$$

$$d_m = 0.42 \times \sqrt{\left(\frac{152}{2}\right)^2 \times \pi - (11 \times 152\pi)^2} = 47.69 mm$$

The maximum allowable diameter of the model is 47.69 mm, through this maximum diameter the maximum allowable front area could be calculated by $\pi d_m^2$, which is 1786.26 $mm^2$, therefore, the total project area of the supporting system and aerodynamic should be less than 1786.26 $mm^2$.

## 3.2 Determine Loading Condition

### 3.2.1 Oblique shock

Oblique shock usually occurs when a supersonic flow passes through a corner that converts the flow into itself and compresses it and the oblique shock wave is irreversible and discontinuous in vibration, flow and aeroacoustics noise,(Ming Hsihsu, 1997). These singularities are essential when analysing the drag force of the supporting system, because these properties increase the drag force when the supporting system is working in the supersonic wind tunnel, (Jorge Luis Garrido Tellez1, 2016), which will influence the maximum holding torque of the actuation system.

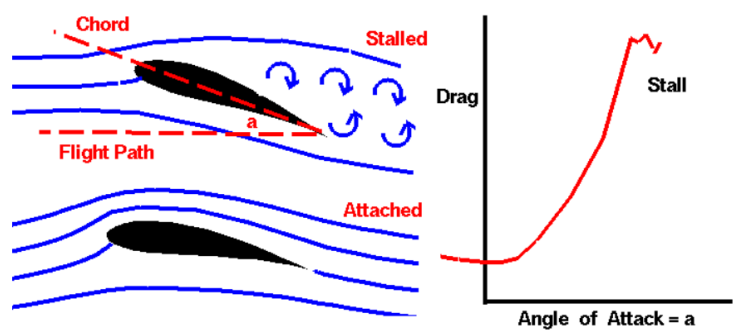

*Figure 16 Relationship between drag force and angle of attack*

#### 3.2.1.1 The $\theta$-$\beta$-$M$ equation

Based on the oblique shock wave theory related to compressible gas dynamics, the $\theta$-$\beta$-$M$ equation can be loosely described as a function of shock wave angle $\beta$, Mach number $M_1$ and the flow deflection angle $\theta$, which is the relationship between the shock wave angle and the flow deflection angle under Mach number $M_1$ as shown in Figure 17.

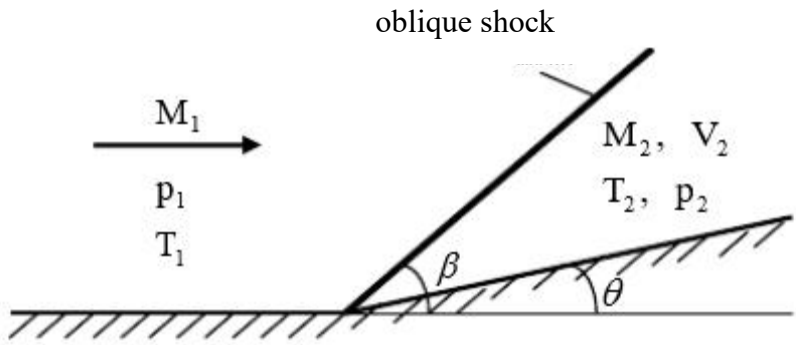

Figure 17 Oblique shock

$$tan\theta = 2cot\beta \left[\frac{M_1^2 sin^2\beta - 1}{M_1^2(cos2\beta + \gamma) + 2}\right] \qquad \text{Equation 3-3}$$

In this project, $M_1$ means Mach number which is 5; $\gamma$ is specific heat capacity ratio, assume $\gamma = 1.4$; $\beta$ is shock angle; $\theta$ is deflection angle, which, here, is 20 degrees. These data are substituted into Equation 3-4; the shock angle can be acquired from Equation 3-5

$$tan20 = 2cot\beta * \left[\frac{5^2*sin^2\beta - 1}{5^2*(cos2*\beta + 1.4) + 2}\right] \qquad \text{Equation 3-6}$$

$$\beta = 29.801$$

Therefore, $M_{1n}$ and $M_{2n}$ can obtain from Equation 3-7 and Equation 3-8 respectively.

$$M_{1n} = M_1 \sin\beta = 5 \times sin(29.801°) = 2.485 \qquad \text{Equation 3-7}$$

$$M_{2n} = \sqrt{\frac{1 + \frac{\gamma - 1}{2}M_1^2}{\gamma M_1^2 - \frac{\gamma - 1}{2}}} = 0.514 \qquad \text{Equation 3-8}$$

Substitute $M_{2n}$, $\theta$, and $\beta$, $M_2$ can be calculated from Equation 3-9.

$$M_2 = \frac{M_{2n}}{\sin(\beta - \theta)} = 3.022 \qquad \text{Equation 3-9}$$

$$\frac{P_2}{P_1} = 7.037; \frac{T_2}{T_1} = 2.123; \frac{P_{02}}{P_{01}} = 0.505$$

As shown in the isentropic flow properties table (Farokhi, 2015), the ratio of $\frac{P_{01}}{P_1}$ is given in the table when $M_1 = 5$ and, from the same table, $P_{01}$ is 890 KPa, which represents the total pressure before the shock wave in the supersonic wind tunnel.

$$\frac{P_{01}}{P_1} = 529.1 \qquad \text{Equation 3-10}$$

$$\frac{P_2}{P_1} = 7.037 \qquad \qquad Equation\ 3\text{-}11$$

$$P_2 = 7.0374 \times 1.682 = 11.84 kPa \qquad \qquad Equation\ 3\text{-}12$$

Therefore, $P_{o1}$ can be substituted into Equation 3-13 to get $P_1 = 1.682$kPa. What's more, the value for $\frac{P_2}{P_1}$ can obtained through the online Compressible Aerodynamics Calculator (William J. Devenport, 2014) and, then, static pressure $P_2$ can be calculated from Equation 3-14, which is equal to 11.84kPa.

### 3.2.2 Drag force

Historically, the term 'drag coefficient' has been used to describe fluid dynamic drag, which usually is influenced by two types of information: form drag, which is caused by the aerodynamic shape and skin friction, which means the friction between the fluid and the aerodynamic shape. The form drag is much more critical in the drag produced and the skin friction can be ignored in this project. Therefore, only form drag is taken into consideration (Gemba, 2007).

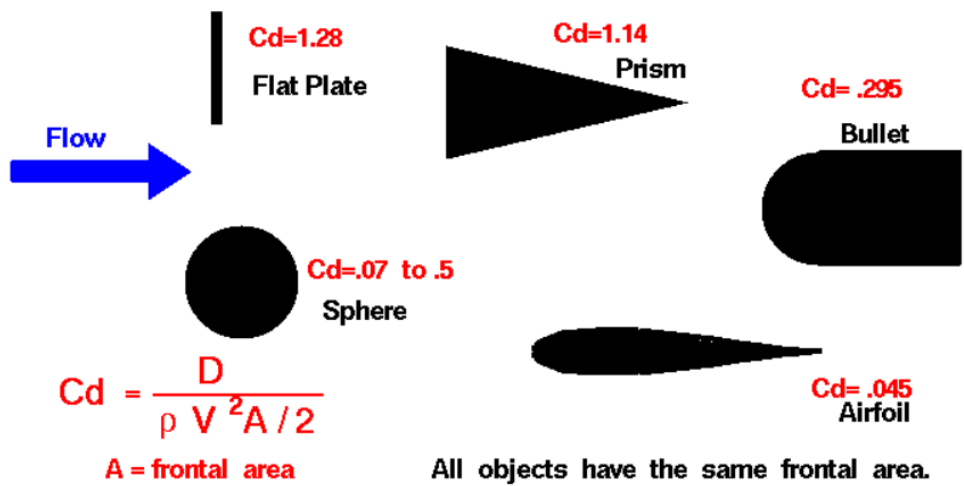

*Figure 18 Drag force coefficient for different shape (NASA)*

The drag of aerodynamic shape is related to lots of factors, such as shapes, dimensions and flow conditions, all of these factors could be represented by a dimensionless number called the drag coefficient, which defined as $C_d = \frac{F_D}{\frac{1}{2}\rho U^2_\infty A}$. $F_D$ can be derived as follows:

$$F_D = 0.5 \times C_d \times \rho \times v^2 \times A \qquad \qquad Equation\ 3\text{-}15$$

Where $F_D$ represents drag force, $\rho$ means the fluid density, $v$ is the free stream velocity, A is the front area and $C_d$ is the drag coefficient.

Generally, $C_d$ varies according to the different body shape and fluid speed, which is related to the Reynolds number. And the drag force coefficient is inversely proportional to the drag force; the drag force increases while the drag force coefficient decreases (Gemba, 2007). Figure 18 shows the drag coefficient for different aerodynamic shapes. The largest drag force coefficient is for the flat plate shape, which is 1.28, and the smallest is the airfoil, which is only 0.045.

$$F_D = F_{drag} = F_{inertia} = \frac{1}{2}\rho v^2 c_d A \qquad Equation\ 3\text{-}16$$

The velocity of airflow after the shock wave is

$$a_2 = \sqrt{\gamma R T_2} \qquad Equation\ 3\text{-}17$$

After the shock wave, the flow velocity is

$$v_2 = M_2 a_2 = M_2\sqrt{\gamma R T_2} \qquad Equation\ 3\text{-}18$$

Substitute $v_2$ and equation $PM = \rho RT$, $F_D$ can acquire from $Equation$ 3-19.

$$F_D = \frac{1}{2}\rho v^2 c_d A = \frac{1}{2} \times \frac{P}{RT} \times \left(M_2\sqrt{\gamma R T_2}\right)^2 \times c_d A = \frac{1}{2} P M_2^2 \gamma c_d A \qquad Equation\ 3\text{-}19$$

$$F_D = 0.5 \times 11.84 \text{Kpa} \times 3.022^2 \times 1.4 \times 1.28 \times 1786 \times 10^{-6}$$

$$F_D = 0.5 \times 11.84 \times 10^3 \times 3.022^2 \times 1.4 \times 1.28 \times 1786 \times 10^{-6}$$

$$F_D = 173.03N$$

According to the experimental data, the maximum drag force is around 200N, and the theoretical value is 173.03N, which is close to the experimental value.

# 4   Model supporting system design

This chapter includes concept design, concept optimisation design and detail design. Firstly, several concept designs are obtained according to the improvement of the existing design in the market, the best design is selected according to the optimisation criteria, and the final design is obtained through optimisation design. This detailed design is carried out based on the final conceptual design, which is divided into structural and electrical design and is the most critical factor to determine the mechanical performance.

## 4.1   Concept design

Concept design should be given with different ideas and then the best one should be chosen by evaluating concepts. Numerical evaluation, benchmarking and quantitative analysis can be used to evaluate the concept design.

### 4.1.1   Former design

Zhang (2019) has already done some work for this project. As shown in Figure 19, his design is based on a plane four-bar mechanism. The structure is simple and easy to assemble: D is the fixed end, a motor drives slider A, which causes slider B to move in the vertical direction. The angle between connecting linkage AB and the horizontal axis changes, which can achieve the adjustment of the angle of attack. However, it is found that the stability of the mechanism is deficient in its practical application. It can only bear a small load and the error of the adjustment angle is relatively large. Based on practical experience combined with theoretical design principles, seven concept designs are given below.

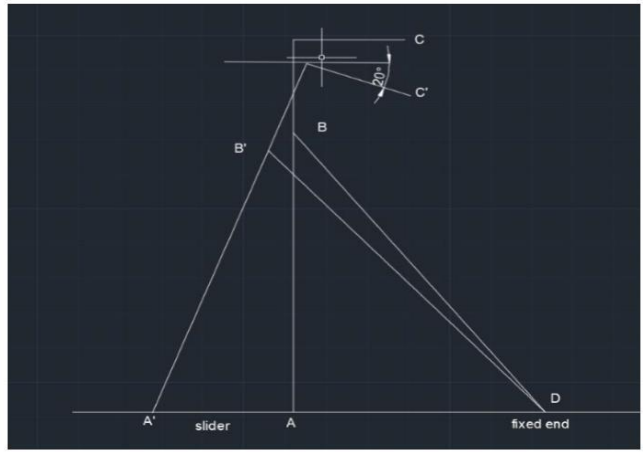

*Figure 19 Former design(Zhang, 2019)*

### 4.1.2 Concept design one

The first concept design is based on Yang (2018) design, as shown in Figure 20. The motor drives the ball screw, which converts circular motion into linear motion, and the screw nut pushes the rod to reciprocate linearly, under the restriction of the linear bearing. When the rod moves to the left, the rod connected by slider A becomes shorter and, at the same time, moves counter-clockwise; the linkage connecting the sliding block rises and the length remains unchanged.

At this time, the change of the angle of attack can be calculated according to the Pythagorean theorem by the distance that the rod moves to the left. The structure of this design is simple, the total cost is low and it also achieves the angle adjustment requirement. The disadvantages are that the stability is low and the inertial force is difficult to balance. When the mechanism performs reciprocating and plane motion, it is easy to generate dynamic loads and accumulate errors. The efficiency is low, the mechanism can only bear small loads, due to its relatively poor rigidity, and, when the wind tunnel works, the error of the actuation system adjustment angle is large.

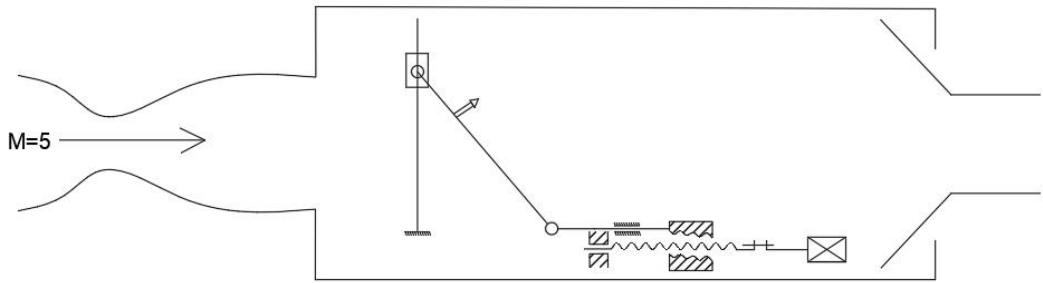

*Figure 20 Concept design one*

### 4.1.3 Concept design two

The second concept design is based on the principle of the suspended support structure. The stepper motor is connected to the coupling and the ball screw to push the trapezoidal slider at the bottom. At the same time, the angle design can be adjusted by the length of the support rod and the angle of the trapezoidal slider, the slider is mounted on a linear guide, the bottom is supported by a column, the linear bearing restricts the connecting rod only to move up and down, which can achieve the angle adjustment.

This design uses a symmetrical structure, which is compact and relatively rigid, so the structure is relatively stable and can bear a large load. What's more, the left and right two connecting rods adjust the angle to ensure there is no gap between the upper plate and the connecting rod, therefore, the error of angle adjustment is small when the wind tunnel works.

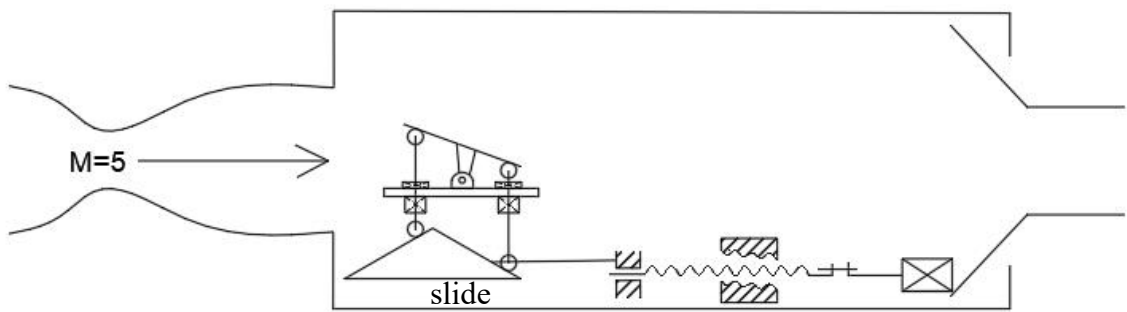

*Figure 21 Concept design two*

### 4.1.4 Concept design three

Concept three applied the gear and rack structure. The design principle is that the motor drives the ball screw motion, the ball screw converts the rotation of the motor into linear motion and uses the rack and gear to convert the linear reciprocating motion into the pitch angle change of the model. Specifically, the rack reciprocates on the sliding track. The reciprocating motion of the rack drives the gear to rotate, matching the shaft with the gear hole to realise the change of the pitch angle of the model.

The advantage of this design is the accurate instantaneous transmission ratio, good transmission efficiency, long service life, and the simple structure. The shortcoming is that the stability of this form is based on the precision of the rack and gear. If the accuracy is not high enough, the actuation system will have a certain error when adjusting the angle. Therefore, the method for eliminating the errors of such parts is to increase the accuracy of the rack and gear and improve the wear resistance and its heat treatment on the material. The transmission error of the high-precision rack and pinion mechanism can reach 0.1mm, but the cost is very high.

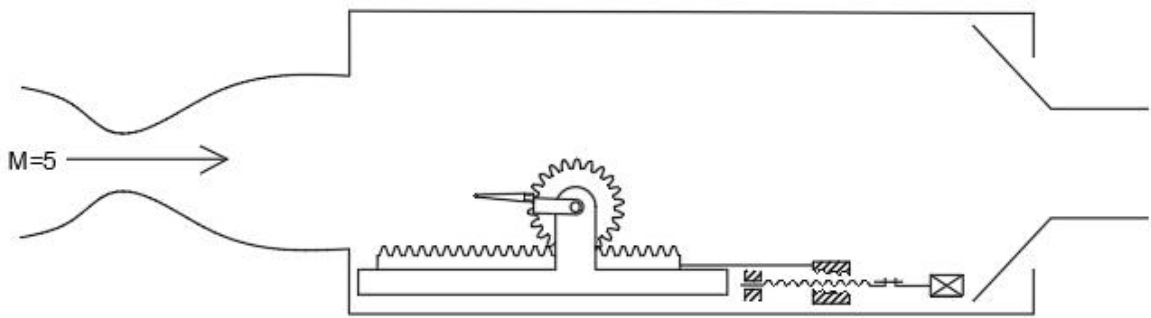

*Figure 22 Concept design three*

### 4.1.5 Concept design four

According to the similar working principle of the goniometer, the concept four is given below. The system is composed of the track, column A, linkage, semi-circular sliders A and semi-circular sliders B. Semi-circular slider B is fixed on the ground, rail is fixed on linkage, column A is connected to semi-circular slider A by a hinge, and the track can constrain column A to move up and down.

The transmission principle is that the motor drives the screw and the screw nut is welded to the linkage, which pushes the track that connected to column A. Column A can move up and down along the track. When the motor rotates clockwise, the screw nut moves to the right, thereby pushing the track and column A to the right. At the same time, because column A is connected to semi-circular slider A by a hinge, semi-circular slider A rotates as column A moves to the right and upward. The model is installed on semi-circular slider A. When the motor rotates in reverse, column A moves to the left and downward and the semi-smooth block rotates in the opposite direction. Thus, adjusting the model angle can be realised by this design.

The advantages of this concept design are its simple structure, the fact that the researchers can more intuitively see the rotation angle and the fact that it has few components. Therefore, there are fewer parts to be manufactured and the cost is relatively low. As for its disadvantages, the vertical degree of the track needs to be ensured, since this will affect the accuracy of the angle of attack of the model. Although this error can also be corrected after the experiment, this will increase friction and shorten the lifetime of the column. Secondly, the linkage and track should be welded together.

The semicircle slider A slides on the semicircle slider B, so the coincidence degree of contact surface should be guaranteed and the friction coefficient should be controlled. Another thing to be noted is that the projected area of this design in the wind tunnel is relatively large.

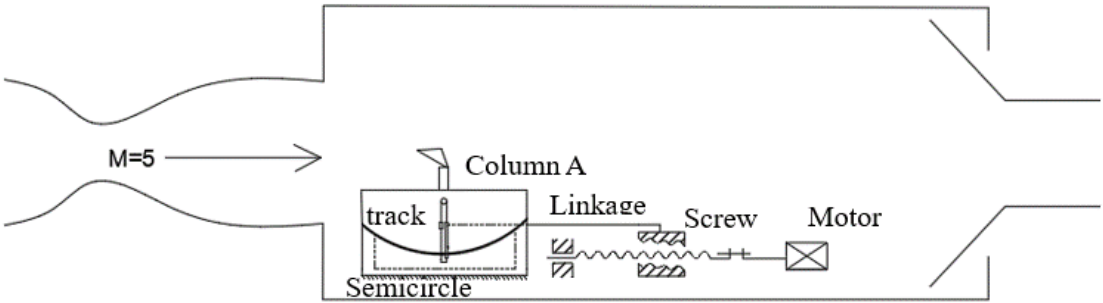

*Figure 23 Concept design four*

### 4.1.6 Concept design five

The principle of this design is based on the Stewart platform, which usually has six actuators to control the pose of the platform. In this design, the two motors drive their respective screws; the screw nut pushes linkage A and linkage B, which can make the angle of the platform change. If the two motors rotate in opposite directions with the same rotational speed, the platform can move to the left and right; when the direction and the speed of these two motors change, the angle of the platform will be adjusted. The model is installed on the platform, which can realise the change of pitching angle. This concept design involves two motors; compared with the motor-driven system, there is an additional degree of freedom and the structure is simple.

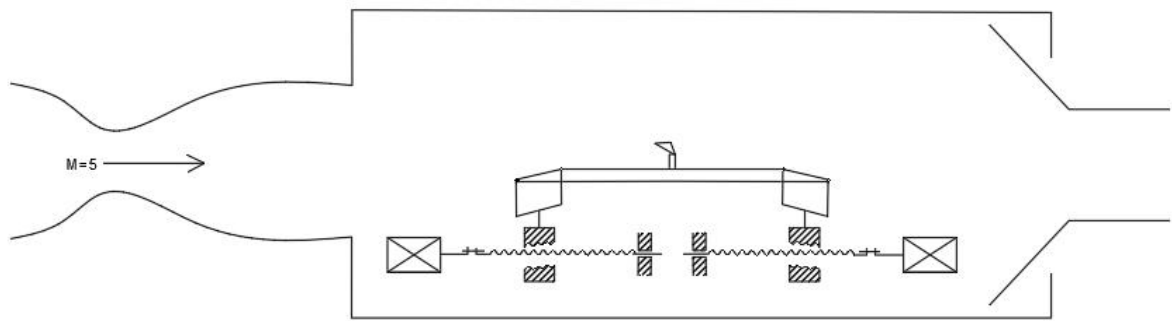

*Figure 24 Concept design five*

However, the required space for this device is large and the cost is high, what's more, the control part needs to be very accurate. Because the stepper motor may lose steps in the working condition, in this situation, the error caused by two motors will be greater. At the same time, the closed-loop control of two stepper motors is required, which requires more expenses.

### 4.1.7 Concept design six

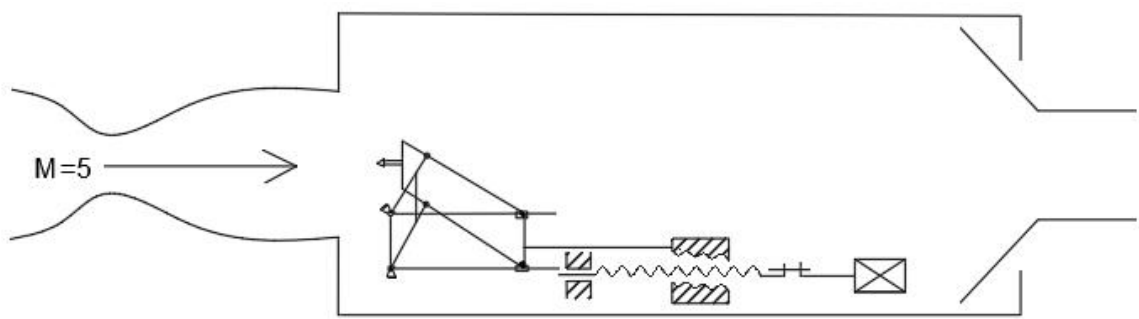

*Figure 25 Concept design six*

The transmission principle of this design is that the motor drives the screw, the screw nut pushes rod A in parallel motion. Rod B is fixed and, when rod A moves, the angle of attack changes. The strengths of the design are its simple structure, low cost, and relative stability. However, when the mechanism works in the wind tunnel, the structural strength of the hinge joints needs to be checked during the process of adjusting the angle.

### 4.1.8 Design selection criteria

1) Design for manufacturing (DFM)

The principle of designing for manufacturing is conducive to subsequent manufacturing(Tien-Chien chang, 1998). During the design process, manufacturing and processing need to be considered, for example, the difficulty of manufacturing parts and manufacturing details, while considering the economic cost of manufacturing and processing methods to choose the most optimised design.

2) Design for assembly (DFA)

Design for assembly (DFA) is a process that makes the product easy to assemble during design. If the product has a large number of parts, it will take more time to assemble, which will increase the cost of assembly (Tien-Chien chang, 1998). In addition, if the parts are easy to assemble, such as being easy to grasp, move, position and fit, etc., it will reduce the difficulty of assembly.

3)Design for reliability and maintainability

Design for reliability and maintainability includes the reliability of machines and procedures, while taking into account the external environment in which the system is used, making sure that the system meets the reliability requirements during the life cycle. At the same time, periodic inspections should be carried out to ensure the stability of the system and it is necessary to ensure that the entire system and its parts are easy to maintain (Taylor, 2014).

According to these criteria, a comparison table can be constructed. The evaluation of concept design involves lots of factors, such as material, structure, mechanisms, reliability, ergonomics and shape/size. As shown in Table 5, concept three is the most suitable choice, the supporting system is relatively stable and the cost is low. The low matching error can increase the dynamic stiffness; therefore, the rigidity of the mechanism is better. Obviously, concept three is the best choice among the three concept designs. The first design is easy to manufacture and assemble and the cost is relatively low, but the stability of the supporting systems is relatively low, due to its poor reliability. The four-bar linkage makes transmission errors large during the process of adjusting the angle. The rack and pinion structure of the third design has a higher installation accuracy and a larger return error.

*Table 5 Comparision for different concept design*

|  | Concept one | Concept two | Concept three | Concept four | Concept five | Concept six | Concept seven |
|---|---|---|---|---|---|---|---|
| Cost | 3 | 2 | 1 | 2 | 1 | 2 | 1 |
| Reliability | 0 | 3 | 2 | 2 | 2 | 1 | 3 |
| Manufacturing | 2 | 3 | 1 | 1 | 3 | 2 | 1 |
| Assembly | 2 | 2 | 3 | 1 | 2 | 2 | 1 |
| Maintenance | 1 | 3 | 2 | 2 | 1 | 2 | 2 |
| Total mark | 8 | 13 | 9 | 8 | 9 | 9 | 8 |

## 4.2  Design improvements

### 4.2.1  Fault tree analysis

System fault tree analysis (FTA) is a deductive failure analysis technique that can be used to specify the root cause of the failure of a system(Kabir, 2018). It uses Boolean logic to combine a series of lower-level events. In this section, FTA is carried out to help unravel and support the design process, as well as assess any potential risk.

Although this design can meet the requirements of this project, there are still many areas that can be improved, such as too many transmission components, leading to large friction resistance and transmission errors. Secondly, due to the influence of wind during the transmission process, the strength of the model interface which is connected to the aerodynamic shape needs to be verified.

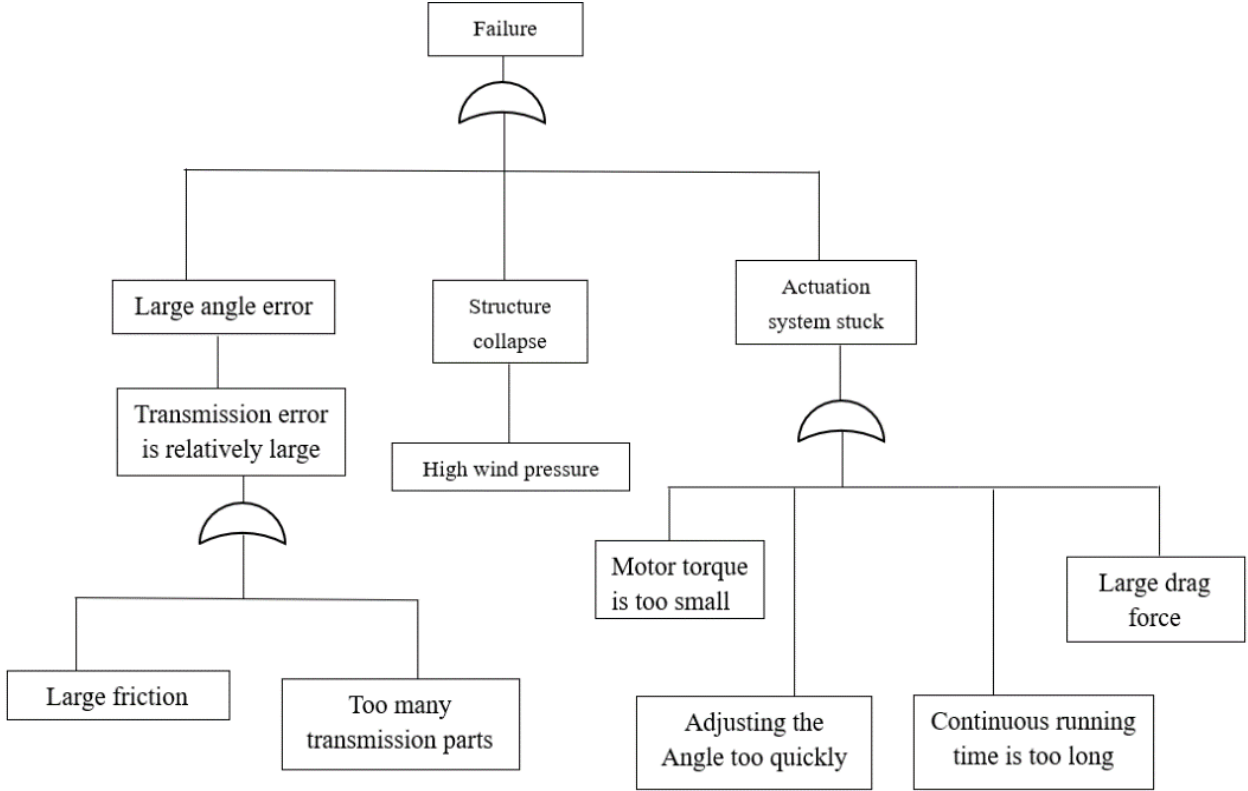

*Figure 26 System fault tree analysis*

The output parameter of an actuation system is the angle of attack of the aerodynamic shape. So, it is a failure condition that the output angle error is out of the permitted range. Besides this, the structural collapse of the supporting parts or the whole actuation system is also a severe failure. Based on these statements, the FTA of our original design was performed. Figure 26 is an illustration of the FTA.

### 4.2.2 Failure modes and effects analysis (FMEA)

Failure modes and effects analysis is a systematic and proactive method for determining a process to identify where and how the design might fail and then assess the relative impact of each different failures. FEMA consists of parts in the device, failure modes, failure causes and failure effects (Kapil Dev Sharma1, 2018). It is performed based on the illustration in the FTA section.

*Table 6 Failure modes and effects analysis*

| Parts | Failure mode | Failure causes | Failure effects | LoC | S | Action to reduce the occurrence of failure |
|---|---|---|---|---|---|---|
| motor | Stuck | Small holding torque and overlong running time | The actuation system cannot move | 4 | 4 | Select the motor with high torque and control the running time of the motor |
| Supporting rod | The four supporting rods are not adjusted simultaneously | Wind force and assembly error | Error in the angle of attack adjustment | 3 | 4 | Reduce the for supporting rod to two to reduce the contact components |
| Soft Switch limit | Broken | Power off | The actuation system cannot stop | 6 | 8 | Adapts hard switch limits |
| Triangular slider | Friction force and transmission error | Triangular slider slides on the floor rails and also work as a slippery to drive the supporting rod | A large error of angle of attack and actuation system stuck | 4 | 5 | Cancel the floor rail and fixed the triangular slider to the screw nut |
| Upper supporting system | Collapse | Structure strength is not enough under the strong wind force | The upper supporting system is broken | 4 | 8 | Do finite element analysis to find the maximum deflection |

Where LoC refers to the likelihood of occurrence, and S refers to the severity

### 4.2.3 Final design

The working environment of this project is a supersonic wind tunnel. At the same time, the size of the motor is also limited due to the overall size of the wind tunnel. Due to the axial and radial forces that the motor will receive in the wind tunnel, the motor should be placed below the nozzle of the wind tunnel as much as possible. To prevent the motor from stepping out or being damaged, the motor with brake should be selected to prevent the motor from overstepping or stepping out when the angle is adjusted to the desired position.

The old design adopts a symmetrical structure and there is a roller slide on each of the four tracks of the trapezoidal block. A trapezoidal slider slides on the track mounted on the floor, which leads to too many sliding-contact components. The screw nut pushes the connecting rod and the connecting rod pushes the trapezoid block to slide on the track.

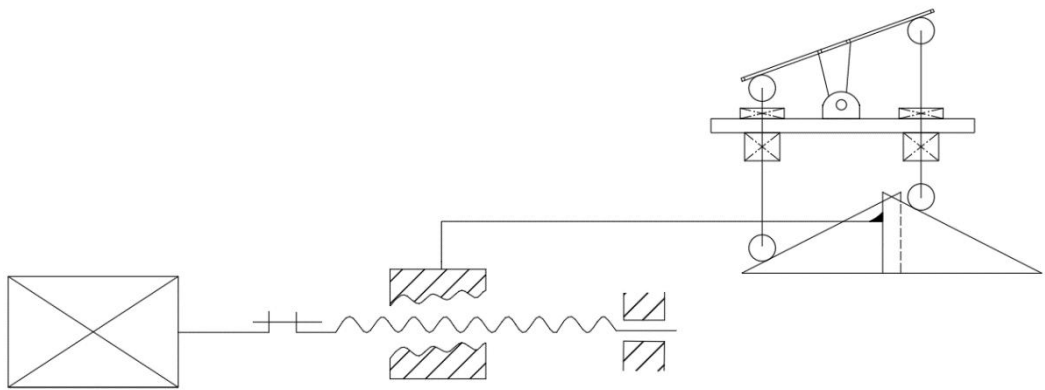

*Figure 27 Final concept design*

The improved design removed two of the four rollers on the diagonal and changing the trapezoidal slider into a triangular slider, which can reduce the sliding contacts. At the same time, the trapezoid block was fixed to the connecting rod, which is driven by the screw nut, therefore, the screw nut, connection rod and trapezoid block are connected together. When the stepper motor drives the screw, the trapezoid block is driven by the screw nut, thus, the track under the trapezoid block can be removed from the design, which reduces the two sliding contacts and friction force caused by the track.

## 4.3  Detailed design of the supporting system

In detailed design, the dimensions of each component are determined by several constraints: limited working space, the limited front area in the working zone, drag force caused by Mach 5 supersonic flow and material properties, etc.

## 4.3.1 Structural design

The first design step of the detailed design is the height design. With the size limitations of the actuation system, the dimensions of the supporting system must be determined first. What's more, as shown in the final concept design, the key part of achieving the angle adjustment is the triangular block, therefore, the height and the base angle should be determined according to the desired limited angle. After this, the upper supporting mechanism and model interface should be designed, and the final 3D model is in appendix Figure 78.

### 4.3.1.1 Height design

The stepper motor will be subjected to forces in three directions in the wind tunnel. However, the motor cannot directly bear the axial force, because the two end covers of the stepper motor are only connected by three bolts and these are deep groove ball bearings, which can only bear small axial loads. Therefore the motor should be mounted under the nozzle, that is $H_m<H_n$ and at the same time, when the model is adjusted to the limit angle, that is, $\pm 20$ degrees, the position of the model needs to be in the area of the nozzle, so the height of the model $H_t$ at the extreme position needs to be less than the height of the top of the nozzle $H_a$.

According to the size of the wind tunnel, the length of the actuation system $L_a$ needs to be less than $L_t$ 575mm and the width $W_a$ should be less than $W_t$ 348mm. According to the restraint condition, the following basic formula can be derived, $L_a = 454mm$, $W_a = 229.5mm$, $H_m = 57.1mm$ , $H_a = 229.46mm$, this is shown in Figure 28.

$$L_a < L_t = 575mm \qquad \text{Equation 4-1}$$

$$95mm < W_a < W_t = 348mm \qquad \text{Equation 4-2}$$

$$H_m < H_n = 95mm \qquad \text{Equation 4-3}$$

$$H_a < H_t = 315mm \qquad \text{Equation 4-4}$$

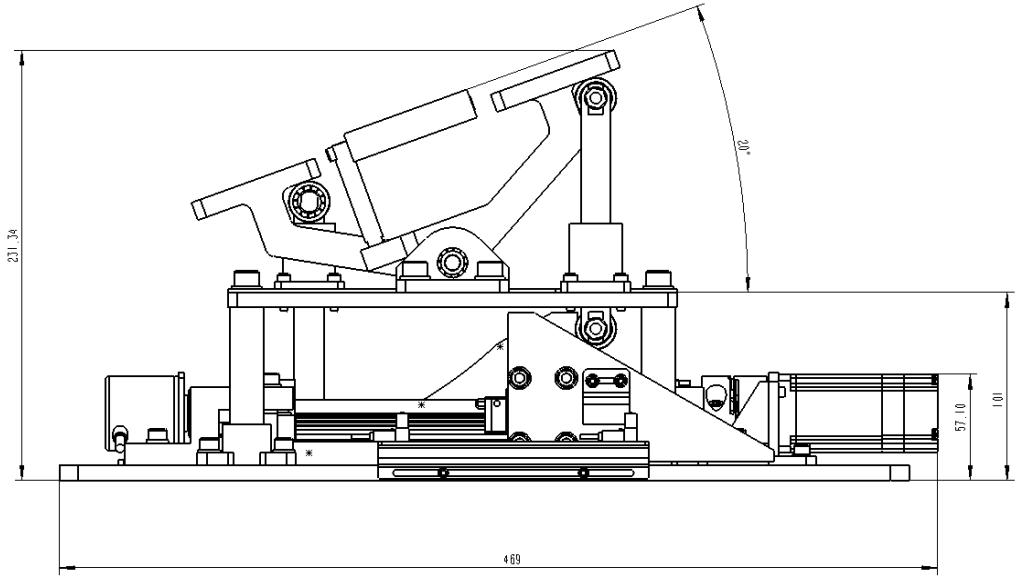

*Figure 28 Height design for the whole system*

### 4.3.1.2 Design of triangular block

The blockage ratio can be reduced by reducing the front area exposed in the nozzle in the wind tunnel. The height needs to be smaller than the height of the middle plate. Because of the limitation of the size of the actuation system, the angle α can be modified by the height and length of the triangle. However, the angle α should not be too large, because this will cause the triangle block to have a steep slope, causing the angle of attack to change too fast. If the angle α is too small, the actuation system fails to meet the demand of 20° degrees, therefore assume angle α is 30°.

$$tan α = \frac{h}{L} \qquad Equation\ 4\text{-}5$$

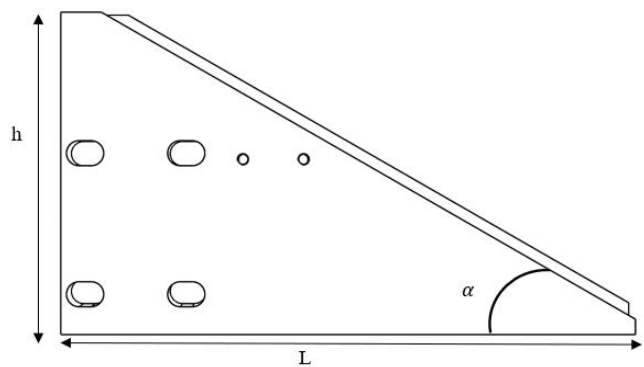

*Figure 29 Triangular block*

### 4.3.1.3 Upper supporting mechanism design

As shown in Figure 30, the picture on the right is the old design, which adopts symmetric structure, and with four supporting rods with the rollers at four corners. In the improved design, there are reduced two supporting rollers. In the beginning, the aerodynamic shape was mounted on the top of the plate, however, when the actuation system moves to the limit position,($\pm 20$ degrees), the aerodynamic shape will exceed the top of the nozzle, so the mounting rod of the aerodynamic shape needs to move down, which is shown in Figure 30.

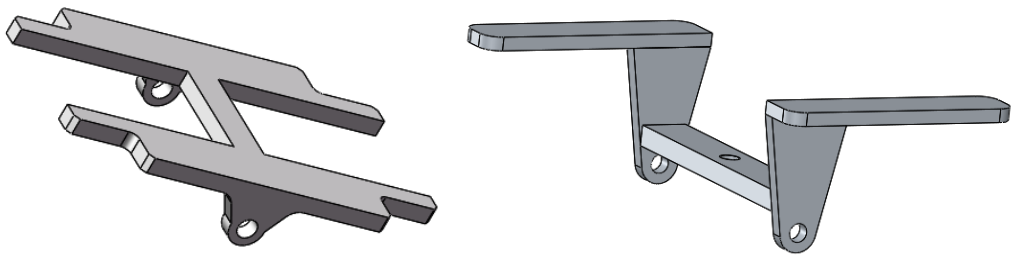

*Figure 30 Original upper supporting mechanism design*

As for the problem that there will be a big board on the side to block the model, based on the former design, the final design is given in Figure 31. The developed design changes the structure of the side plate to make the model higher than the side link so that the light can go through.

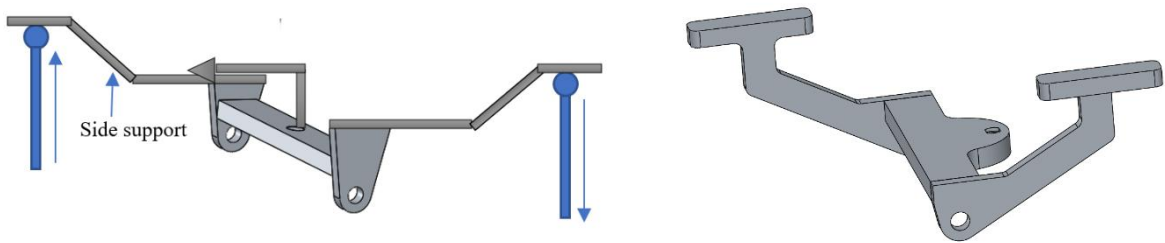

*Figure 31 Final upper supporting plate design*

### 4.3.1.4 Software overtravel limit switches

There is a metal piece connected to the triangle block, which is in the groove of the photoelectric switch. During formal work, the metal piece will not enter the groove of the photoelectric switch. When the work is completed and reset, the metal piece will automatically enter in the photoelectric switch groove.

As an initial sensor, a hall switch is installed as a position sensor at the zero position. When the stepper motor returns to the zero position, the sensor gives a detection signal. When the control circuit detects this signal, the motor stops at the zero position. This method of 'homing' is accurate and reliable, but it increases the complexity of the circuit and has certain requirements for installation.

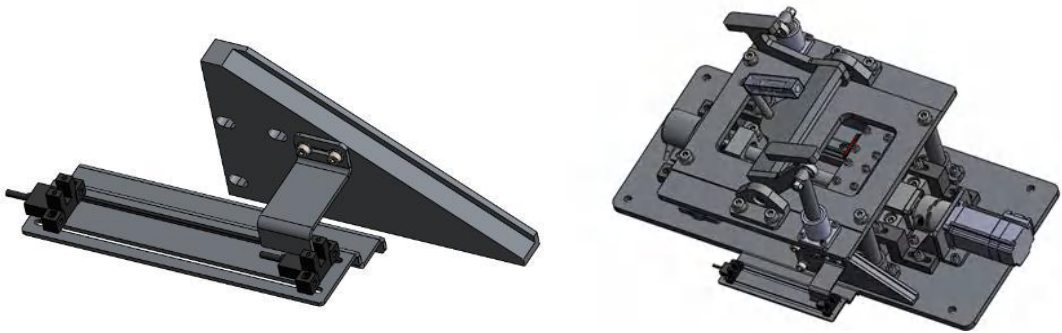

*Figure 32 Software overtravel limit switches*

### 4.3.1.5 Hardware overtravel limit switches

Hardware overtravel limit switches use machined parts to limit the position of the device. Generally, these should be used in combination with software overtravel limit switches. The software overtravel limit switch needs to be set in front of the hardware limit switch. Under the normal operation of the device, the position of the moving parts should be within the area set by the soft limit switch; the hardware overtravel limit switch does not work when the software limit switch is in good condition. Only when the software limit switch fails does the hardware limit switch start the limit position protection and the motion system fault alarm effect.

The hardware limit switch prevents the actuation system from having a workbench collision, which is caused by the failure of the stepping motor or the software limit switch. Secondly, it acts as power-off protection. When the drive system moves into the limit position and the power goes off suddenly, because of inertia, the motor will not stop immediately. However, the photoelectric switch cannot stop the stepper motor, which will drive the actuation systems to exceed the limit position and cause the damage to it. In case of power failure, the hardware limit switch can interrupt the movement of the actuation system until the stepper motor stops moving, which can prevent a collision.

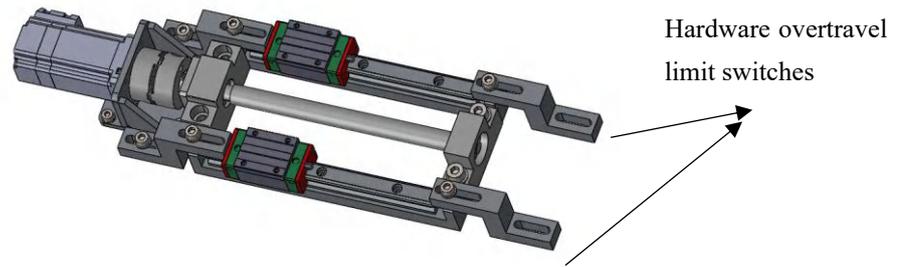

*Figure 33 Hardware overtravel limit switches*

### 4.3.1.6 Model interface

The mounting hole is a hole designed for part installation. The mounting hole of the model of this project can be connected with M4 thread or keyway. The design is as shown in *Table 7*, which used an M4 screw to connect with the aerodynamic shape. The maximum deflection is around 0.002mm under 200N force.

*Table 7 FEA for connecting rod*

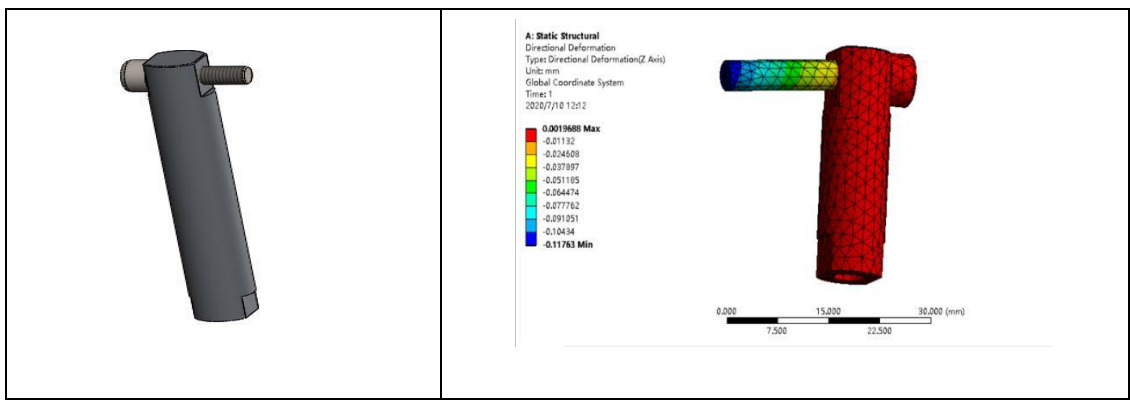

## 4.3.2 Electrical systems

### 4.3.2.1 Selection of motors

There are three ways of actuation for the supporting system, according to the literature review, which are hydraulic, pneumatic and electric. The electric is the best choice and can use either a DC, Stepper or Servo motor. There is no speed requirement when adjusting the angle: the most suitable motor is selected by comparing the advantages and disadvantages of these different types of motors and combining the project requirements.

The brushless DC motor is controlled by an alternating power supply provided by hall element positioning. The stepper motor is driven directly by a single pulse voltage, which does not require the positioning of the hall element. It can precisely locate the rotation angle by controlling the number of pulses to the motor.

A stepper motor is an open-loop control motor, which is a good choice for many occasions where precise positioning is required. It can be loosely described as a DC motor that can rotate at a precise increment or step. Another strength a stepper motor has over a DC motor is the ability to move workpieces very slowly. However, its weakness is that stepper motors easily skip steps: because of the low top-speeds of the motor, this is called the lost steps phenomenon.

Servo motors are motors that can perform very precise motion control. The servo motor system is a closed-loop control motor. It can feedback the difference between actual and expected position or speed to the controller, which can adjust the output to correct any drift that deviates from the target position. Compared with stepper motor, a servo motor knows its position and can rotate to the desired angle, even if the motor shaft is moved by an external force. But the servo motor is really expensive: almost ten times as much as the stepper motor.

Therefore, the best choice is combining a stepper motor with a rotary encoder and hall switch to achieve the closed-loop control, which achieves the accuracy requirements and position control. In addition, the cost is much less than the expense of servo motors.

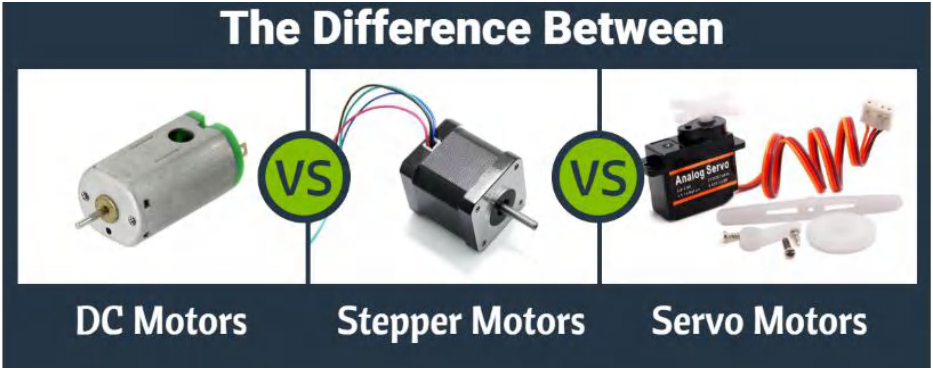

*Figure 34 Comparison of different types of motor(from seeed)*

### 4.3.2.2 Selection of the stepper motor

There are three kinds of stepper motor: variable reluctance, permanent magnet and hybrid.

For variable reluctance stepper motors, the stator has windings and the rotor is composed of soft magnetic materials. The structure is simple and the cost low and the smallest step angle is 1.2°. However, the dynamic performance of variable reluctance motors is poor. It also has low efficiency and large motor heat, therefore, it's hard to guarantee its reliability.

The rotor of the permanent magnet stepper motor is made of permanent magnetic material and the number of poles of the rotor is the same as that of the stator. The advantages of this motor are good dynamic performance and large output torque; the shortcomings are poor precision and the step angle is large (generally 7.5 or 15 degrees). This could cause errors in the process of angle adjustment.

Hybrid stepping motors combine the strengths of variable reluctance stepping motors and permanent magnet stepping motors. There are multiple small teeth on the rotor and stator to improve the stepping accuracy, the rotor is made of permanent magnet material and the stator is multi-phase winding. The advantages of this type of motor are large output torque, good dynamic performance, small step angle and good reliability. The disadvantages are that the structure is complex and the cost is relatively high. Hybrid stepping motors are the most commonly used stepping motors on the market: they combine the low noise of permanent magnet stepping motors with the high resolution of reactive stepping motors.

For this actuation system, since it is working in a supersonic wind tunnel, the stability of the motor needs to be guaranteed. The accuracy and stability of the hybrid stepper motor are pretty good. The large step angle leads to adjustment errors and the price of the hybrid stepper motor is acceptable, so the hybrid stepper motor was selected.

### 4.3.2.3 Transmission ways for stepper motor

There are two ways to drive the motor, one is direct rotation and the other is the indirect transmission. The indirect transmission is belt, screw or simple gear to change the transmission ratio and transmission direction. Direct drive is a direct connection between the shaft of the motor and the shaft of the production machine. Therefore, direct transmission should be a better choice for machines without speed and transmission direction requirement, if the speed ratio needs to be adjusted in the transmission, other transmission modes should be selected.

In this project, there is no requirement for the movement speed of the machine but the rotary motion of the stepper motor needs to be changed into a linear motion, there are mainly three ways to change transmission direction, Rack and pinion, synchronous belt and lead screw, generally speaking, the screw rod is used for accurate transmission, if maintenance requirements are not high, it is better to use belt drive, therefore, the leas screw is the best choice.

In general, the applied force is 50N and the maximum is 200N, which means the required motor torque is small, the driving method of the motor is shown in Figure 35. The stepping motor, coupling and ball screw connection are used to convert the rotary motion into a straight line.

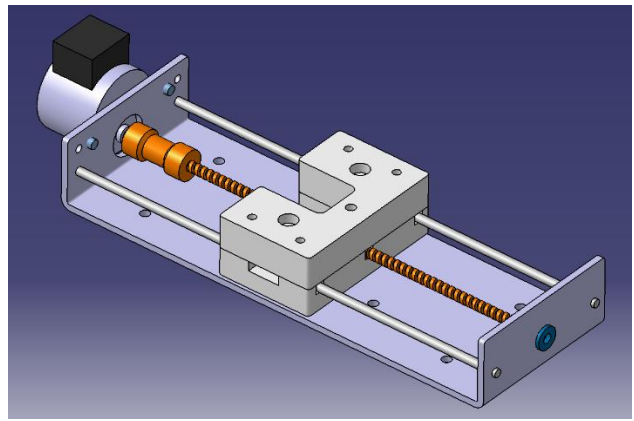

*Figure 35 Transmission structure*

### 4.3.2.4 The motor torque calculation

Since the motor works in a supersonic wind tunnel, the torque of the motor needs to be considered when selecting the motor. They can generate the holding torque required to drive the models to the desired angle when working in a supersonic wind tunnel. The motor torque should be large enough so that the motor can overcome the resistance torque of the load and start quickly and the resistance can be calculated through the formula below:

$$F_a = (G + F_{drag})\mu \qquad \text{Equation 4-6}$$

Where $F_a$ is the total axial drag force for the model, $\mu$ means the coefficient of friction and

$G$ represents the maximum weight of the aerodynamic shape can be moved

In this project, the system slides on the guide, according to the user manual, the coefficient of friction for a linear guideway is equal to $1/50$, the theoretical friction coefficient of the rolling elements in the ball guide is about $0.003 - 0.005$. After assembly, the coefficient of friction is about $0.01 - 0.05$ due to the weight of the workpiece and the preload.

Due to the use of the slider and the impact of installation, the friction coefficient is theoretically between $0.01$ and $0.02$. However, in practical applications, due to the parallelism of the installation, the pretension used to eliminate the gap and the fact that the curve of the ball returner is distorted (reflected at high speed), factors such as the consistency of the inner and outer raceways are too uncontrollable. It is recommended to calculate by using the value $0.15$. Therefore, the total axial drag force for the model $F_a$ could be calculated by $Equation$ 4-7.

$$F_a = 0.15 \times (7 * 9.81 + 173.03) = 36.255N \qquad Equation\ 4\text{-}8$$

The choice of motor installation specifications is mainly related to torque requirements. The required holding torque generated from electric motors is determined by the loading condition acting on the supporting system and the gravity of model and wind tunnel balance.

The maximum thrust required when the motor is running is during acceleration. This project does not require acceleration when adjusting the angle. Therefore, the maximum torque required by the motor should be during the starting and stopping of the motor. It is necessary to accelerate the motor smoothly as much as possible. The driving torque at constant speed is:

$$T_a = \frac{(F_a \times I)}{(2 \times 3.14 \times n_1)} \qquad Equation\ 4\text{-}9$$

Where $T_a$ means holding torque of the stepper motor, $F_a$ means axial load (N), which is equal to the axial friction force, $F$ is the axial cutting force of screw (N), $\mu$ represents comprehensive friction coefficient of the guide, $m$ is the moving object weight kg, $g: 9.8$, $I$ means Lead screw (mm), and $n_1$ represents Positive efficiency of the feed screw.

According to Equation 4-7 and Equation 4-8, assume $n_1 = 0.94$, therefore $T_a =$

$\frac{(36.255\times5)}{(2\times3.14\times0.94)} \approx 30.71 N.mm \approx 0.031 N.M$. Assume the maximum coefficient of friction is 1, the maximum drag force including the dead weight is 241.7N, which is consistent with the experimental maximum drag force that is around 200N, and the motor power can be selected, generally speaking, 1.5 times the theoretical torque required

$$T_{amax} = 241.7 * 5/5.9032 \approx 204.7 N.\text{mm} = 0.2047 N.M$$

After getting this value, the holding torque can be determined, according to engineering experience, the safety factor is generally 2, therefore, the holding torque required for the chosen stepper motor is given below:

$$T_a = 1.5 \times 0.2047 N.M = 0.307 N.M$$

For this project, the requirement of the stepper motors is that they can generate the holding torque required to stop the models from moving in the middle of a run. And the required torque is 0.254 N.M, at the same time, in order to ensure the accuracy of angle adjustment, the step angle of the system cannot be large, and the motor type selected is RS PRO Hybrid, Permanent Magnet Stepper Motor, 42SH47.

*Table 8 Specification of 42SH38-1A stepper motor*

| Properties | Value |
| --- | --- |
| Rated Voltage | 12 |
| Number of wires | 4 |
| Step angle | 1.8 |
| Holding torque | 0.44 |
| Stepper motor type | Hybrid |
| weight | 0.28 |
| Number of leads | 4 |
| Winding arrangement | Bipolar |
| Current rating | 600mA |

# 5 Wind tunnel balances

To decompose the forces and moments into different directions, the measuring element structure of the wind tunnel balance is consists of axial force element, compound combination element and rolling moment element.

## 5.1 Typical rod balance measuring element structure

### 5.1.1 Axial force element

For strain balances, the axial force is the most challenging load component to measure. On the one hand, to perform a high angle of attack force test, the effect of the model weight on the axial force element must be considered. Therefore, axial force elements are required to have higher rigidity and a more comprehensive range. On the other hand, to improve the sensitivity of axial force measurement, it is necessary to reduce the stiffness of the axial force component of the balance. At the same time, it's required to be insensitive to the effects of other component loads to reduce the interference of other components with the axial force. Therefore, there is a contradiction between the sensitivity and stiffness of the balance. How to solve this contradiction is the key to the design of the axial force element (Boutemedjet et al., 2018).

The axial force element is an independent balance element, generally composed of a measuring element and a supporting piece, which transmits force through a parallelogram structure. Both the measuring element and the supporting sheet are elastic elements and the strain gauge is pasted on the measuring element. The function for support piece is given below:
1. Plays a supporting role and improve the stiffness of the balance;
2. Plays a disturbance-removing role and reduce the interference of the balance;
3. Plays the role of force transmission and transmits the aerodynamic load acting on the model to the measuring element.

Sometimes, the strain gauge can also be directly pasted on the support sheet to act as a measuring element to increase the stiffness of the balance. (Heidelberg, 2011)

There are many structural forms of axial force elements, including tension and compression beam, horizontal beam, eccentric beam, cantilever beam and vertical beam types (Figure 36). Generally, the axial force element mainly adopts the vertical beam type. This type can be provided with more support pieces on the opposite side of the parallelogram structure, which is beneficial to improving the measurement accuracy of the axial force (Heidelberg, 2011). According to the shape and setting method of the vertical beam, there are many different forms of the vertical beam-type axial force element. The commonly used ones are the 'I'-shaped and 'T'-shaped vertical beam-type axial force elements, which can be seen in Figure 37.

5.1.1.1 **The structural form of vertical beam type axial force element**

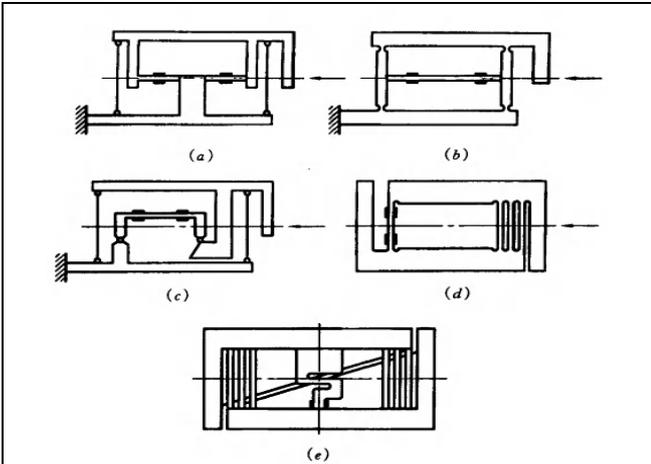

*Figure 36 Structural form of axial force element(Dexin, 2001)*

(a) Tension and compression beam;(b)Horizontal beam;(c)Eccentric beam;(d)The vertical beam;(e)The cantilever beam

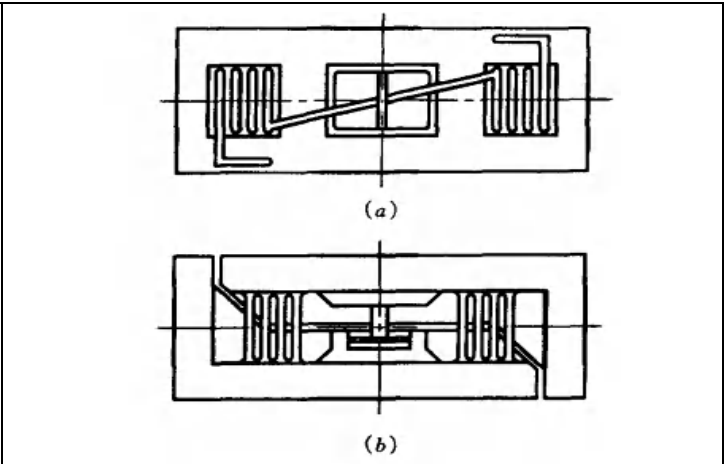

*Figure 37 Structural form of vertical beam type axial force element(Dexin, 2001)*

(a)'I'shape; (b)'T' shape

a. 'I'-shaped vertical beam-type axial force element

Figure 37 (a) is a typical 'I'-shaped vertical beam-type axial force element structure. The measuring element is composed of two vertical beams, which are symmetrically arranged on both sides of the longitudinal symmetry plane of balance at the centre of the balanced design. The supporting piece consists of 12 vertical beams, where three are connected in parallel and arranged symmetrically on both sides of the plane of balance and before and after the centre of the balance. Both the measuring element and the supporting piece can be simplified into a statically indeterminate beam, which exhibits a double bending deformation (also known as an S-shaped deformation) under the action of axial force.

The advantage of the axial force element of this structure is its good rigidity, which can reduce the interference of the normal force and the pitching moment to the axial force and is suitable for rod balances with a wide range of sizes.

b. 'T'-shaped vertical beam-type axial force element

Figure 37(b) is a typical 'T'-shaped vertical beam-type axial force element structure. The measuring element is composed of two vertical beams, which are symmetrically arranged on both sides of the longitudinal symmetry plane of balance. The supporting piece consists of 16 vertical beams, where four beams are connected in parallel and are arranged symmetrically on both sides of the plane of balance and before and after the centre of the balance. The measuring element can be simplified into a cantilever beam, which exhibits single bending deformation under the action of axial force; the support piece can be simplified into a statically indeterminate beam, which shows a double bending deformation under the effect of axial force.

The advantages of the axial force element of this structure are substantial rigidity and little interference, which are suitable for rod balances with larger dimensions.

## 5.1.2 Compound combination element

Normal force element, pitching moment element, transverse force element and yaw moment element can be combined to form a composite combination element, which is set symmetrically before and after the design centre of the balance.

According to the relative size between the thickness and width of the elastic beam of the measuring element, the structural form of the composite element can be divided into the split beam and column beam types. Among the split-beam types, generally, there are two split-beam types and three split-beam types (Miguel A. González, 2011). The column beam-type typically has the single column beam type, the three-column beam type, the four-column beam type and the multi-column beam type.

When the combined element is symmetrically arranged before and after the balance design centre, whether it is a split beam or a column beam, the measuring element can be simplified into a cantilever beam under the action of force and moment and the measuring principle for each component is the single bending deformation.

When the combination element is set in the centre of balance design, the measuring element can be simplified into a statically indeterminate beam under the action of the force. And the force measurement can be made by using the principle of double bending deformation. At the same time, under the effect of the moment, the measuring element can be simplified into a cantilever beam and the moment measurement can be made by using the principle of single bending deformation.

a. Split-beam combination element

Figure 38(a) and Figure 38(b) are typical two-piece beam type and three-piece beam-type combination elements, which are generally used to measure the normal force and pitching moment. The three-piece beam-type combination element is generally used to measure the normal force with an intermediate beam and the pitch moment with two beams set symmetrically on the top and bottom.

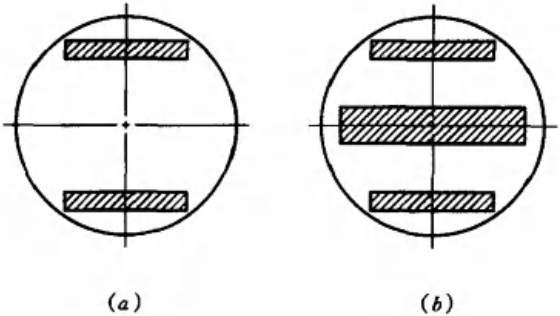

*Figure 38 The structural form of a split-beam combination element(Dexin, 2001)*

(a) two beams ;(b) three beams

The rigidity of the three-beam composite element in the normal direction is weak, but the ability to resist the interference of the pitching moment $M_y$ is greatly increased. Because the strain gauge that measures the normal force $F_x$ is attached to the middle beam of the three beams, the strain produced by the pitching moment $M_y$ on the beam is very small, when the beam type composite element is set vertically, the same can be used to measure the lateral force $F_y$ and the yaw moment $M_x$.

b. Column-beam composite element

Figure 39(a) is a single-column beam-type composite element (also known as a rectangular cross-section beam-type composite element). A column beam is used to measure the normal force $F_x$, lateral force $F_y$, pitching moment $M_y$ and yaw moment $M_x$. The advantages of this composite element structure are simple structure, high rigidity, easy processing and high sensitivity; the disadvantage is that it is difficult to match the load during design.

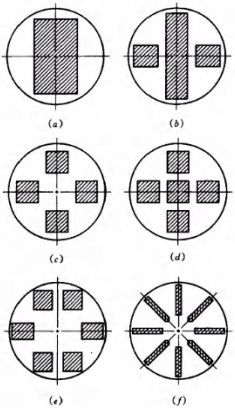

*Figure 39 The structural form of a column - beam combination element(Dexin, 2001)*

(a) Single column beam;(b)Three beam;(c)Four pillars beam

(d)Five beam;(e)Six column beam;(f)M beam

Figure 39(b) and,(c) show typical three-column beam type and four-column beam type combined elements. Generally, two symmetrical column beams are used to measure lateral force $F_y$ and yaw moment $M_x$. The normal force $F_x$ and the pitching moment $M_y$ are measured with one column beam in the middle (three-column beam type) or two upper and lower column beams (four-column beam type).

A column-beam combined element is a commonly used structural form of strain balances. The advantages of this form are large balance rigidity, small interference and large output. According to the requirements of the wind tunnel balance design, the column-beam composite element can also be a five-column beam component or a beam component with even more than five columns, which can be seen in Figure 39(d), (e), (f).

### 5.1.3 Rolling moment element

For the six-component rod strain balance, the design range of the rolling moment $M_z$ is small. And the stiffness of the rolling moment element is comparatively large to withstand the load of other components. Therefore, it is difficult to obtain an ideal rolling moment signal output. The structure of the rolling moment element can be composed of an independent element, like the axial force element, using a four-column beam or M-beam element (Figure 39(f)). However, it is generally combined with the normal force element, pitch moment element, lateral force element and yaw moment element to form a combined element (Miguel A. González, 2011).

In a rectangular cross-section beam-type composite element, the rolling moment $M_z$ can share a rectangular cross-section beam as a measuring element with other components. The strain gauge is placed in the centre of the long side of the rectangular cross-section at a 45° angle with the axis of the balance (Miguel A. González, 2011). The rolling moment $M_z$ is measured according to the principle that the principal stress in the 45° direction of the beam is equal to the shear stress when the beam is twisted.

In the column-beam type composite element, two symmetrically arranged column beams on the left, right or upper and lower can be used to measure the rolling moment $M_z$. Under the action of the rolling moment $M_z$, the composite deformation of bending and torsion is generated. The bending deformation is used to measure the rolling moment $M_z$. The middle column beam can also be used to measure the rolling moment $M_z$ in the three-post beam composite element.

## 5.2 Design specification of balance

(1) The diameter of the wind tunnel balance $D = 12mm$

(2) The design scale of the wind tunnel balance

Normal force: $F_x = 200N$      pitch moment: $F_y = 10N.M$

Axial force: $F_z = 200N$      roll moment: $F_x = 5N.M$

Side force: $F_y = 100N$      yaw moment: $F_z = 5N.M$

## 5.3 Dimensions of strain gauge balance

The dimensions of the strain balance depend on the design range of the balance, the geometric size and internal dimensions and internal structure of the model. For a rod balance, the overall design mainly determined by the diameter $D$ of the balance, the length $L$ of the balance, and the end of the balance model $l_1$ and the length of the cone of the balance rod end $l_2$.

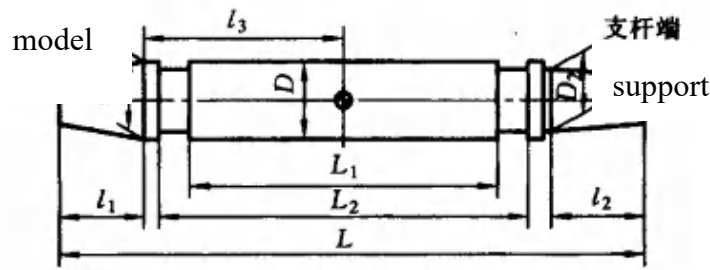

*Figure 40 Strain gauge balance*

In general, the mechanical design and processing of the force sensor, such as the bearing components, need to be particularly careful, which cannot appear greater stress concentration. Depending on the operating condition, the allowable stress must also be less than a certain value to ensure safety. The minimum area of strain gauge is approximately $1mm \times 2mm$. In this case, the patch area should be at least greater than this area.

Because of the size limitation of the wind tunnel, the length of the balance generally proportional to diameter, therefore, the most suitable way is to minimise the diameter of the balance. In this project, it was assumed the diameter of the balance was 12mm. Generally, the ratio of the length L of the strain balance to the diameter D of the strain balance is around 6–

10. Therefore, the basic size is: diameter 12mm, the component element length is 70 mm.

$$L = 12 \times 6 = 72 \text{ mm, assume } L = 70 \text{ mm}$$

## 5.4 Selection of balance element structure

The balance adopts a mature integral rod balance structure. The axial force element is set at the balance design centre (moment reference centre) and two three-column beams are installed symmetrically at the front and rear of the balance design centre to measure the remaining five components except the axial force. Based on this size, the bending element has enough space to arrange the strain gauge. The axial force component can also be measured with a simple 'I' beam in the middle, the design for wind tunnel balance is given below.

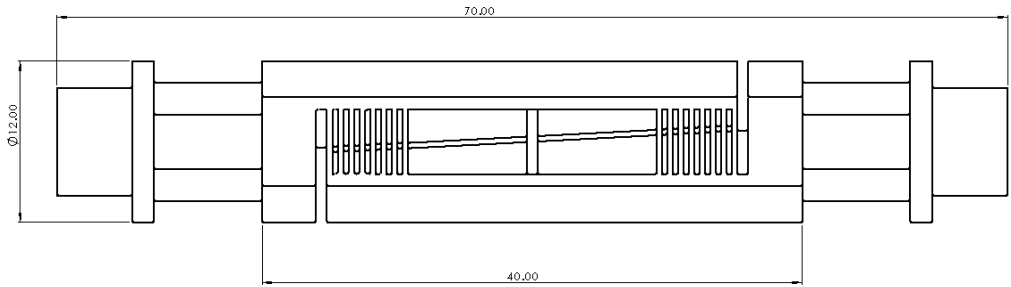

Figure 41 Dimensions of wind tunnel balance

## 5.5 Measure circuit

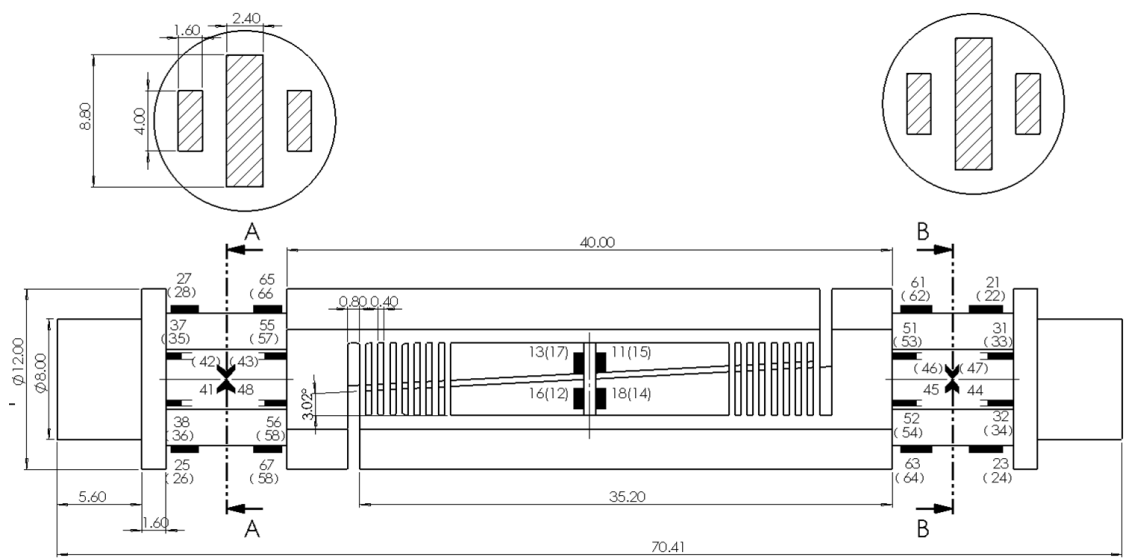

Figure 42 Structure and size for wind tunnel balance

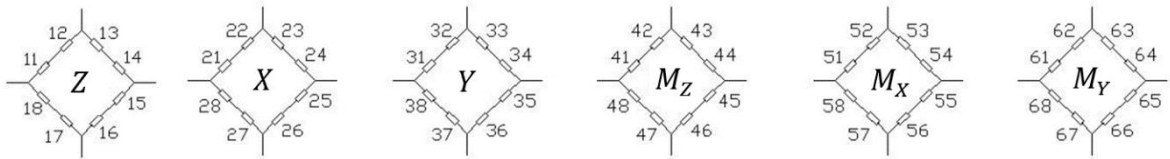

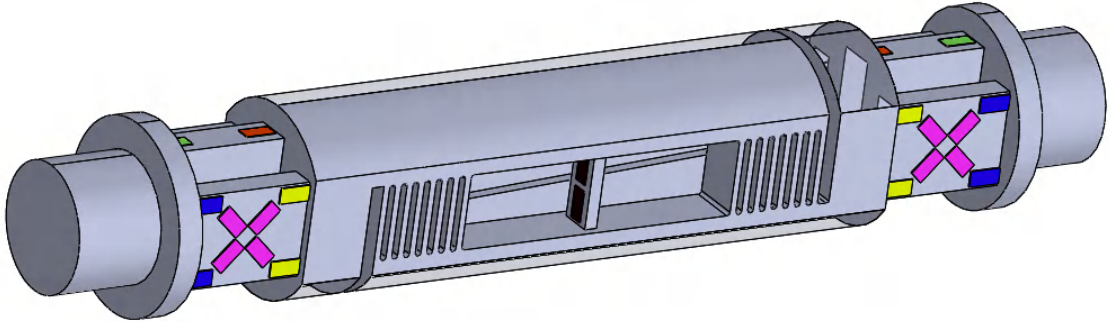

*Figure 43 Design of measuring circuit of balance*

The measuring element of each component force and moment is generally composed of the measuring plate and supporting plate and the force is transmitted through the parallelogram structure. The measuring sheet and the supporting sheet are both elastic elements, appearing in the form of cuboid vertical beams on balance, the number of which is determined according to the design needs. The strain gauge is pasted onto the measuring element. The measuring elements of the balance are shown in Figure 42. Each component of the balance has two sets of measuring elements and each measuring sheet is pasted with strain gauges. The strain gauge is 1.5mm away from the root of the measuring gauge. In order to improve the sensitivity of the output result of the force measurement, the four-arm, full-bridge wiring method is adopted to form a Wheatstone bridge.

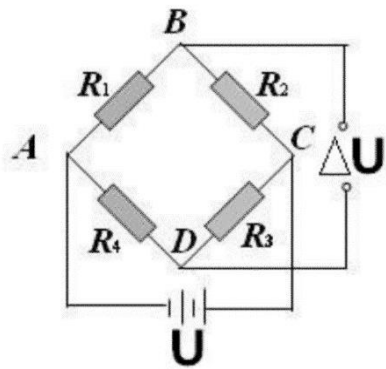

*Figure 44 Wheatstone bridge circuit*

The output formula of $\Delta U/U$ can be obtained for the Wheatstone bridge, which is $\frac{\Delta U}{U} = \frac{1000k(\varepsilon_1 - \varepsilon_2 + \varepsilon_2 - \varepsilon_2)}{4}$. Two strain gauges in the same branch have opposite polarity, that is, one is stretched and the other is compressed. The corresponding resistors of different branches have opposite polarity. The resistor of the branch is stretched while the first resistor of the other branch and compressed, therefore, $\varepsilon_1 - \varepsilon_2 + \varepsilon_2 - \varepsilon_2$ is equal to $4\varepsilon$. Deduced from the above formula, the output result of the axial force and normal force applied to the balance under the full-bridge measuring circuit is consistent with the output result of the axial force applied alone. It shows that the bridge laid by the balance has the ability to reduce the interference between the loads (Tintoré, 2018).

The deformation of the strain balance under aerodynamic load is linear to the external force. By deforming the strain gauge affixed to the balance, its resistance value is changed and the change in resistance is converted into an electrical signal, that is a change in voltage, by a device consisting of a strain gauge and a measuring circuit. This change in electrical signal is converted by A/D and then input into the computer to solve the aerodynamic load acting on the balance.

In addition, the proper sticking position of the strain gauge and the design of the full-bridge measuring circuit can ensure that the deformation caused by other component loads will not change the balance state of the bridge circuit, so that the purpose of distinguishing aerodynamic force and aerodynamic moment can be achieved (Nouri et al., 2014).

## 5.6 Finite element analysis for wind tunnel balance

In order to predict the performance and sensitivity of the balance, the stiffness, total deformation and strain of each measuring element under different load conditions are observed in Ansys to test whether the force and moment can be decomposed well. According to the calibration load, apply forces and moments in different directions to the balance in different steps. In the last step, all forces and moments are applied to the balance, which is shown in Table 9, the boundary condition is that one end is fixed and the other end is loaded.

*Table 9 Steps and loading condition*

| Steps | 1 | 2 | 3 | 4 | 5 | 6 | 7 |
|---|---|---|---|---|---|---|---|
| Loading condition | $F_y$ side force | $F_z$ drag force | $F_x$ lift force | $M_y$ pitch moment | $M_z$ roll moment | $M_x$ yaw moment | Total force |
| X (N) | 0 | 0 | 200 | 0 | 0 | 0 | 200 |
| Y (N) | 100 | 0 | 0 | 0 | 0 | 0 | 100 |
| Z (N) | 0 | 200 | 0 | 0 | 0 | 0 | 200 |
| X (N.mm) | 0 | 0 | 0 | 0 | 0 | 5000 | 5000 |
| Y (N.mm) | 0 | 0 | 0 | 10000 | 0 | 0 | 1000 |
| X (N.mm) | 0 | 0 | 0 | 0 | 5000 | 0 | 5000 |

The maximum deflection of strain balance when it only suffers the axial force can be seen in Figure 45, which is $1.28 \times E^{-2}$. Table 10 indicates the deflection under the action of each component force and moment, respectively, and the deflection after combining them together. However, because of the limitation of the diameter of the wind tunnel balance, the balance cannot measure the drag force very well, which can be solved can by increasing the diameter of the scales but, at the same time, it also increases the length of the balance and the strength of the supporting rod connected to the aerodynamic shape also needs to be validated. In the subsequent design, further discussion can be carried out through parametric design and optimisation algorithm.

*Table 10 Deflection of strain balance*

| Steps | 1 | 2 | 3 | 4 | 5 | 6 | 7 |
|---|---|---|---|---|---|---|---|
| Maximum deflection(mm) | 0.359 | $2.20 \times E^{-2}$ | 0.534 | 0.658 | 0.163 | 0.289 | 0.892 |
| Average deflection(mm) | 0.126 | $1.23 \times E^{-2}$ | 0.278 | 0.201 | $4.12 \times E^{-2}$ | $8.93 \times E^{-2}$ | 0.37 |

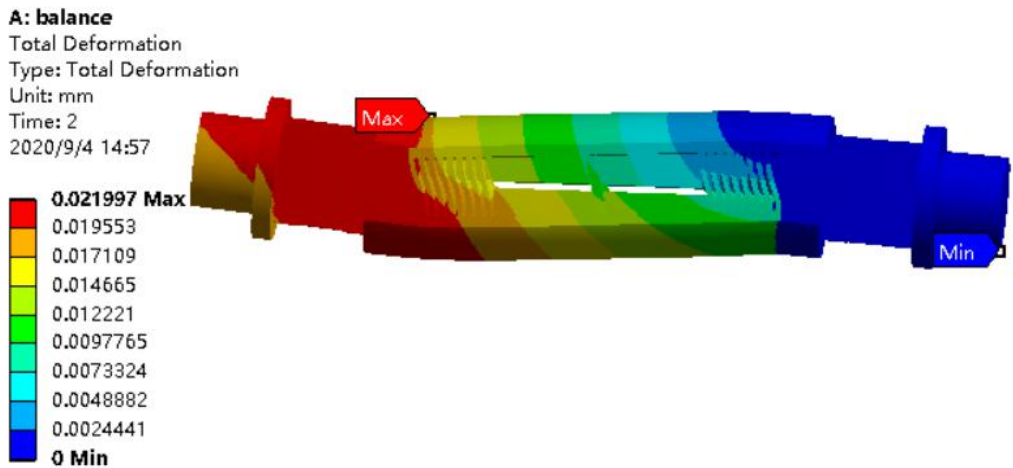

Figure 45 Deflection for balance

For the material of the wind tunnel balance, from Figure 46, the maximum normal stress is 3106.3MPa，General alloy steel can also meet the requirements of allowable stress.

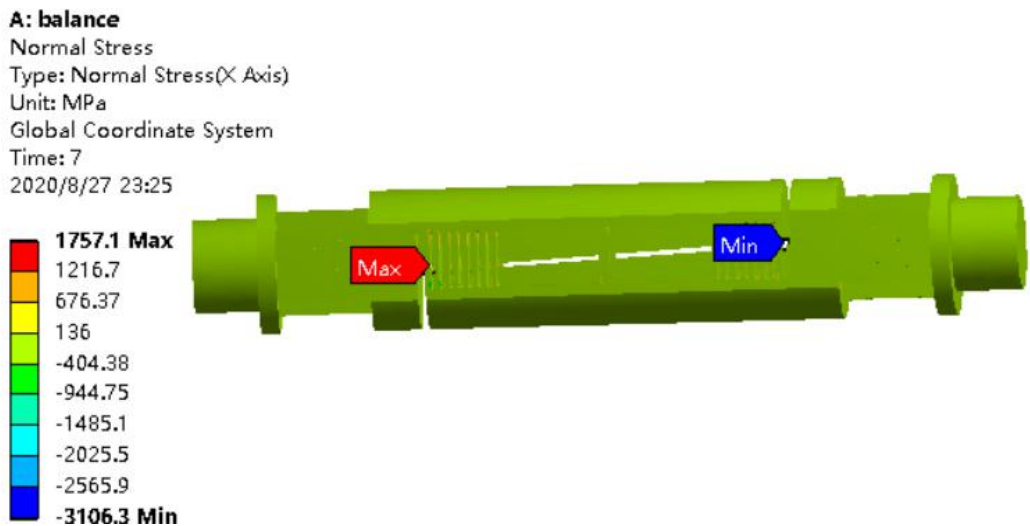

Figure 46 Normal stress for balance

### 5.6.1 Simulation results

Finite element analysis can help us know the sensitivity of the strain gauge balance in advance. The average strain in the full-bridge circuit of each component is the key value for sensitivity analysis.

### 5.6.2 The effective strain of each measure component

1. Axial force $F_z$ effective output

The effective strain surface of the axial force is the two inclined planes of the 'I'-shaped element.

The measuring element connects the mainframe of the balance with the floating frame. In addition, there are 18 support plates between the two frames. According to the theory of material mechanics, it is known that the slope of the 'I'-shaped element deforms into S under the action of axial force. When the X-axis force is a compression (tension), a small amount of strain superimposition (offset) occurs on the S-deformed compression surface and a small amount of offset (superimposition) occur on the stretched surface. The effective strain at the corresponding patch is shown in Figure 47. The following are all taking the parameter distance 0.5 (centre position of the strain gauge) as the average strain, which is $26.5 \times E^{-6}$.

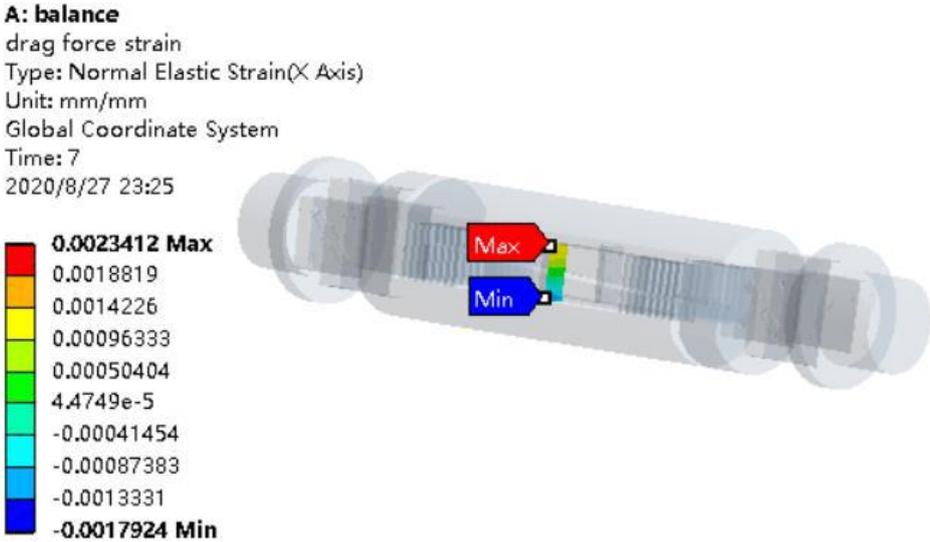

*Figure 47 Strain for axial force*

2. The effective output of lift force $F_x$

The lift measurement is completed by the combined element and its structure is a pair of three beams, which are symmetrically arranged on both sides of the centre of the balance. The lift value is relatively large compared to other values, so the strain gauge application surface is selected as a pair of composite elements three beams. These are located in the upper and lower sides of the main beam and the ends of the support rod. According to the theory of material mechanics, it can be known that the balance deforms into S under the lift force and the effective strain at the corresponding patch is shown in Figure 48. The strain gauge is attached near the model end and the support end, where the strain is around $490 \times E^{-6}$.

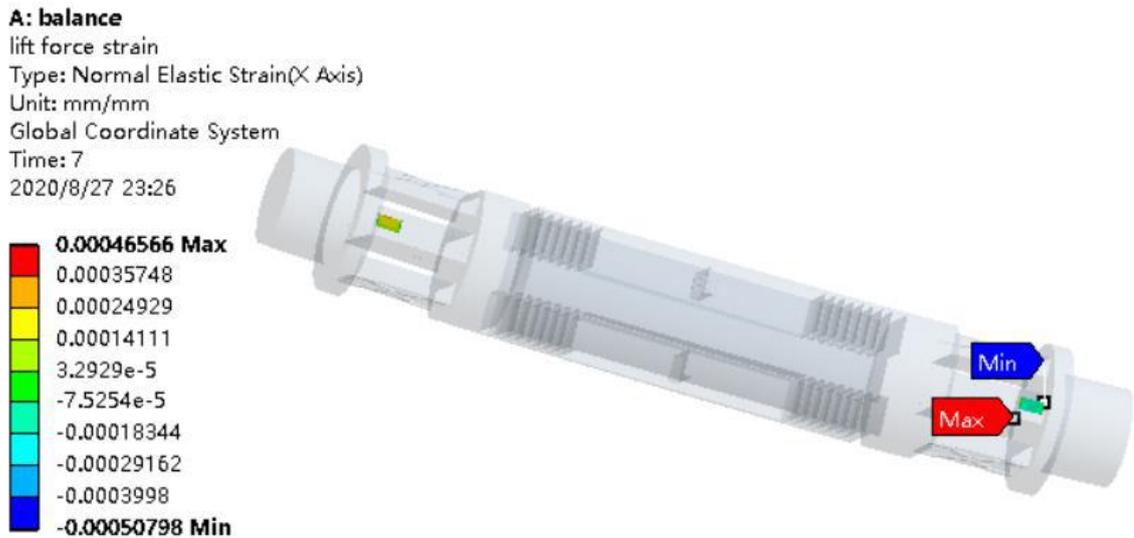

Figure 48 Strain for lift force

3. Lateral force $F_y$

The effective output lateral force measurement is completed by the combined element. When the balance is subjected to the lateral force, the balance becomes S-deformed. The strain gauge is applied to the centre of the outer surface of the outer beam of a pair of composite elements; the strain gauge shall be located on the outer surface of the external beam of the component near the model and the support rod, as shown in Figure 49, which is around $97.9 \times E^{-6}$.

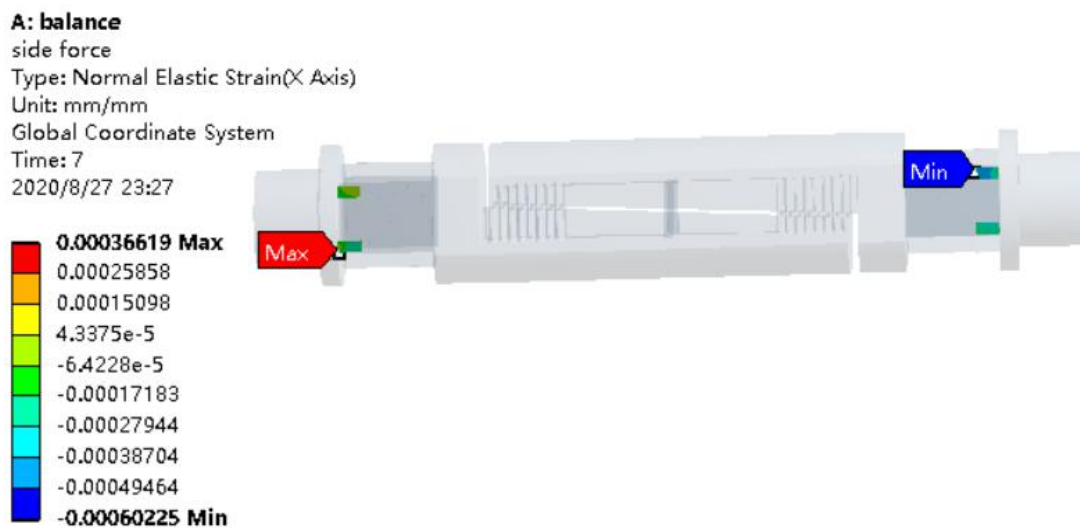

Figure 49 Strain for lateral force

4. Rolling torque $M_z$ effective output

Rolling torque measurement is completed by the combined element, the strain gauge is applied to the centre of the outer surface of the outer beam of a pair of combined elements and the patch direction is 45 degrees from the horizontal; the average strain is $295 \times E^{-6}$.

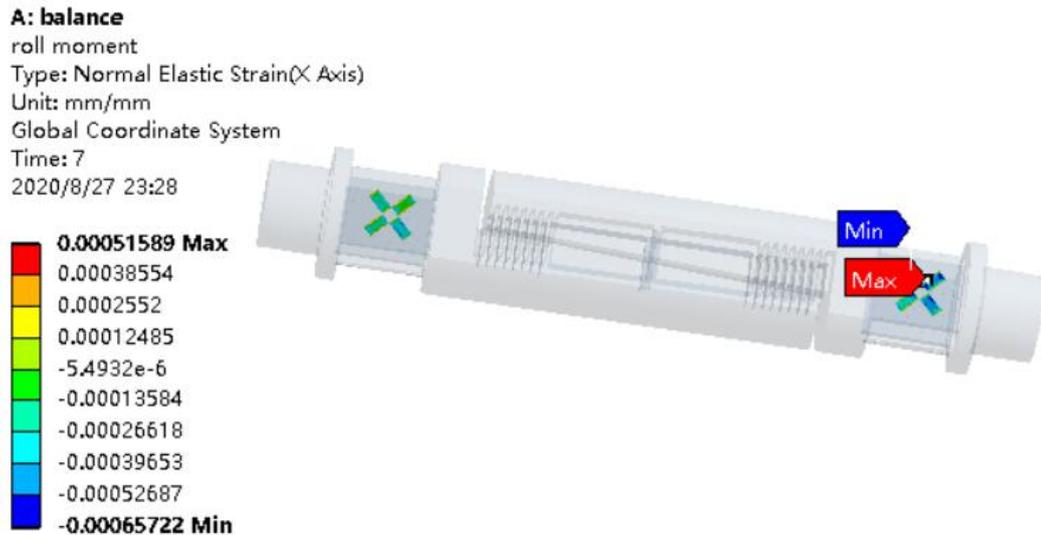

*Figure 50 Strain for roll moment*

5. Yaw moment $M_x$

The measurement of the effective output yaw moment is completed by the balance combination element. The strain balance becomes pure bending deformation when subjected to the yaw moment load. The applied position of the strain gauge to measure the component is the outer surface of the outer beam of a pair of combined components near the centre of the balance. The overall strain cloud diagram of the strain balance and the effective strain of the patch position are shown in Figure 51, according to the equal 5-1, the average strain is $512 \times E^{-6}$.

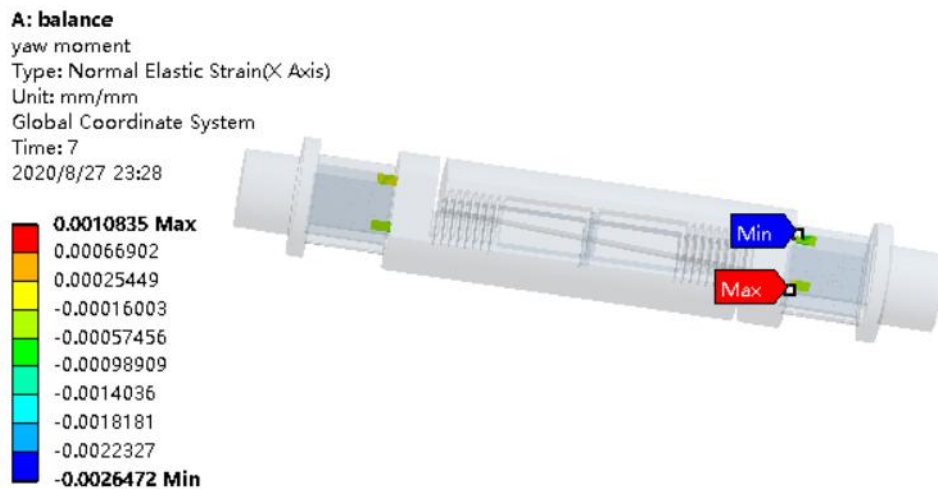

*Figure 51 Strain for yaw moment*

6. Pitching moment $M_y$

The effective output pitching moment measurement is completed by the combined element. The balance becomes purely bent when subjected to the pitching moment. Strain gauges are applied to the upper and lower surfaces of a pair of combined element main beams near the centre of the balance. The effective strain situation is shown in Figure 52, which is $467 \times E^{-6}$.

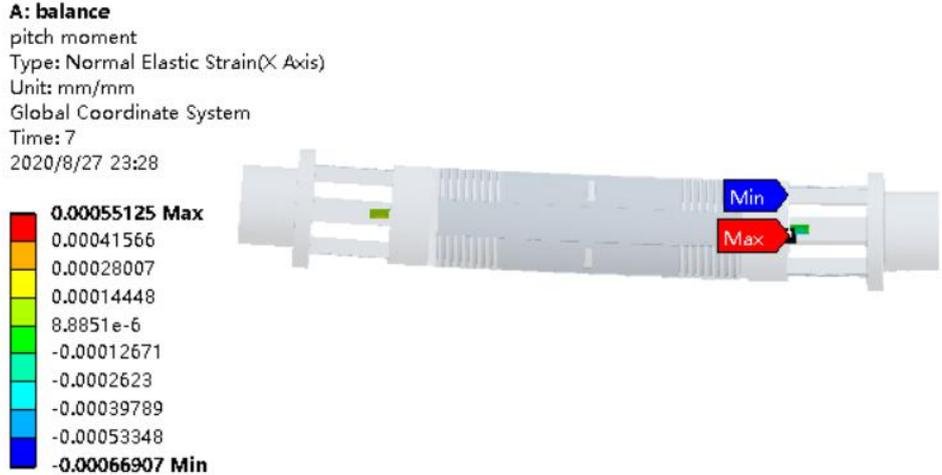

Figure 52 Strain for pitch moment

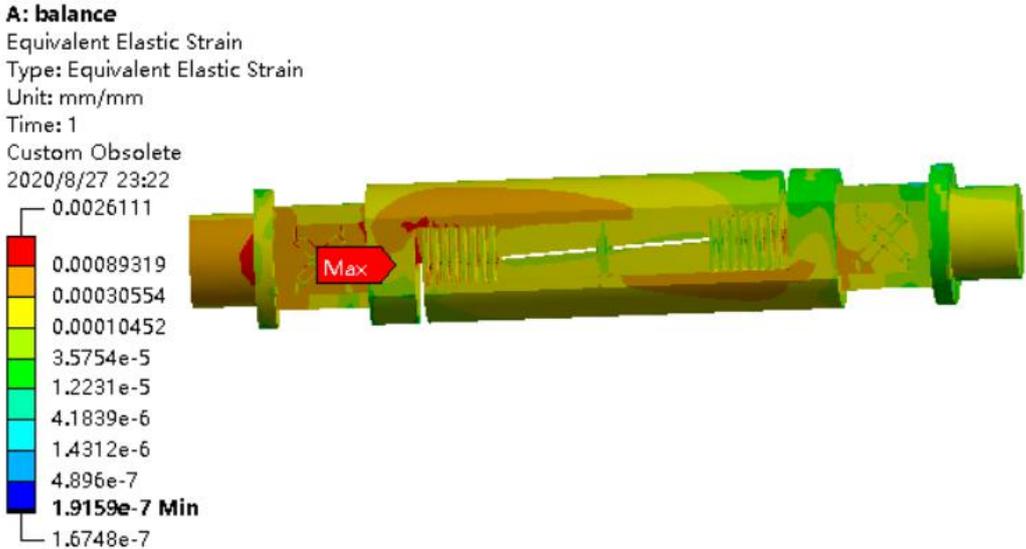

Figure 53 Equivalent elastic strain of balance

Table 11 displays the results of the seven-step analysis, which means applying the three direction forces and three component moments in their six respective steps and then applying the total forces and moment together in the last step.

Table 11 FEA steps and results

| step | Component | Calibration load | maximum deflection | Maximum stress (MPa) | The average strain of measured beam(e-6) | The average strain of the measured position |
|---|---|---|---|---|---|---|
|  |  |  |  |  |  |  |

| 1 | $F_y$ side force | 100N | 0.36mm | 791 | 174 | 97.9 |
|---|---|---|---|---|---|---|
| 2 | $F_z$ drag force | 200N | 0.02mm | 327.7 | 47.2 | 26.5 |
| 3 | $F_x$ lift force | 200N | 0.53mm | 3881.7 | 448 | 490 |
| 4 | $M_y$ pitch moment | 10N.M | 0.73° | 3572.1 | 459 | 467 |
| 5 | $M_z$ roll moment | 5N.M | 0.18° | 561.6 | 318 | 295 |
| 6 | $M_x$ yaw moment | 5N.M | 0.32° | 661.7 | 174 | 512 |
| 7 | Total force | | 0.892mm | 3320.3 | 742 | |

From the third Component Balance and Arc System Operations Manual for the University of Manchester, according to the balance deflection for the calibration loads which is given in Figure 54, compared with the six-component balance deflection for the calibration loads, which is given in Table 9, the deflection of them under the same calibration load is almost same, therefore the designed six-component balance performs well.

| Component | | Calibration Load | Measured Deflection | Stiffness |
|---|---|---|---|---|
| Fx | Drag | 20 kg | 0.02mm | $9.8 \times 10^6$ N/m |
| Fz | Lift | 20 kg | 0.33mm | $6.0 \times 10^5$ N/m |
| My | Pitch Moment | 1 kgm | 0.75° | $7.5 \times 10^2$ Nm/rad |

*Figure 54 Balance deflection for the calibration loads (Aerotech)*

By extracting the strain of each component separately, the sensitivity can be calculated roughly. Of course, sensitivity can be improved and further optimised and calibrated, which requires more information about design requirements. By comparing the static calibration data of the six-component balance and the three-component balance already given, and by combining it with the sensitivity analysis, it can be seen that the designed six-component balance is reasonable and can effectively measure and decompose the six components. The preliminary design shows that the six-component force sensor with a small diameter is technically feasible.

# 6  Control system design

This project is a remote-controlled angle adjustment device in a supersonic wind tunnel; my task is designing a remote angle adjustment robot in the supersonic wind tunnel, which involves a model support system and a strain gauge balance in the wind tunnel. A wind tunnel balance is intended to decompose the force and feed these data into simple models of support (to calculate deflection). The control part involves closed-loop control and a strain gauge balance feedback. The system can achieve the precise transmission of the supporting system by improving the stability of the mechanical structure and reducing the transmission error.

## 6.1  Detailed design

Arduino is a typical example of open source hardware. It can sense the environment through various sensors, which can provide feedback and affect the environment by controlling lights, motors and other devices. The microcontroller onboard can write programs through the Arduino programming language and compile it into a binary file, then transfer it to the microcontroller. Projects based on Arduino can not only be implemented by Arduino but also can be accomplished by communicating with Arduino and some other software running on PC.

LabVIEW is an example of graphical programming. It uses graphical programming language for software design and can build a virtual working interface.

Combining LabVIEW with Arduino to construct the upper and lower system could control the whole angle adjustment system. And the communication between them could be achieved by LINX. Arduino is used as a lower computer to collect data and output tasks; as the host computer, LabVIEW could carry on more multiple complex calculations and display functions.

The system uses the Arduino UNO as the lower machine of the angle adjustment system, LabVIEW is used as a host computer to control the Arduino. In the LabVIEW program, the rotation speed of the stepper motor is controlled by the time of the loop; what's more, the direction and the amount of step could be controlled through building Sub VI in the LabVIEW. The limit sensor and homing position could implement the closed-loop control of the stepper motor. What's more, the six-component force sensor collects the running deflection state in real-time and the obtained data are fed back to the Arduino to build a closed-loop real-time feedback control system.

### 6.1.1 Control system design plan

The system hardware is composed of Arduino UNO, a six-dimensional force sensor, a stepper motor, a power supply module, a drive circuit, a supporting system and a model. The Arduino UNO processor is the core of the angle control system. The wind tunnel balance collects the angle deviation of the model due to wind force in real-time to implement a closed-loop control system. The Arduino UNO, combined with the rotary encoder and Hall sensor, performs a closed-loop control on the stepper motor, thereby realising the adjustment of the operation of the motor. The logical procedure diagram of the control system design is given below.

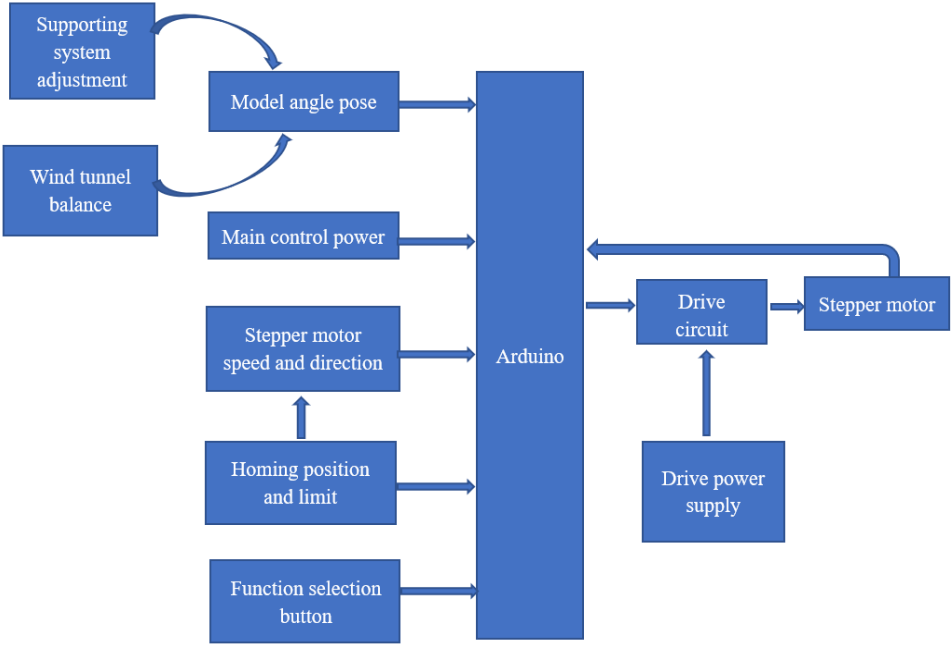

*Figure 55 Logic diagram*

## 6.2 System hardware circuit design

### 6.2.1 Circuit diagram

The circuit of the actuation system is given below

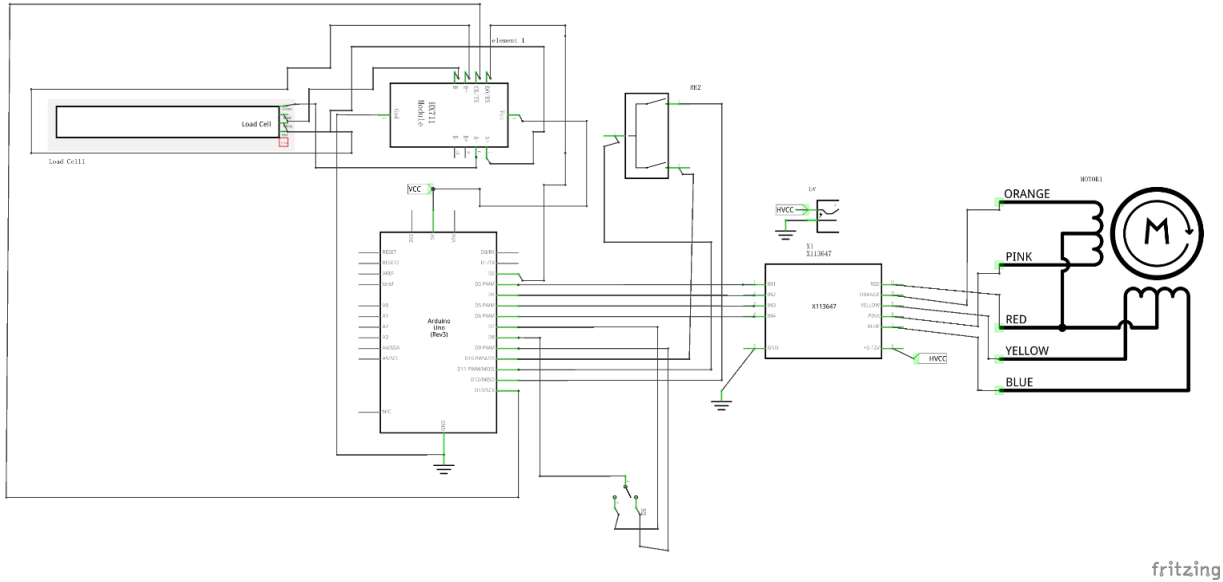

Figure 56 Schematic diagram

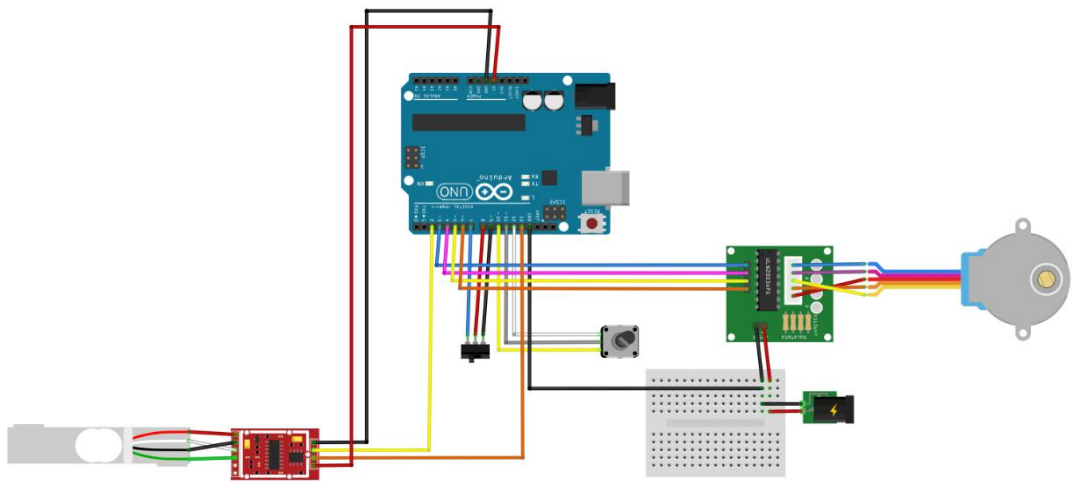

Figure 57 Breadboard wiring diagram

### 6.2.2 Selection of the motor driver

The motor used for the program simulation is a 28BYJ-48 Stepper motor, which is driven by a ULN2003 four-phase, five-wire driver board electronics stepper motor.

In the actual system design, according to the motor selected by holding torque, the L298N motor drive module was chosen as the motor driver. The L298N chip has a high operating voltage, a continuous operating current of 2A and a rated power of 25W. It contains two full-bridge H-bridge type drives, which can drive a two-phase stepper motor.

*Table 12 Stepper motor and motor driver*

|  | Stepper motor | Stepper motor driver |
|---|---|---|
| simulation | 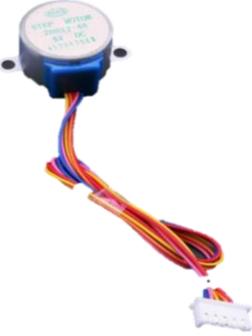 28BYJ-48 Stepper motor | 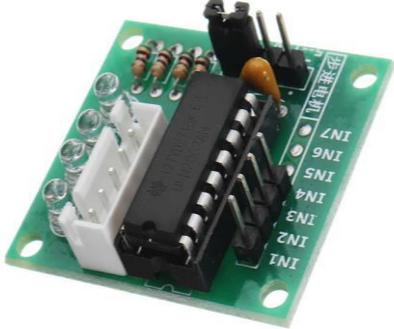 ULN2003 Stepper Motor Driver Board |
| practical | 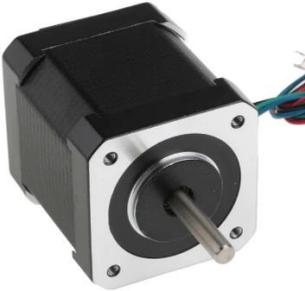 Hybrid stepper motor 42SH47 | 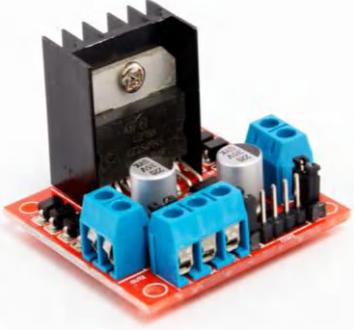 L298n dual H-bridge |

### 6.2.3 Power supply module

The motor drive and control circuit used in actuation system programming simulation need a 9V power supply and the power module uses a power supply module and an adapter to provide a 9V power supply.

### 6.2.4 Selection of limit and 'homing' sensors

The stepper motor operates in an open loop, which means that the motor may be out of step when it is working (such as encountering a large resistance or some special circumstances) and, after a power failure, there is no locking torque in the stepper motor and the position will change after the power is on, so the system needs to calibrate the zero position. The use of photoelectric switches to control stepper motors is a function of position control. The signal of the photoelectric switch sent outputs a pulse signal to the stepper motor driver, thereby controlling the origin return, endpoint stop and position control of the stepper motor.

Generally, there are three pins, the VCC power supply is positive, the GND power supply is negative and OUTPUT means output voltage. An infrared photoelectric switch power supply of voltage DC5V, when there is no object blocking, outputs a high-level DC5V. When there is an object blocking, it outputs a low-level 0V, so the signal of the OUTPUT pin could be sent to a microcontroller or other controller to detect whether an object is in its position.

## 6.3 Closed-loop control

### 6.3.1 Lost steps and overshoot of stepper motors

Stepper systems are designed to provide position control without expensive feedback where precise motion is a requirement. Exceeding the available torque or other limits causes lost steps failures to advance in the position that goes undetected. Overshoot is relatively rare.

The stepper motor converts the pulse signal into a displacement signal, which gives the motor a pulse degree and the motor rotates by an angle. If the pulse frequency is too fast, the motor cannot rotate to the specified angle, which is equivalent to throwing away some of the pulse signals, which is called lost steps. Overshoot refers to the braking torque after the motor receives the stop command. When the stop position of the motor exceeds the specified position, it is called overshoot.

### 6.3.2 Rotary encoder and homing position

#### 6.3.2.1 Rotary encoder

To solve the problem of the lost steps and overshoot of the stepper motor, the rotary encoder could be mounted on the output shaft of the stepper motor, which could feedback how many steps are rotated in practice. An incremental encoder is a position sensor that converts displacement into a periodic electrical signal after conversion and then converts the electrical signal into count pulses. The number of pulses represents the magnitude of the displacement. The output of the incremental rotary encoder in this project is a series of square wave pulses.

The rotary incremental encoder is installed on the output shaft of the motor, which rotates with the rotation of the motor and outputs pulses during rotation. The number of pulses rotated can be calculated in LabVIEW. When the encoder stops rotating or is powered off, the internal memory of the counting device can be used to remember the position. In this way, when the power is turned off, the encoder cannot move. When the power is turned on, the encoder will not lose steps and be disturbed when outputting pulses. Otherwise, the zero point of the counting device will shift. Once the zero point shifts, the offset cannot be known before the error result appears.

### 6.3.2.2 Homing position

The homing position ensures that each motion's initial position is the same because the step motor is 'blind,' which can achieve angle and position control based on the homing position. There is no deviation caused during reverse adjustment. The other two halls are used for soft limits, and when the step motor moves to the limit position, it can stop automatically.

The homing position could be achieved by a photoelectric sensor. Every time the encoder passes the reference point, the reference position is corrected into the memory position of the counting device. Before the reference point, the accuracy of the position cannot be guaranteed. What's more, the reference point could know the position of the stepper motor. Therefore, in combination with the zero position of the stepper motor, the accuracy of the position of the stepper motor can be guaranteed and the practical steps rotate of the motor can be known.

## 6.3.3 Wind tunnel balance

As mentioned in Chapter Six, the wind tunnel balance could feedback the forces and moments to the supporting system to achieve the accurate angle of attack adjustment.

# 7 Virtual interface design and Program implementation

LabVIEW and Arduino are interactively programmed by LINX in this project. There are three hall switch sensors, one to determine homing position, the other used for limit switch and one for realising the closed-loop control of stepper motor with the rotary encoder. Meanwhile, the force sensor was used for feedback, and three are about five functions for this programme. Firstly, remote control the motor without input angle, secondly, inputting a specific angle to reach the motion position automatically, thirdly, limit switch and homing position, fourthly, Motor closed-loop control is realized by rotary encoder. Lastly, using a six-component wind tunnel balance to achieve feedback. The flow diagram of the program is shown in Figure 58.

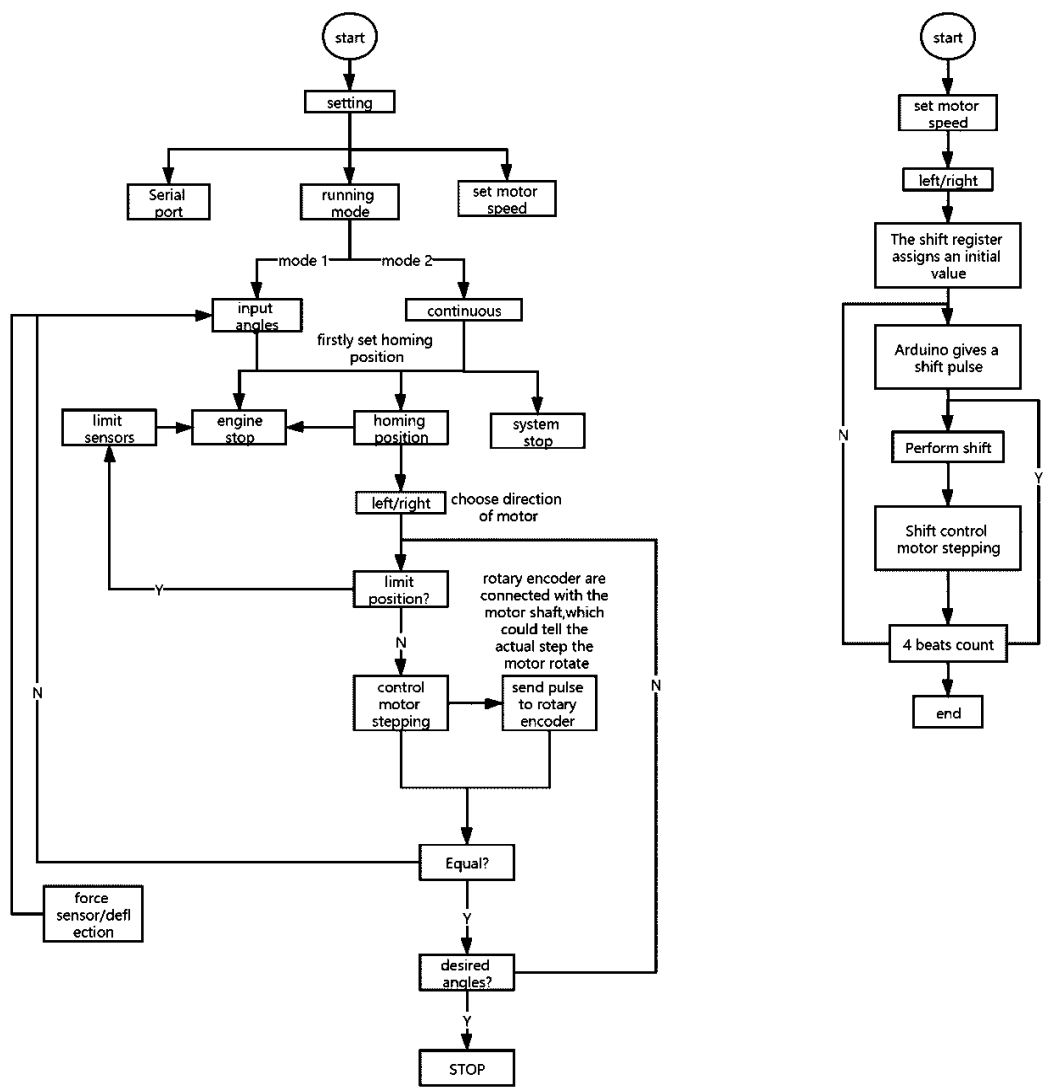

*Figure 58 Flow diagram of the program*

## 7.1 User interface introduction

As shown in Figure 59, In the operation panel, there are two tabs in which one corresponds 'operation' mode and another one is 'setting' mode, which can be seen in Figure 60. In the 'operation' mode, researchers can choose the direction of rotation of the motor, in this case, we have left, right and engine stop. And, finally, there is the homing position button, to set the zero point of the step motor, the speed indicator and the stop button to stop the system. On the left of the user interface, the 'deflection angles' indicate the angles of attack change caused by the drag force, the 'encode pulses' show the real steps of the stepper motor rotate, the 'remaining out' shows how many steps rotate theoretically, the 'lost steps count' shows the difference between the theoretical steps and actual steps of the stepper motor, the 'stepper motor steps' means the revised steps that the motor needs to rotate in order to reach the desired angle, the pressure gives out the drag force in real-time in the wind tunnel.

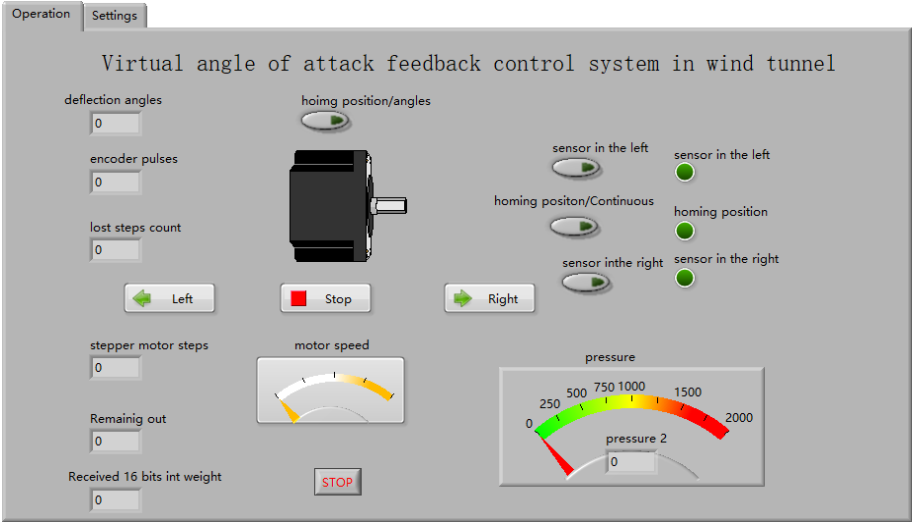

*Figure 59 Operation interface design*

In the setting panel, there is the communication port of the Arduino board and the serial port is COM3.In operation mode including the 'input angle' mode and 'continuous' mode operation. When 'input angle' mode is selected, the desired number of steps of the stepper motor needs to be given, then, the stepper motor will move to the desired angle. Also, we can indicate the speed of the motor.

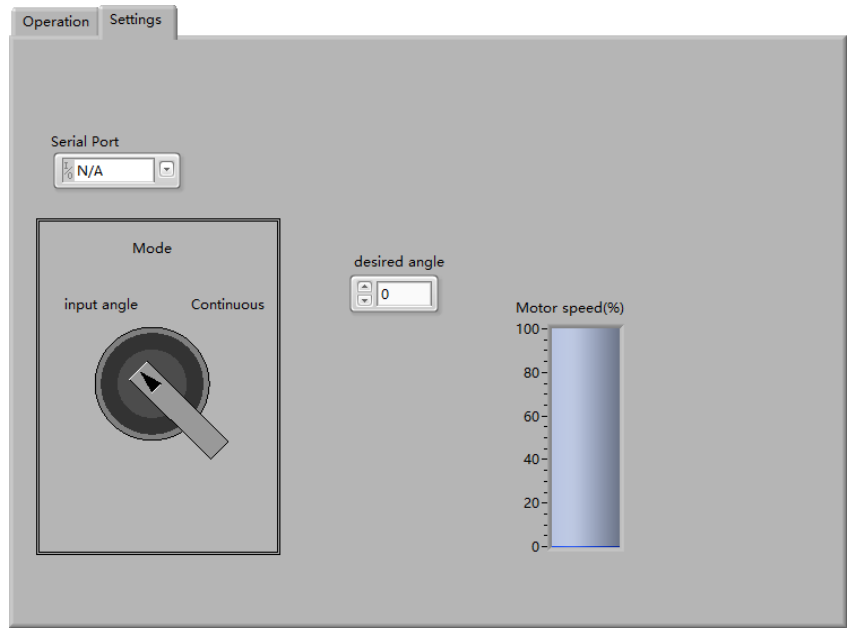

*Figure 60 Actuation system setting*

The program was created by Arduino and LabVIEW interactive programming and was tested and verified by Arduino UNO. The program can run normally and the functions of each part can be realised. The specific program of each control module is as follows and the complete block diagram is in the appendix.

*Table 13 Input channels*

|  | Digital input | | | |
|---|---|---|---|---|
| Channels | 3,4,5,6 | 7,8,9 | 10,11,12 | 2,13 |
| Function | Stepper motor | Hall sensor | Rotary encoder | Force sensor |

## 7.2 Stepper motor control

The stepper motor and motor driver for simulation are the 28BYJ-48 Unipolar Stepper motor and the ULN2003 driver board, respectively.

When continuous pulses are applied to the stepping motor, it can rotate continuously. Each pulse signal can change the energisation state of the two-phase winding of the stepper motor, corresponding to a certain angle of the rotor (a step angle). When the change of the energised state completes a cycle, the rotor rotates one pitch. The four-phase stepper motor can be operated under different energising methods; the energising method is dual (dual-phase winding energisation), four beats (AB-BC-CD-DA-AB-...), as shown in Figure 61.

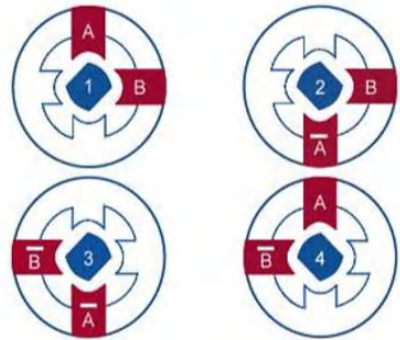

| sequence | A | B | A' | B' |
|---|---|---|---|---|
| 1 | 1 | 1 | 0 | 0 |
| 2 | 0 | 1 | 1 | 0 |
| 3 | 0 | 0 | 1 | 1 |
| 4 | 1 | 0 | 0 | 1 |

Figure 61 Stepper motor rotate sequence

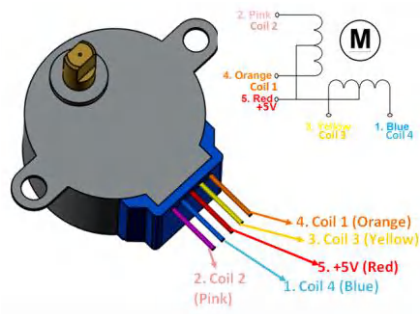

Figure 62 Principle of stepper motor

Table 12 28BYJ-48 Stepper Motor Technical Specifications

| Stepper motor | Rated voltage | Number of phases | Stride angle | Pull in torque | Insulated power | Coil |
|---|---|---|---|---|---|---|
| 28BYJ48 | 5V DC | 4 | 360/(64*32) | 300 gf.cm | 600VAC/1mA/1s | Unipolar 5 lead coil |

## 7.2.1 Motor speed

The first step is to initialise the serial port and communication and then enter the whole loop,

where we have three ways to move the motor in different directions and to be able to stop it. The speed control can control the time of moving or executing our program; the waiting time of the loop can determine the speed for the stepper motor. The wait command can be used to set the time interval of each cycle of the whole loop, which can control the speed of the stepper motor when the waiting time is set at 0 seconds. This corresponds to the maximum speed of stepper motor and, then, the slowest speed of the stepper motor, that is, the longest waiting time of the cycle, must be set as 0.2 seconds.

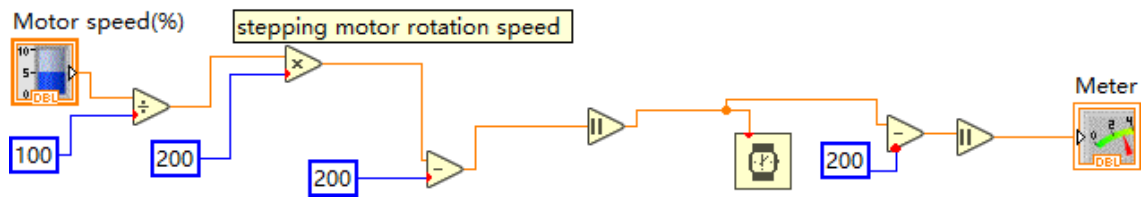

*Figure 63 Speed control*

### 7.2.2 Mode of operation

We have a case structure with two states, one is the false state and the other is the true state, which corresponds to the two modes of the stepper motor in the operation interface. In the beginning, we can select the mode of the operation, where one is going to allow the motor to move continuously and another is going to allow the motor to move in steps.

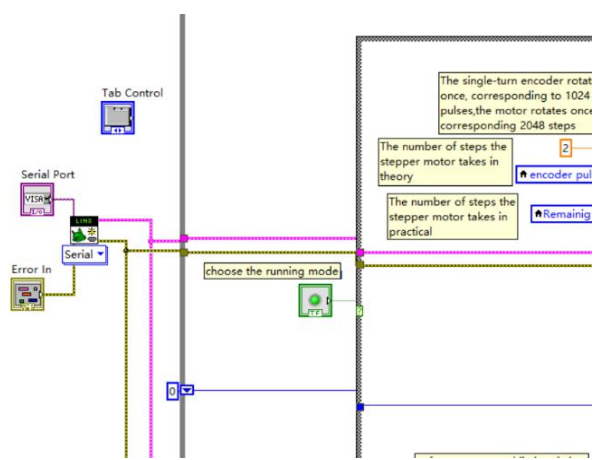

*Figure 64 Mode of operation*

#### 7.2.2.1 Continuous mode

In the continuous mode, there are three buttons 'left'、'right' and 'stop' to control the direction and start of the stepper motor respectively. The direction can be controlled by exciting the coil of the stepper motor in the sequence shown in Figure 61. Digital pins 3, 4, 5 and 6 of Arduino are used in this program. There is a control that lives on the front panel and is an input indicating the direction of rotation. This control is a Boolean type control: in case 1, this control is false, the stepper motor is going to move to the left; in case 2, the control is true, in the true mode, the stepper motor will move to the right. The program is in appendix Figure 109.

1 Sub vi of stepper motor direction control

There is a case structure to indicate the direction of rotation, either left or right. The 28BYJ-48 Unipolar stepper motor has a step sequence as follows: 1-3-2-4. This information will be used to drive the motor with the pin sequence of 3, 4, 5, 6, as shown in Figure 65. The sequence of excitation coils is given: in the program, pulses can be sent to the next frame, followed by the given sequence of excitation, then, the motor will move to the left continuously. When reversing the sequence of the coil excitation, the motor will move to the right.

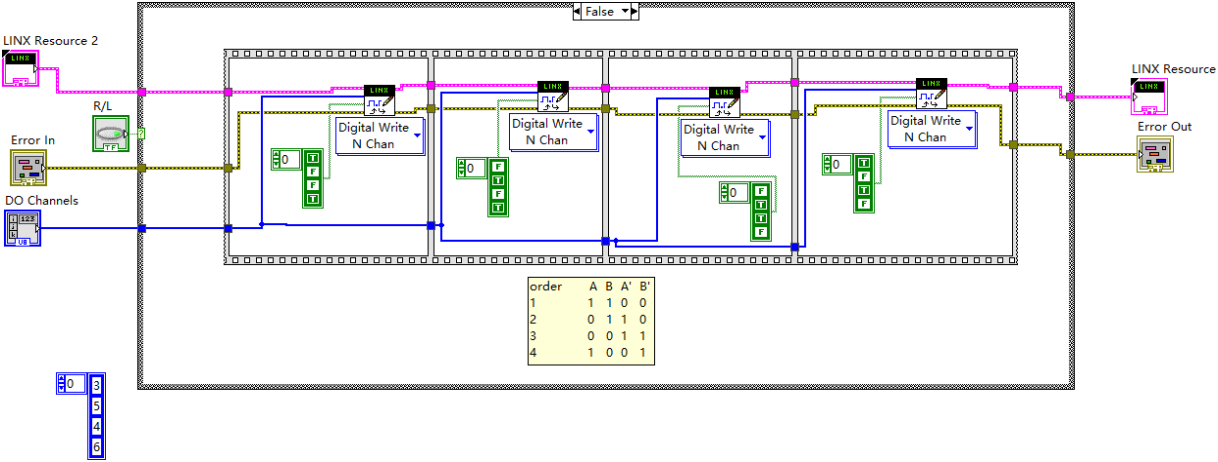

Figure 65 Sub vi for continuous motor direction control

### 7.2.2.2 'Input angle' mode

In the false case, the program will run in the step mode, with the desired angle of attack and entry of the steps that have already been moved. After the transmission calculation, the number of steps for the motor to move can be transferred to the sub VI, which can be seen in the appendix Figure 106.

2 Sub vi of stepper motor steps control

Similar to the continuous mode, there is a case structure, false/true, which corresponds to left/right, respectively. In this case, the desired steps that the stepper motor needs to rotate could be given by researchers and gives out the initial steps of the stepper motor. When the stepper motor rotates one step, on the basis of the initial number of steps, it will compare the current number of running steps with the target number of steps until the stepper motor reaches the specified position. The engine of the stepper motor will stop automatically.

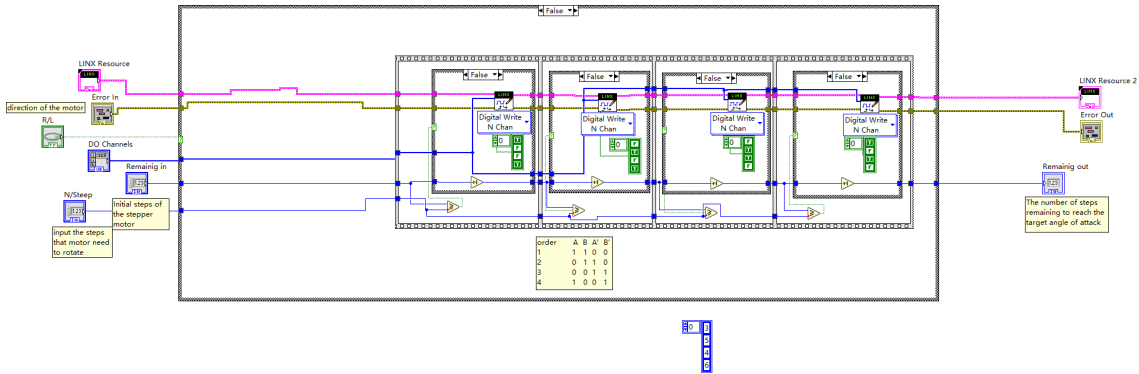

*Figure 66 Direction and angle control of step mode*

### 7.2.3 Stop the engine

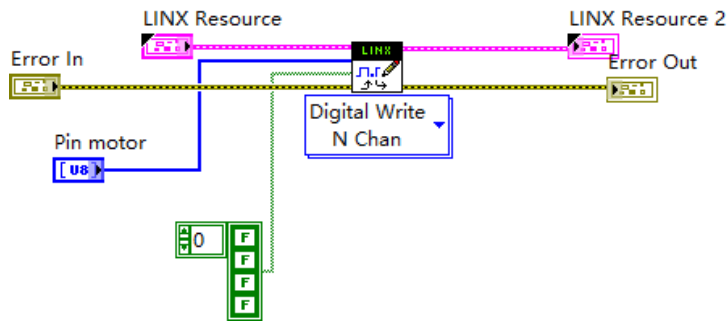

*Figure 67 Engine stop*

As shown in Figure 67, in case 3, the engine of the stepper motor will stop. Also the stop button stops the system, which is given in Figure 68.

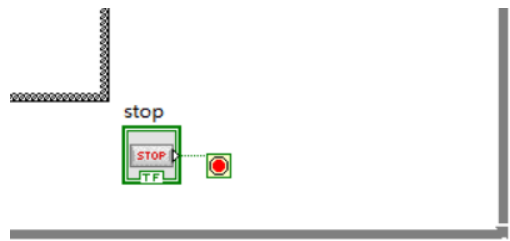

*Figure 68 System stop*

## 7.3 Homing position of the stepper motor

The homing position ensures that each motion's initial position is the same because the step motor is 'blind', which can achieve the angle and position control based on the homing position. There is no deviation caused during reverse adjustment.

When the system starts, the zero position of the stepper motor needs to be determined. The subsequent angle adjustments are based on the 'zero position'. In this project, the 'zero position' of the stepper motor is set on the left and a photoelectric sensor is installed on the leftmost side of the actuation movement system.

In the program, if the actuation movement system is started in continuous mode and the left button is pressed, the drive system will automatically run to the left when it reaches the zero position. The photoelectric sensor will detect a high-level signal and the stepper motor will automatically stop. At the same time, the 'homing position' button should be pressed to record the position of the zero position, which means that the initial steps are set to 0. After completing this series of preparations, the operation mode could be set to the step mode and then the angle of attack could be adjusted.

*Table 14 Homing position*

| Homing position determination | Record the homing position |
|---|---|
| 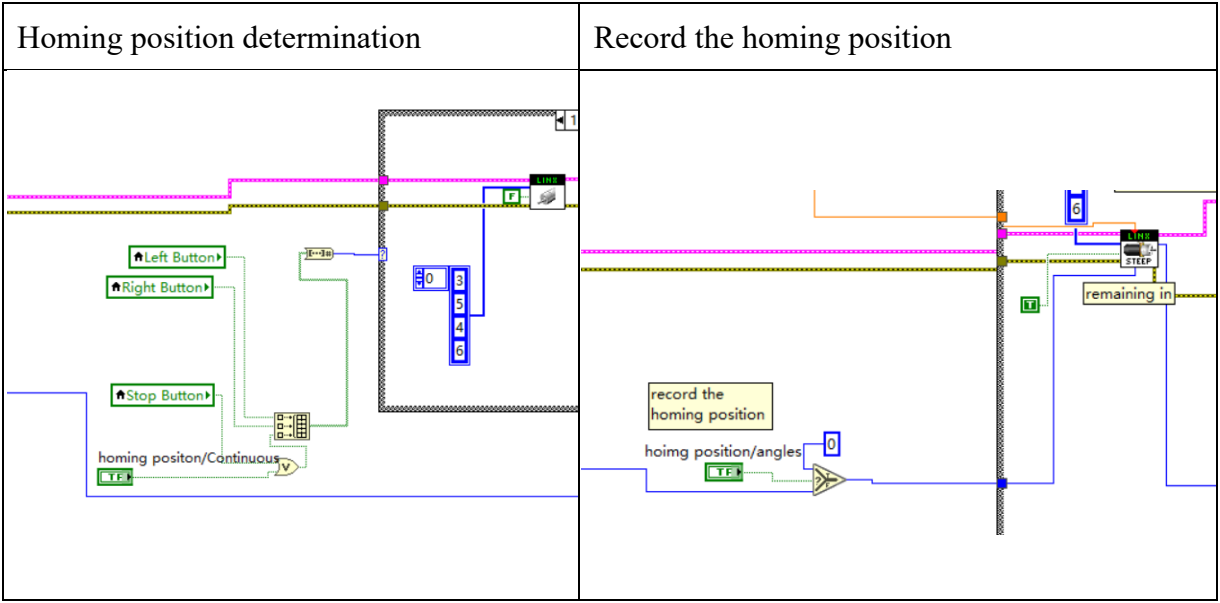 | |

## 7.4 Limit switch

The two hall sensors are used for software overtravel limits, when the step motor moves to the limit position, it can stop automatically. When the left button is pressed and the stepper motor runs to the left limit position, the photoelectric sensor at the left limit position senses the stepper motor and the stepper motor enters the right-turn mode. When the left button and the right-turn mode are in progress at the same time, the stepper motor engine will stop running, which indicates that the limit position has been reached.

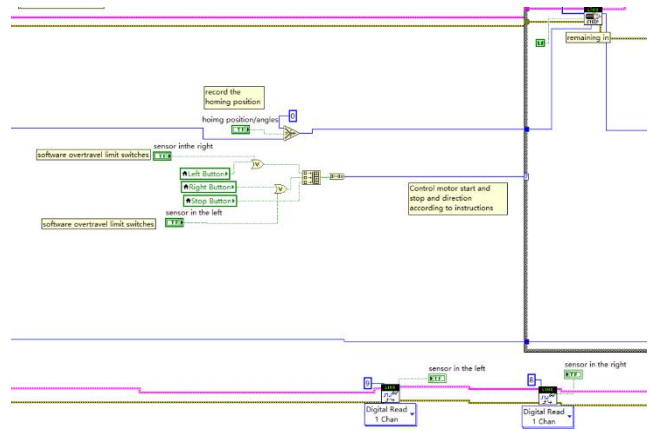

*Figure 69 Limit switch*

## 7.5 Rotary encoder

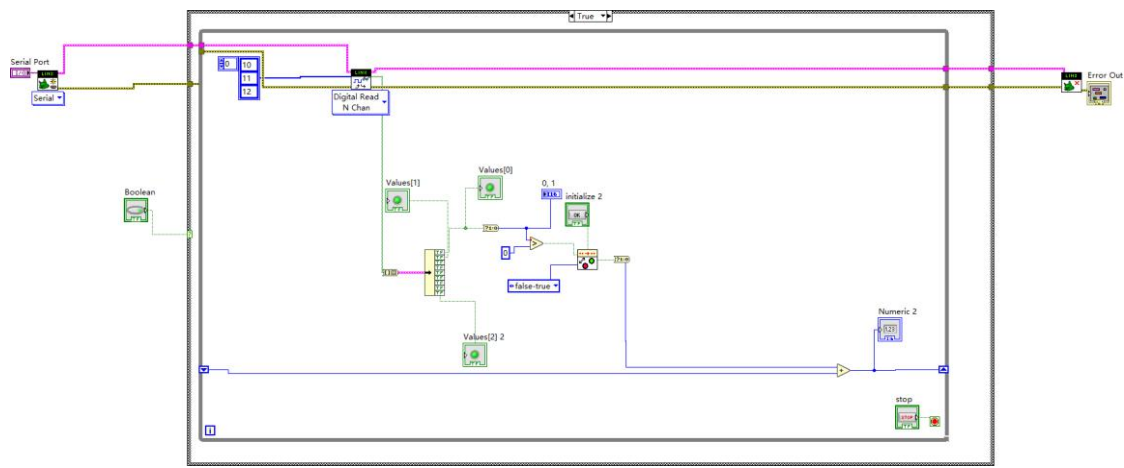

*Figure 70 Rotary encoder*

When the rotary encoder rotates by one pulse, the digital input will change from the false to the true and then back to the false state. Installing an incremental encoder on the motor shaft can record the number of steps the motor actually rotates, which can prevent the lost steps or overstep. As shown in Figure 70, the Boolean Crossing VI which can detect transitions of input and then multiply the sum of the number of pulses by the transfer coefficient, the result is equal to the number of steps the stepper motor has rotated in practical. The number of pulses for the rotary encoder in this project is 1024, while the number of pulses for a stepper motor is 2048, so the transfer coefficient is $\frac{2048}{1024} = 2$.

## 7.6 Six-component strain gauge balance

Each measuring component contains two strain gauges connected in a Wheatstone bridge. When wind tunnel balance suffered the aerodynamic forces, the strain gauges change resistance. With excitation on the bridge, the change in resistance is proportional to voltage.

Because of the small change of the strain gauge balance, the signal needs to be amplified. There are two ways to achieve this. One is to amplify the mV-level signal output of the strain balance to $5V$ for Arduino to read, but the A/A module suffers more interference, including power supply interference, wire length interference, wire position change interference, radio interference, etc.

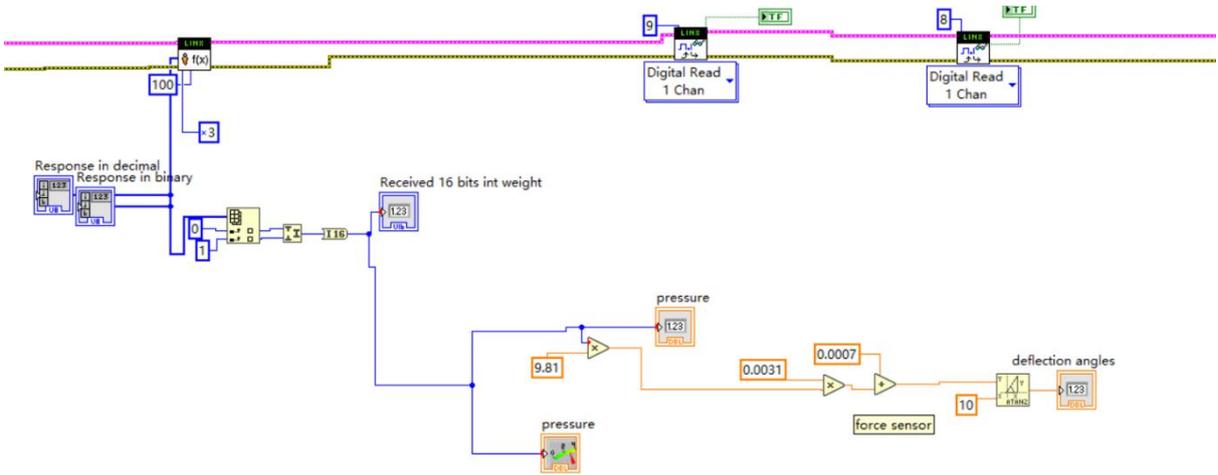

*Figure 71 Deflection angles measured by wind tunnel balance*

The second is to use the A/D module to convert the weight into a digital signal for transmission. The A/D module is fixed at the nearest place to the balance and the output digital signal is transmitted to the Arduino with a wire, which can greatly reduce the kind of interference.

That sensor is not incompatible with LINX, only bit-banging (communicating by sending digital true-false signals from LabVIEW directly) is compatible. The easiest option is, indeed, to add one (or two) custom functions, one for the initialisation of the chip and one for sending data.

The custom command could be added to the LINX Arduino-firmware, which communicates with the HX711 locally (using existing libraries) and then sends the weight (as a 16-bit INT) to LabVIEW. This could be implemented based on the HX711 library in the Arduino IDE, which defaults to reading out channel A, with an amplification factor of 128. The code for an Arduino UNO (the custom command is nr3) can be seen in the Appendix. The LabVIEW vi, which reads out and visualises the amplified signal, is shown in Figure 71.

The clock data pin and data pin are currently hardcoded as digital 2 and digital 13, respectively. Finally, only a 16-bit number is sent to LabVIEW. In theory, we should send a 24-bit number, however, its value varies from one reading to another; this happens with the last 8 bits. Therefore, the lowest 8 bits are not accurate and the whole range of this six-component force sensor can be covered with 16 bits. In short, a 16-bit resolution is enough and the whole weight range should be measured with 16 bits with a 5V supply and amplification by 128.

The relationship between aerodynamic force/moment and deflection is linear and this relationship can be acquired through Finite element analysis. The drag force of the model can be detected through the six-dimensional force sensor and the deformation of the model under different force conditions is simulated through finite element analysis. Data fitting is performed through MATLAB and indicates the relationship between the drag force and deflection of the model, which can be seen in the appendix Figure 115 and Table 15, and the relationship is given in Equation 7-1.

$$d = 0.0031 F_{loads} + 0.0007 \qquad \text{Equation 7-1}$$

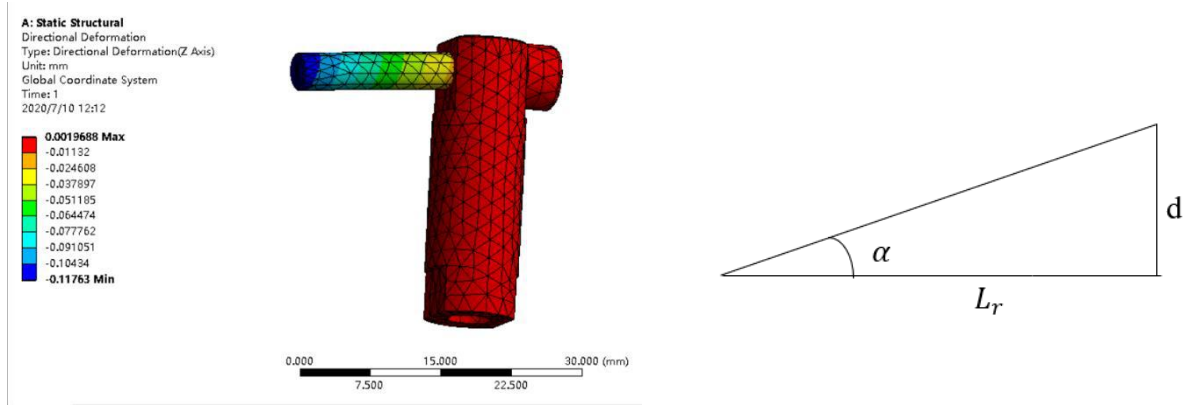

*Figure 72 FEA of the model interface*

Therefore, in LabVIEW programming, after knowing the relationship between displacement and drag force, the displacement should be fed back to the stepper motor to correct the angle of attack. The deflection angle of attack is the arctangent of $\frac{d}{L_r}$, d is the deflection of the model and $L_r$ is the length of the supporting rod. The deflection angle of attack is:

$$\alpha_d = arctan\frac{d}{L_r} \qquad \text{Equation 7-2}$$

## 7.7 Transmission function

The 28BYJ-48 is a 5-wire unipolar stepper motor that moves 32 steps per rotation internally, therefore, the stepping angle of the stepper motor is 360/64, but this motor has a gearing system that moves the shaft by a factor of 64. The result is a motor that spins at 2048 steps per rotation, which is $\frac{360}{2048} = 0.176°$.

$$\frac{s}{L} = \frac{N_S}{n \times \frac{360}{\gamma}} \qquad \text{Equation 7-3}$$

The variables are as follows:

1) $N_S$: The number of steps of the stepper motor

2) $s$: Running displacement; $L$ : lead

   $\frac{s}{L}$: Ts the relationship between displacement and lead, which is the number of turns required to move the screw for this distance.

3) $\gamma$: step angle

$\frac{360}{\gamma}$ —The number of steps the motor takes in one turn.

4) Number of subdivisions

   Driving in subdivision mode, which can control the motor to turn only half a step angle in one step, this value is equal to 1 in this project.

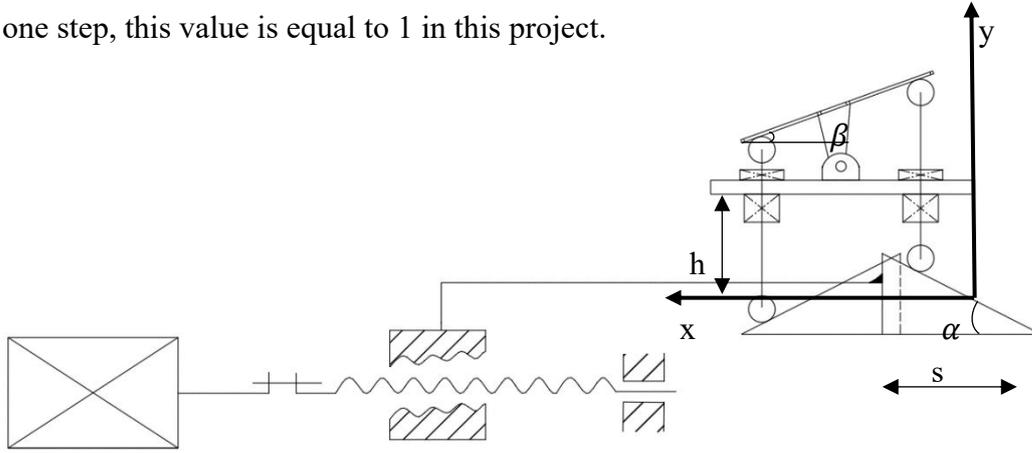

Figure 73 Transfer function between motor steps and angle of attack

Assume that when the angle of attack is 0, the origin of coordinates is the centre of the lower right corner roller, which is shown in Figure 73. When the motor rotates forward, the slider moves to the left. The left direction is the positive direction and the distance the slider moves from the origin is $s$ and makes the height change from the origin $h$. $\alpha$ is the base angle of the triangular block, which is 30°; $r$ represents the length of the model interference which connected to the model, equal to 77.2 mm.

$$s = \frac{h}{tan\alpha} \qquad \text{Equation 7-4}$$

$$tan\alpha = \frac{h}{r} = \frac{h}{77.2mm} \qquad \text{Equation 7-5}$$

Combine with the Equation 7-3, Equation 7-4 and Equation 7-5, the relationship between the number of steps of the stepper motor $N_S$ and desired angle a can obtained.

$$N_S = \frac{tan\alpha \times 77.2 \times 2048}{5 \times tan30°}$$

$$N_S = 54770 tan\alpha \qquad \text{Equation 7-6}$$

Where $s$ means the displacement of the triangular block on the $x$ axis; $L$ is the lead of the rolled ball screw, which is 5 mm. and $a$ is the angle of attack.

$$\alpha_m = tan(\alpha - \alpha_d) \times 54770 - (2 \times N_{encoder} - N_{motor}) \qquad \text{Equation 7-7}$$

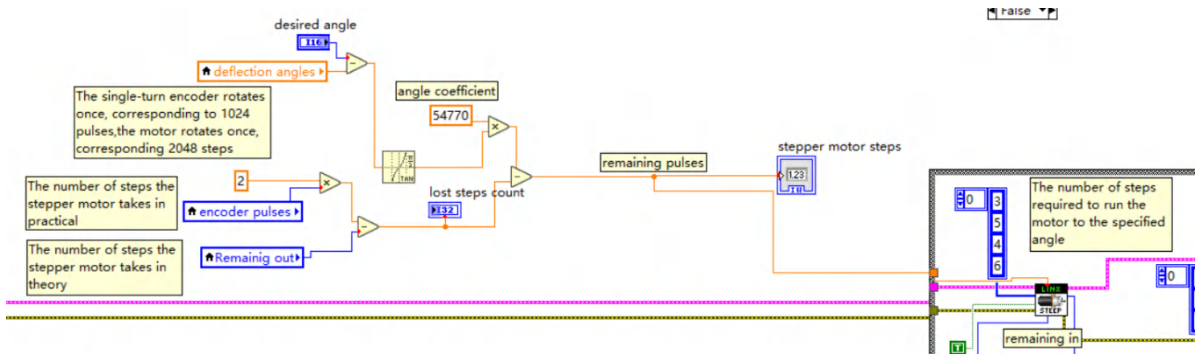

*Figure 74 Transmission function*

Where $\alpha$ is the desired angle and $\alpha_d$ means deflection angle and $C_{encoder}$ represents the transfer coefficient of the encoder, which is equal to 2 in this project. What's more, $N_{encoder}$ means encoder pulses, $N_{motor}$ is the remaining out steps of the motor and $\alpha_m$ represents the modified desired steps of the stepper motor.

The variation of the angle of attack caused by the drag force is given, the modified desired angle of attack is the desired angle of attack minus the angle offset caused by the drag force, the angle coefficient and the transfer coefficient of the encoder has already been mentioned before. Therefore, the lost steps of the stepper motor can be calculated by comparing 'encoder' times encoder transfer coefficient and the 'remaining out', which is the difference between the theoretical number of steps and the actual number of steps. This difference value is fed back to the stepper motor and the target step number is adjusted according to the difference value.

## 7.8 Programming results and analysis

Assume the measured pressure is 38N, and given the input desired angle is 15°, which can be set as the homing position to test whether the function of homing position works well, it can be done by Arduino and some sensor module, the results and setting parameters can be seen in Figure 75 and Figure 76 respectively. Figure 76 shows real-time data detection during the movement of the angle of attack adjustment. And Figure 76 shows the setting of the actuation system. From the Equation7-1 and Equation 7-2, the deflection angle is 0.68°.which can be seen in Equation 7-9.

$$\alpha_d = arctan\frac{d}{L_r} = \frac{arctan 0.0031 F_{loads} + 0.0007}{L_r} \qquad Equation\ 7\text{-}8$$

$$\alpha_d = arctan\frac{0.0031 \times 38 + 0.0007}{10} = 0.68°  \qquad Equation\ 7\text{-}7$$

From the front panel, the lost steps are 22 and the corrected desired motor steps can be calculated through Equation7-6 and Equation 7-7, which is equal to 13981, as calculated in Equation 7-9. This result is the same as shown in Figure 75, therefore, the virtual operation interface works well.

$$\alpha_m = tan(15 - 0.68) \times 54770 - 22 = 13981 \qquad Equation\ 7\text{-}9$$

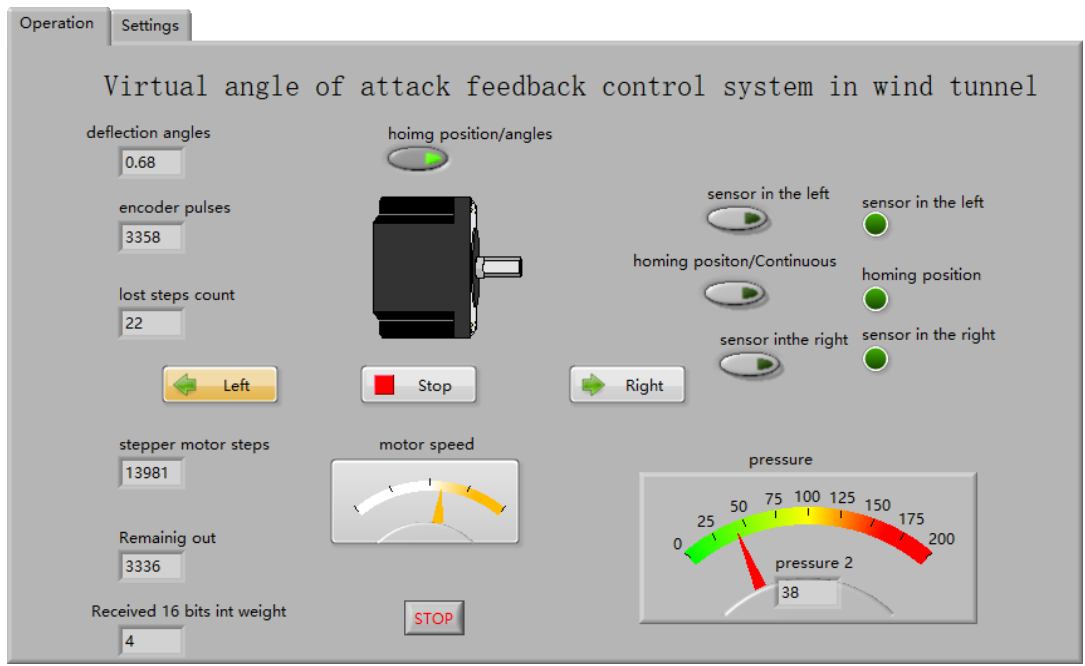

Figure 75 Front panel for real-time operation interface

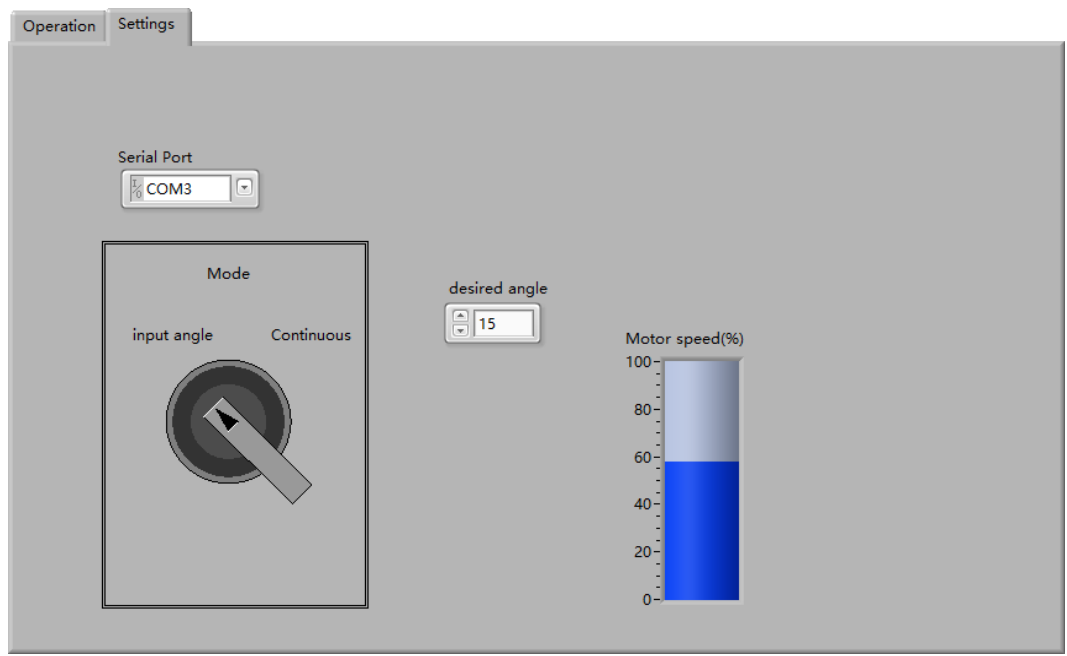

Figure 76 Front panel for setting

# 8 Conclusion and further work

The purpose of this research was to design an actuation system in the supersonic wind tunnel to save researchers' time and improve their working efficiency. This chapter provides a brief summary of this actuation system, a discussion of the limitations of this design and areas for future research in this field.

## 8.1 Summary

First of all, for the design of the mechanical structure, through the comparison of several conceptual designs, the optimal design was selected. The angle of attack is adjusted from -20 to 20 degrees and the mechanical support system of this project has no interference on the experimental results and aerodynamic shape imaging. In addition, finite element analysis was performed to ensure that the structure would not fail. In addition, the blockage ratio of the aerodynamic shape and the support system was calculated and the largest projected area was obtained. The projected shape of the model support system in this project is three rectangles. The area of the control model and the support system of the model is less than the required maximum projection area. If the projection area of the aerodynamic shape is increased, the thickness of the three plates of the support system can be appropriately reduced (without affecting the normal operation of the support system). Therefore, the design of the mechanical system of the support system is more flexible and solves the research problems raised.

The second is the design of the control system. The control design requirement of this project is to be able to achieve accurate elevation angle changes of the model. The main control object is a stepper motor and the input is the required angle of attack given by the researcher. By controlling the rotation of the stepper motor, the output of the control system is the change of the angle of attack of the aerodynamic shape. In order to achieve precise control of the angle of attack, a closed-loop control system is constructed, mainly for the closed-loop control of the stepper motor and the closed-loop feedback of the wind tunnel balance. In this project, a closed-loop feedback control system is constructed based on Arduino. The precise adjustment of the model support system is realised.

The third part of this project is a six-component strain gauge balance. In a supersonic wind tunnel, a wind tunnel balance is usually equipped to measure the force received by the model. Based on the size limitations of the supersonic wind tunnel in the University of Manchester and the mounting location of the support system, a six-component strain balance was designed and, through finite element analysis, this six-component strain balance can measure the force and moment in three directions and has good sensitivity.

The fourth is the virtual operation interface design of pitch movement for researchers and programming implementation. The project aims to save researchers' time and improve the efficiency of their experiments. Therefore, the control effect display, human-computer interaction interface and dynamic data display were completed based on LabVIEW. The project was based on the interactive programming of LabVIEW and Arduino. The model pitch angle control simulation platform was developed based on LabVIEW and the pitch angle control simulation was realised by Arduino UNO and other control modules. Each control module reached the expected control effect.

## 8.2 Limitations and Future work

While all aspects of this project were successfully completed, there are some limitations and some aspects that can be optimised, as follows.

In this project, only the programming control module has been verified and the finite element analysis has been done for the mechanical structure to make sure that it would not fail under the load in the wind tunnel. However, a prototype for the mechanical design and the whole system testing needs to be finished in the future.

The designed wind tunnel balance only carried out finite element analysis to predict the sensor performance but it hasn't been calibrated in the wind tunnel experiment. Meanwhile, the overall structure of the balance can be further optimised. In addition, the analysis of the active vibration suppression for the wind tunnel balance could be a potential research point, which could help us better obtain the aerodynamic parameters of the model.

The error analysis and optimisation of the mechanical structure haven't been finished yet, due to our inability to run a test in the wind tunnel, therefore, subsequent optimisation analysis could be carried out in the future.

The closed-loop control of this system is the stepper motor, combined with the photoelectric sensor as the zero point and the rotary encoder for step count. The overall control effect is still not as good as the servo motor and the servo motor can be purchased for control if funds allow.

The overall mechanical structure can also be further optimised, including the friction force and the optimisation of the model workspace.

The cabling for the actuation system in the University of Manchester is ethernet port, however, in this project, the cablings are a USB cable, F-M wires and M-M wire, therefore, this problem can be optimised in the future work.

# Appendix

## Drawing

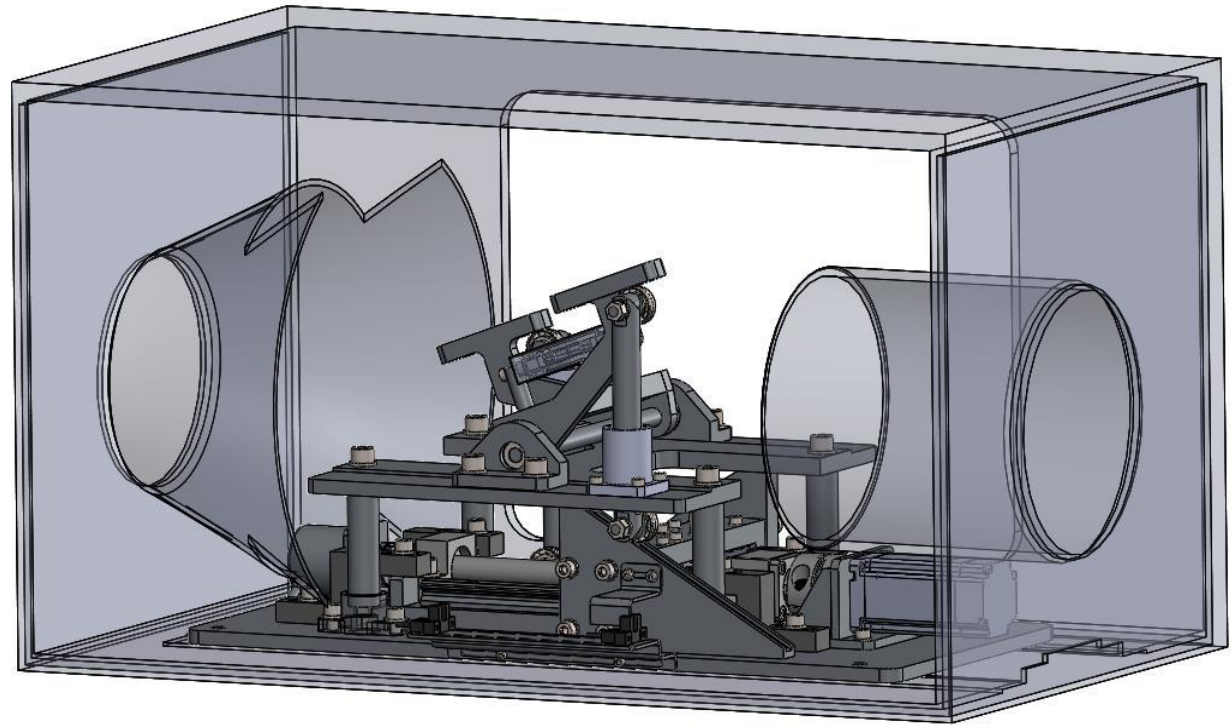

*Figure 77 Three-dimensional model of the actuation system*

*Figure 78 Final assembly*

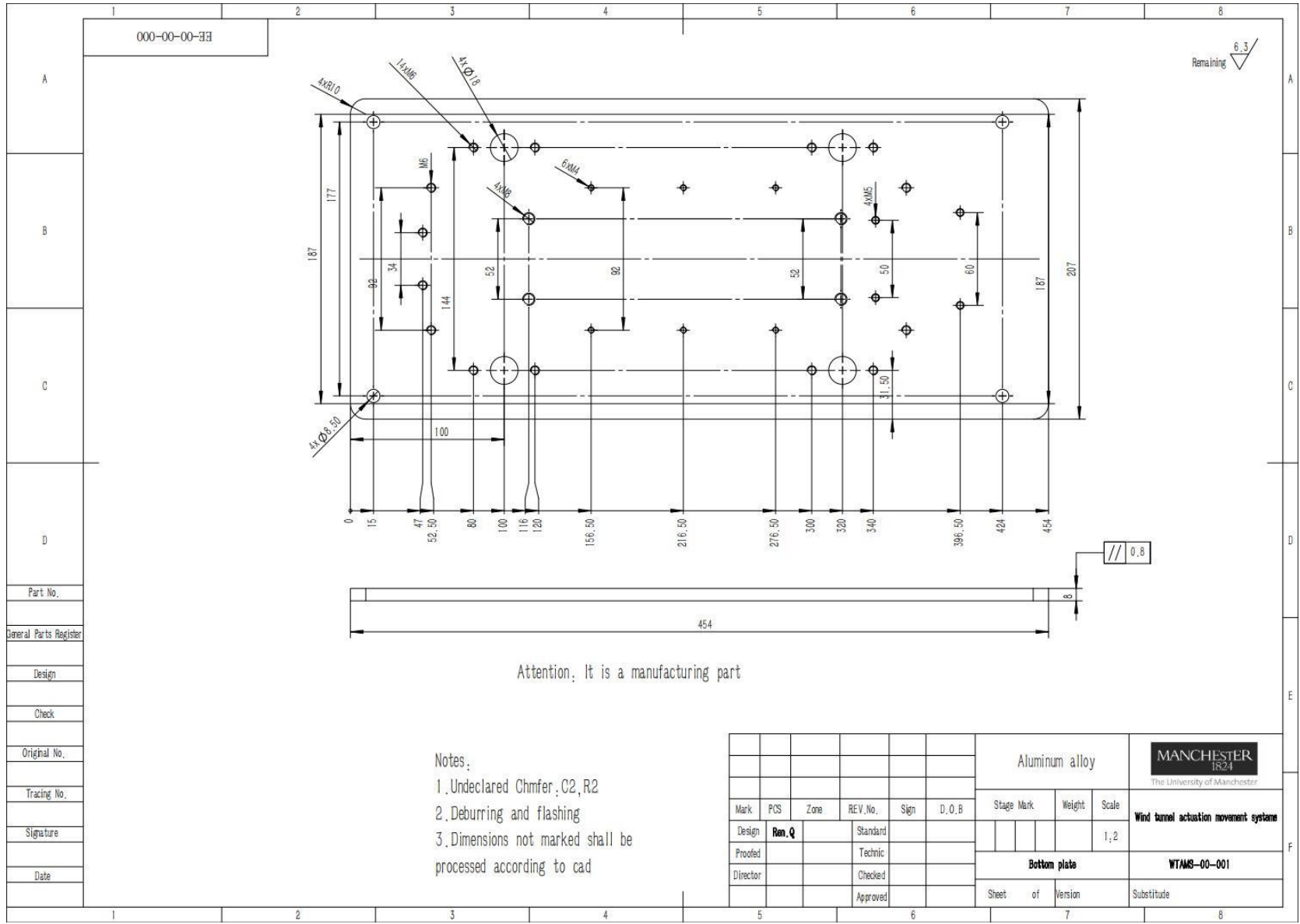

*Figure 79 Bottom plate*

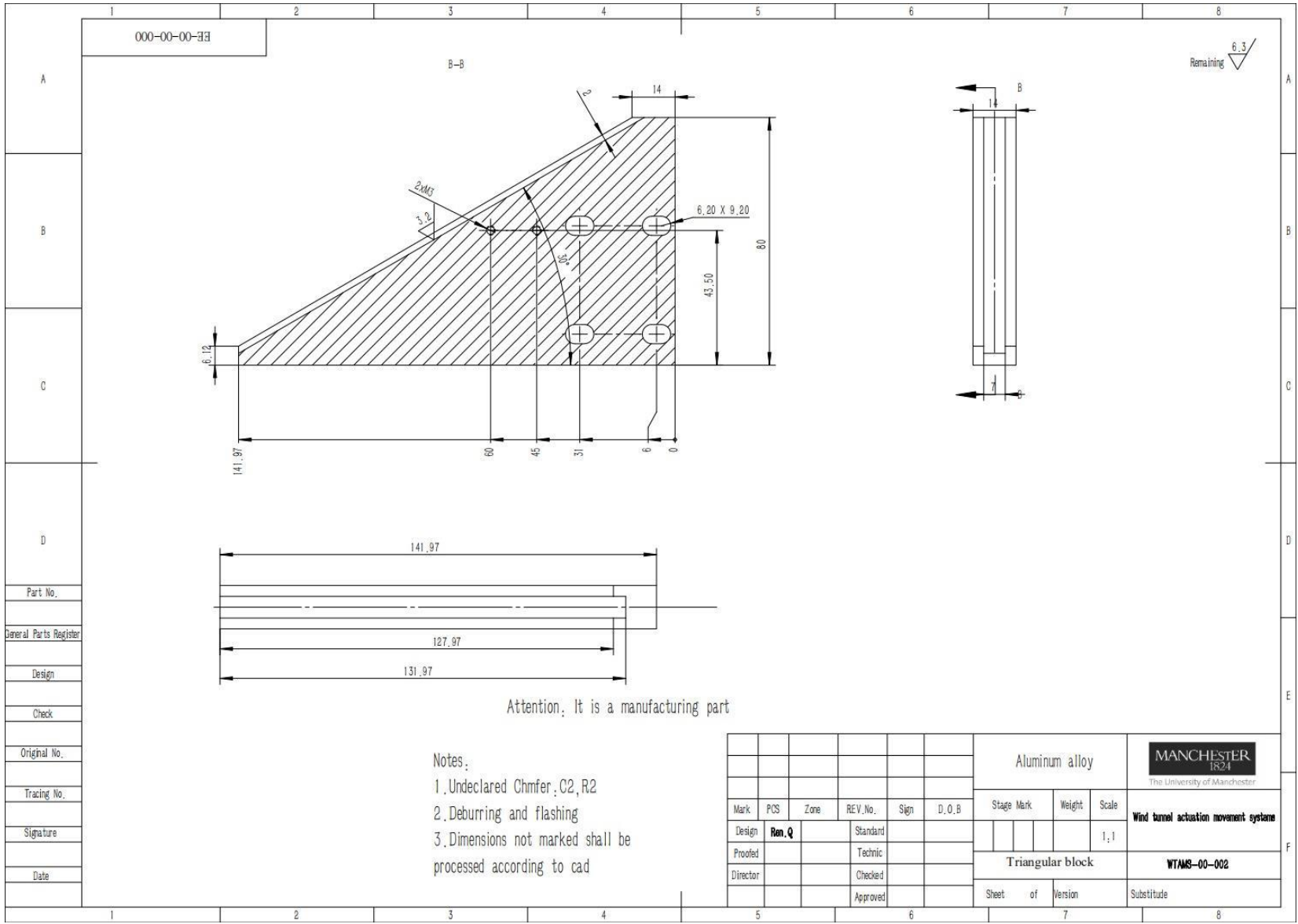

*Figure 80 Triangular block*

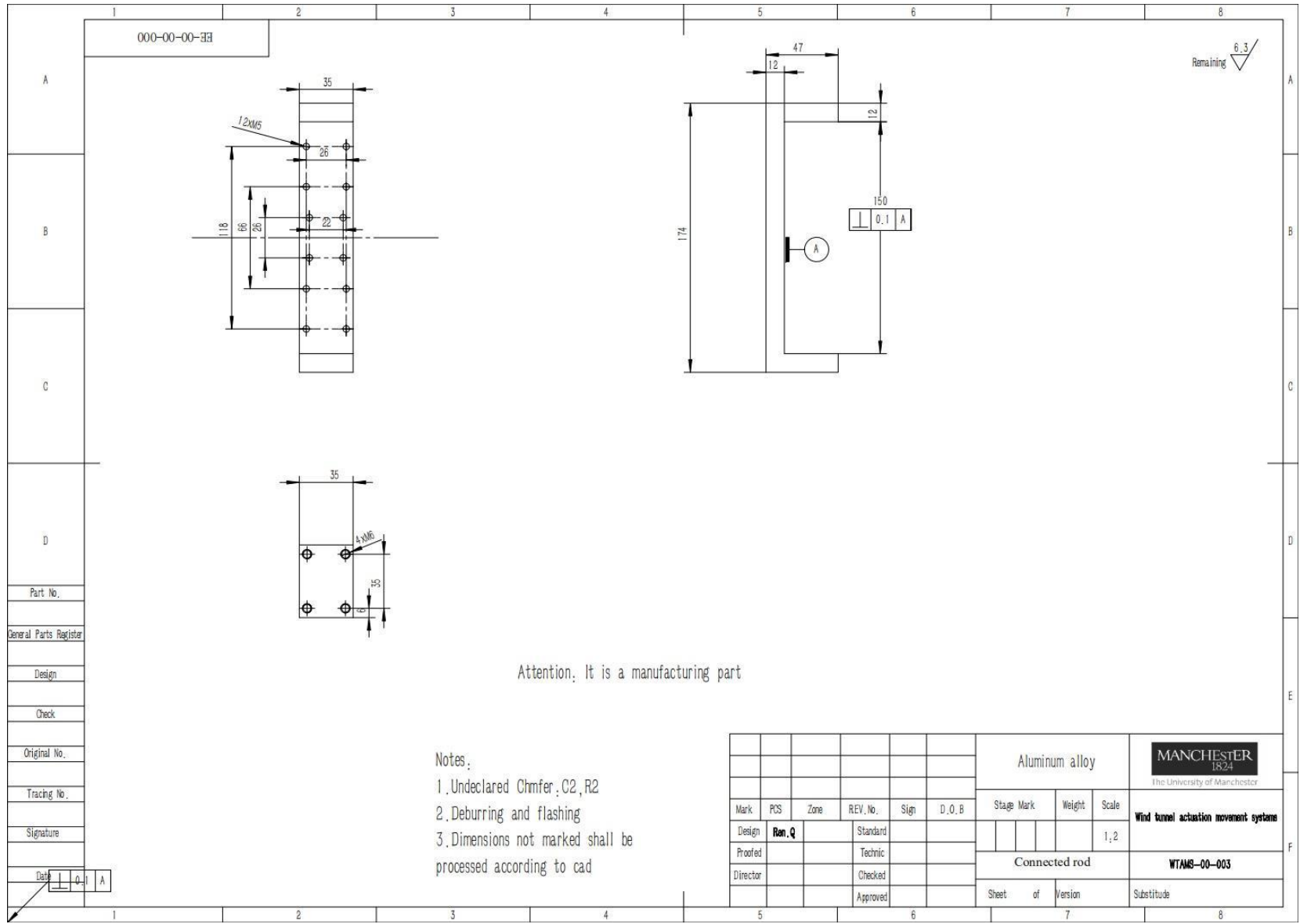

*Figure 81 Connected rod*

*Figure 82 Fixed column locking part*

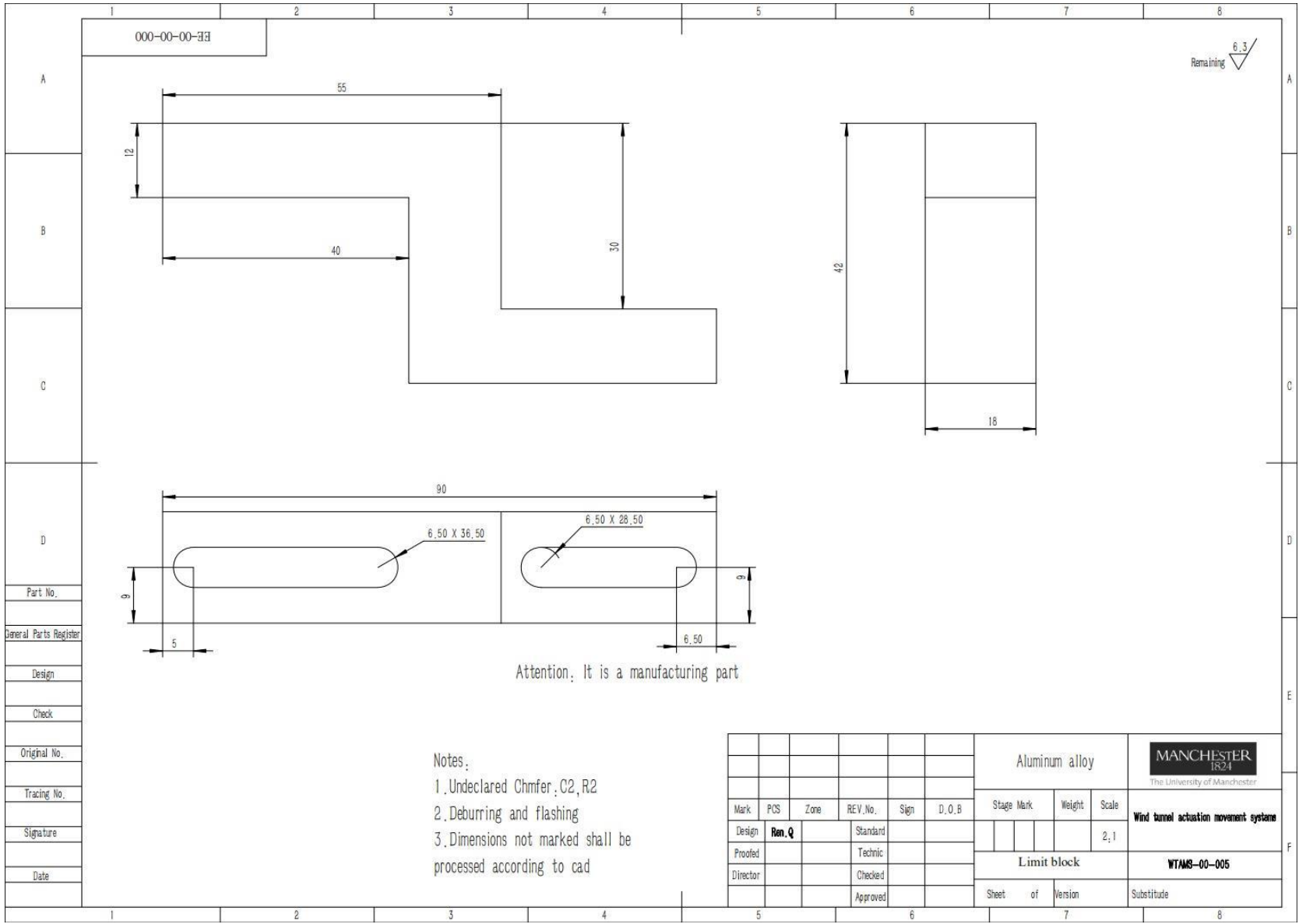

Figure 83 Limit block

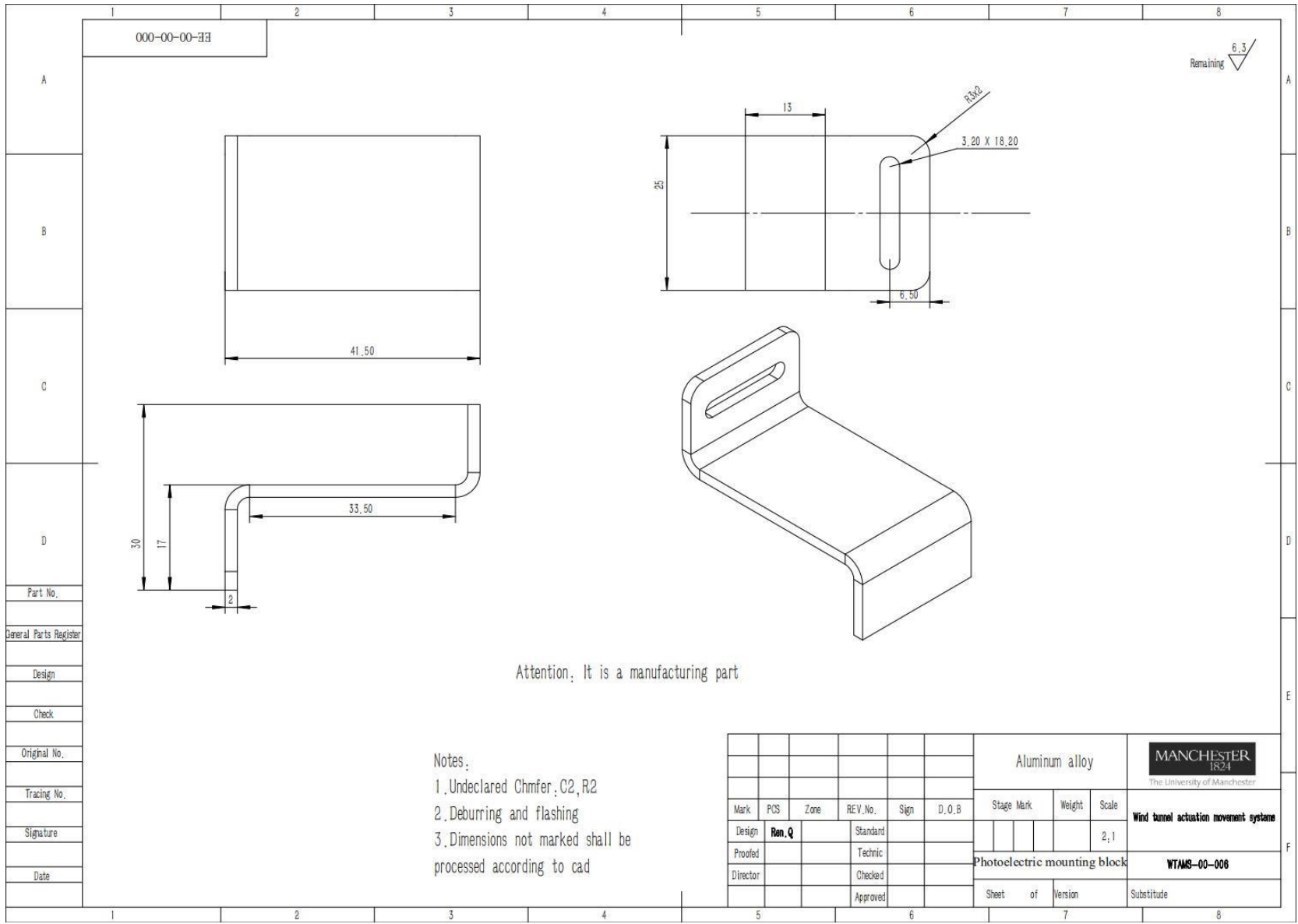

*Figure 84 Photoelectric mounting block*

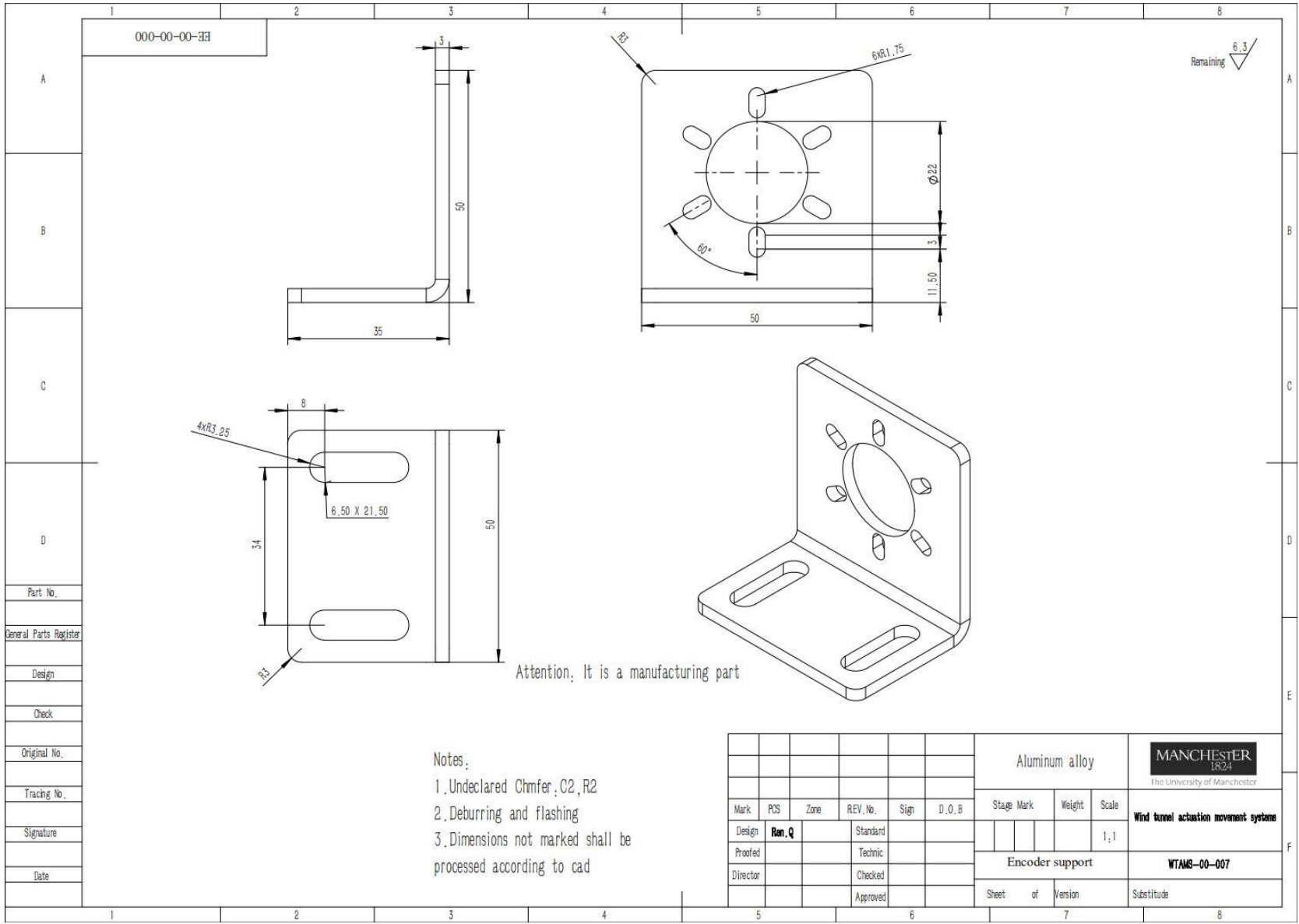

*Figure 85 Encoder support*

*Figure 86 Photoelectric mount*

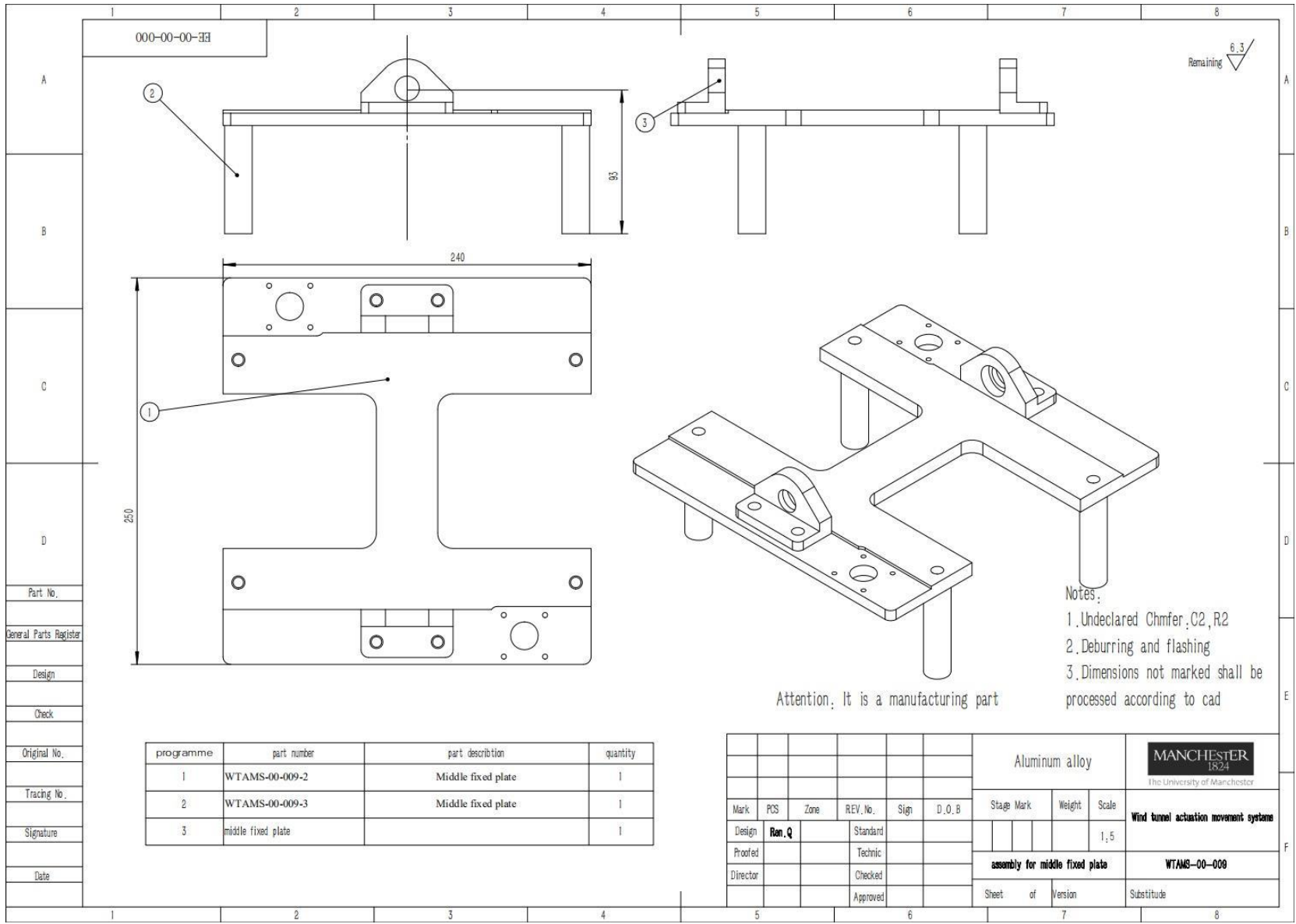

*Figure 87 Assembly for middle fixed plate*

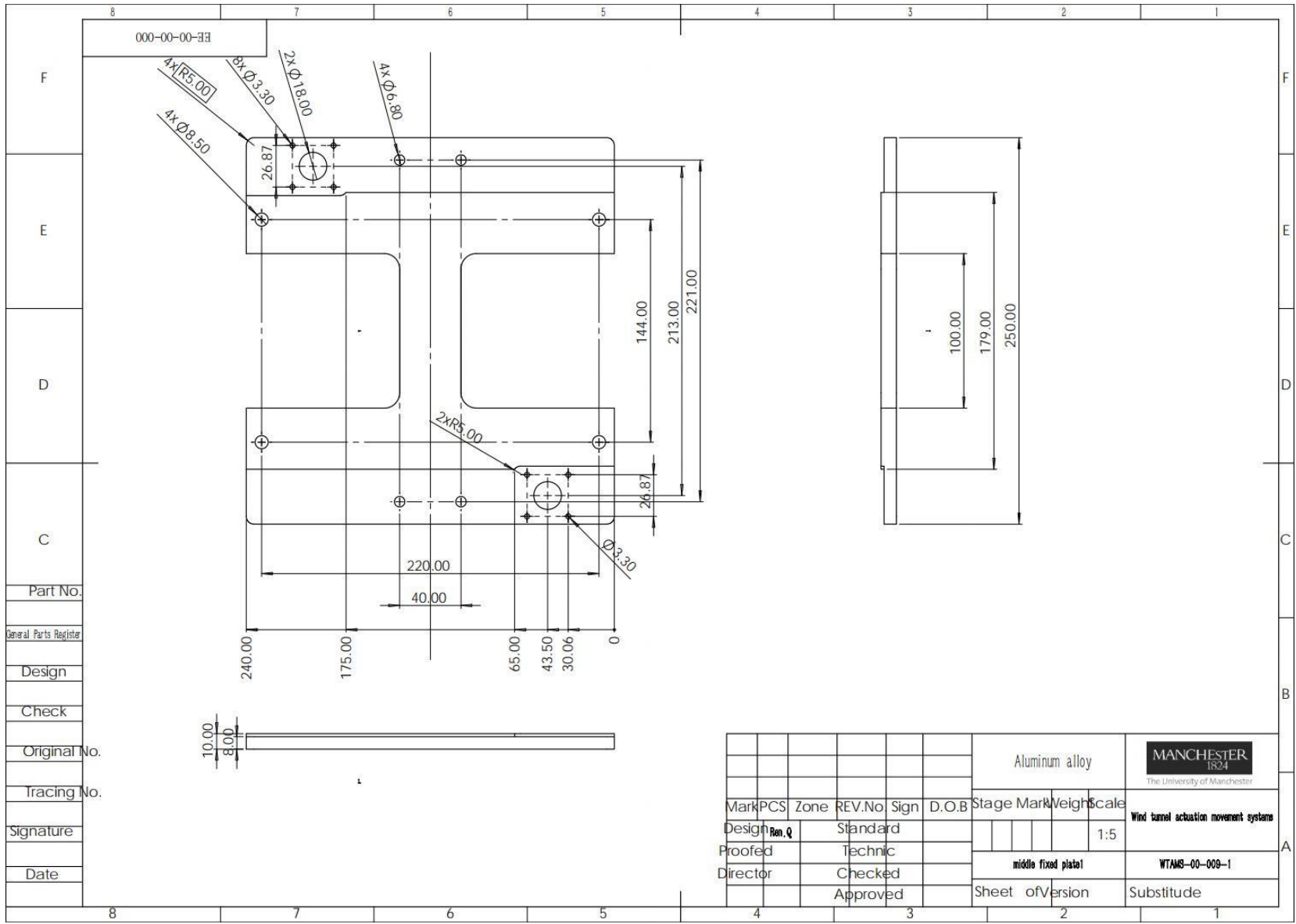

Figure 88 Middle fixed plate1

Figure 89 Middle fixed plate part2

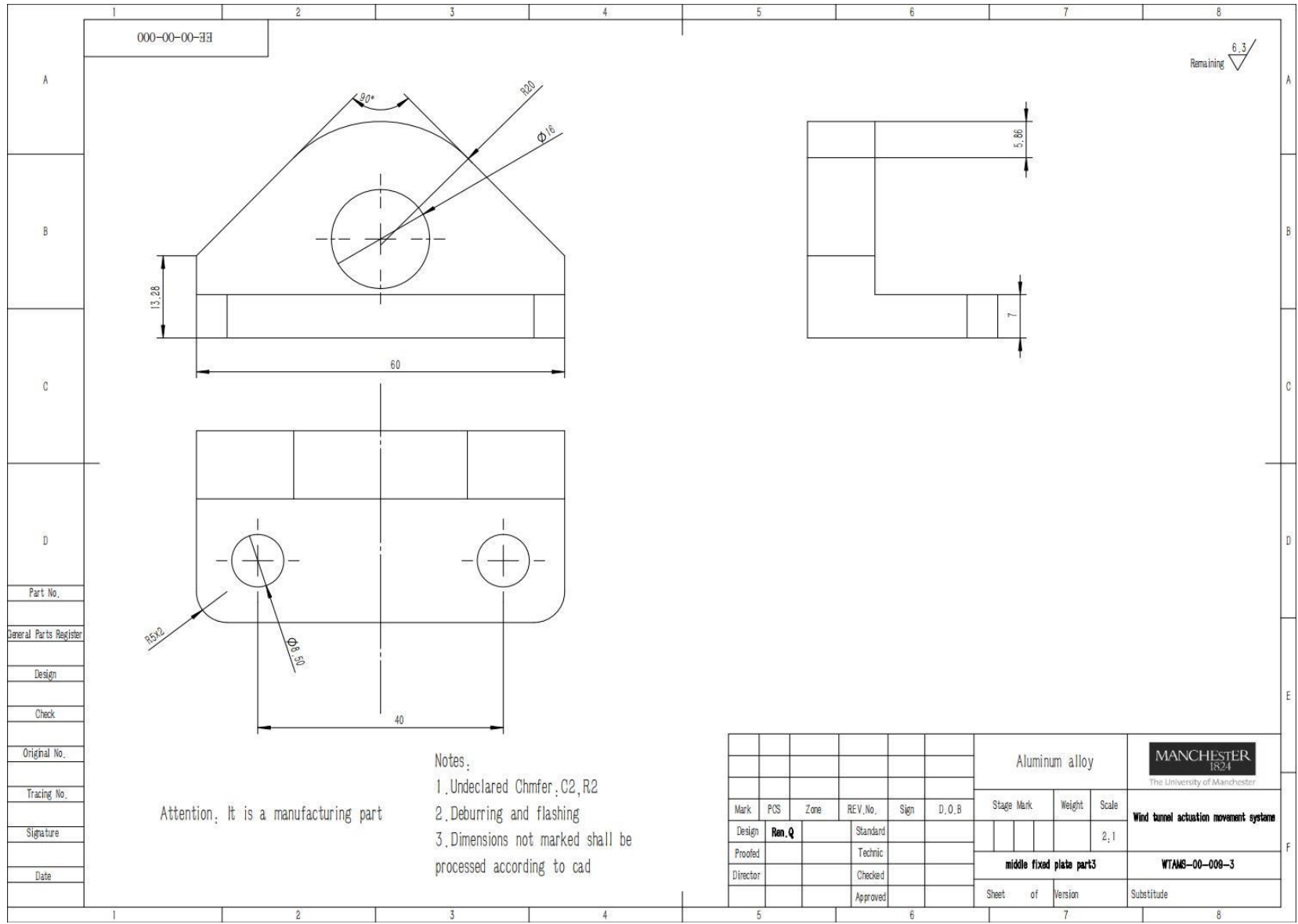

*Figure 90 Middle fixed plate part3*

*Figure 91 Sliding shaft*

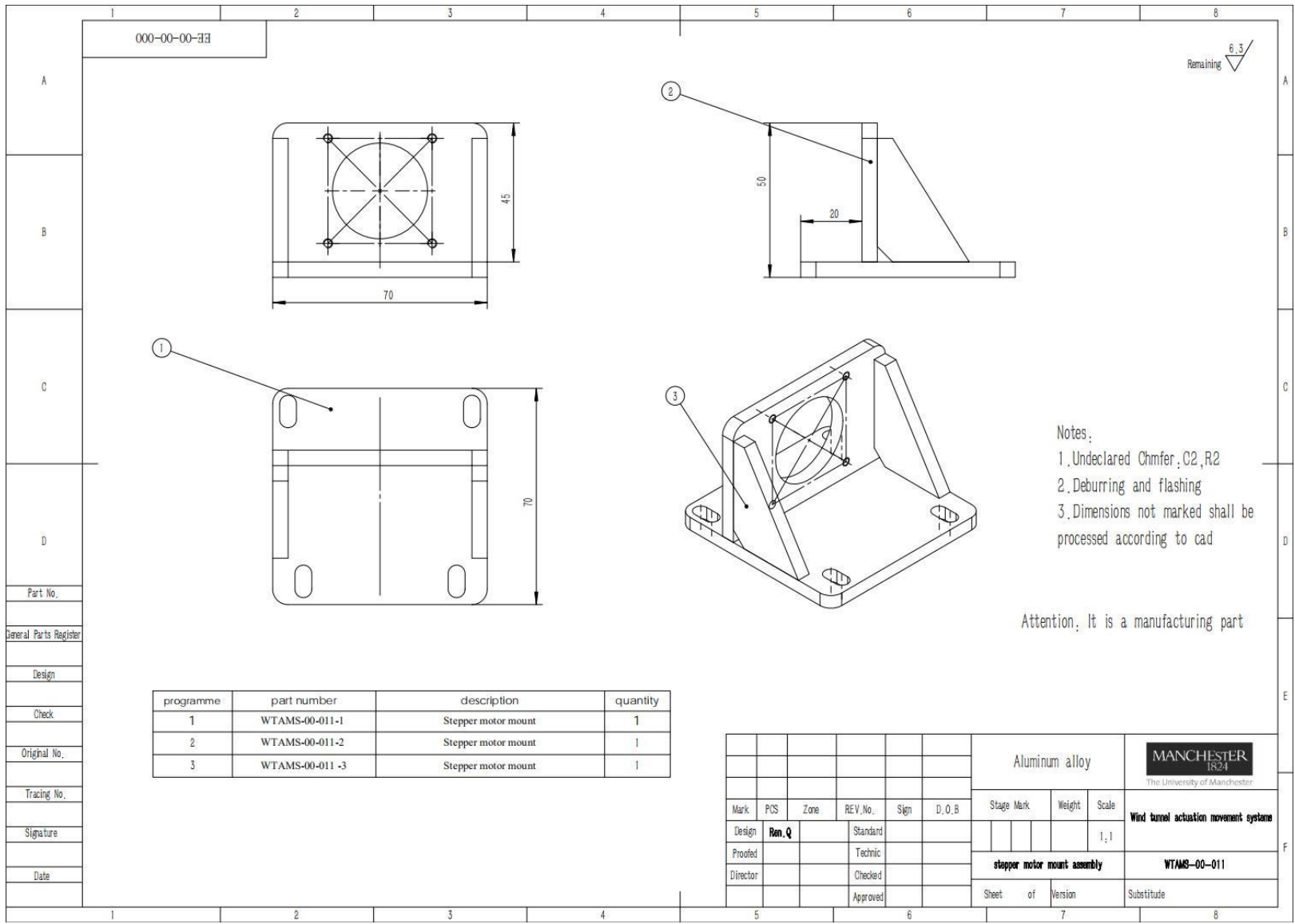

*Figure 92 Stepper motor mount assembly*

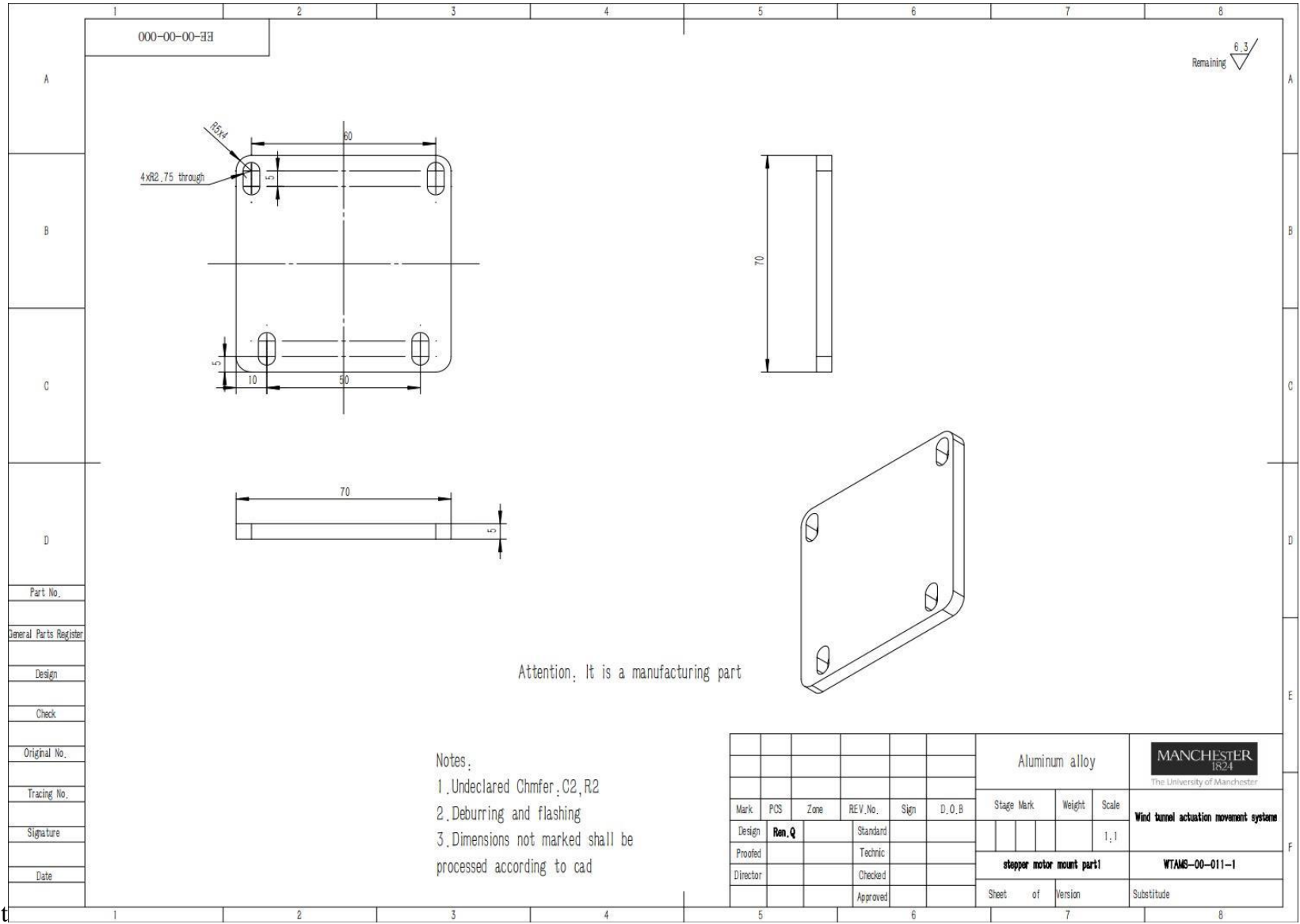

*Figure 93 Stepper motor mount part1*

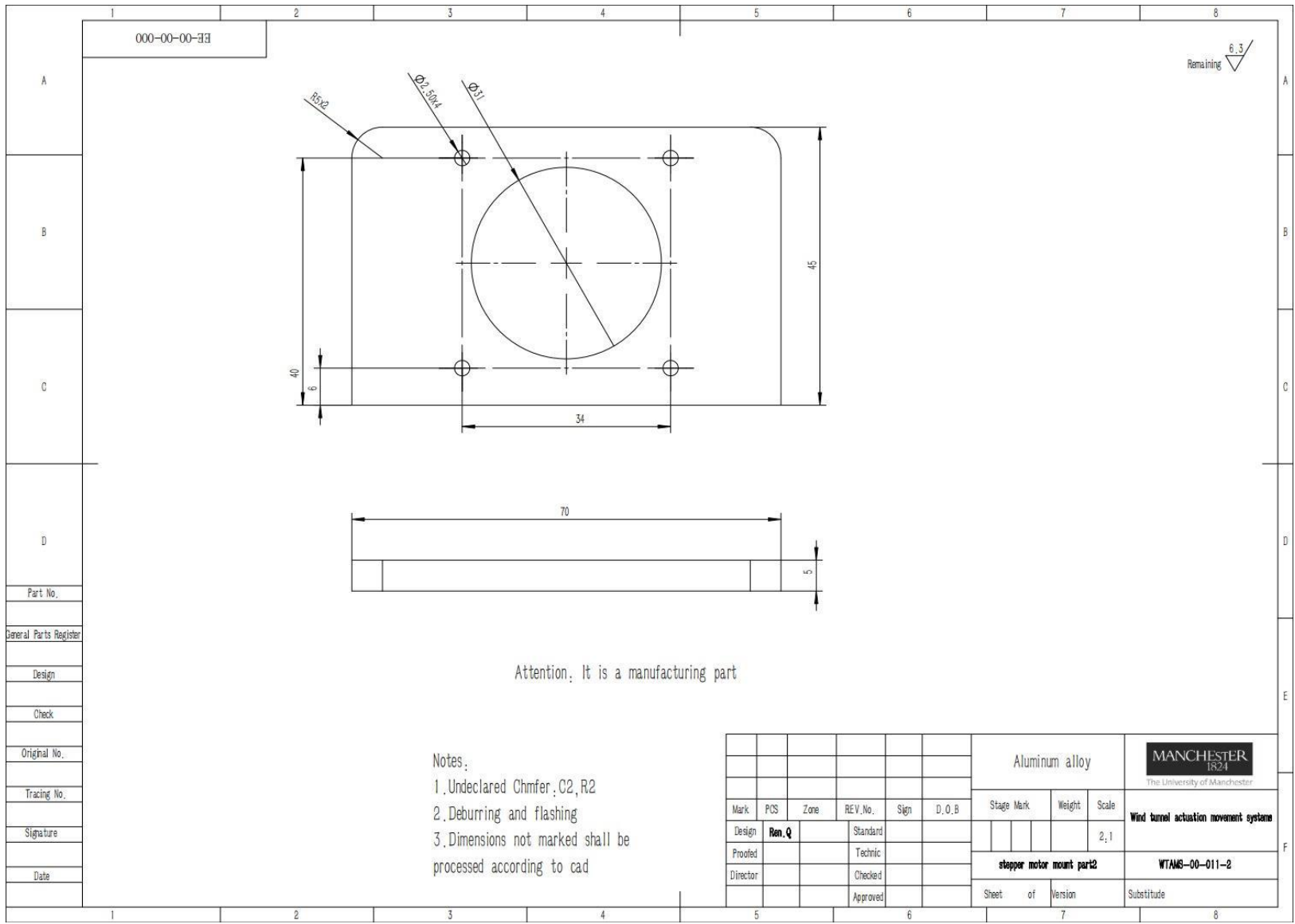

*Figure 94 Stepper motor mount part2*

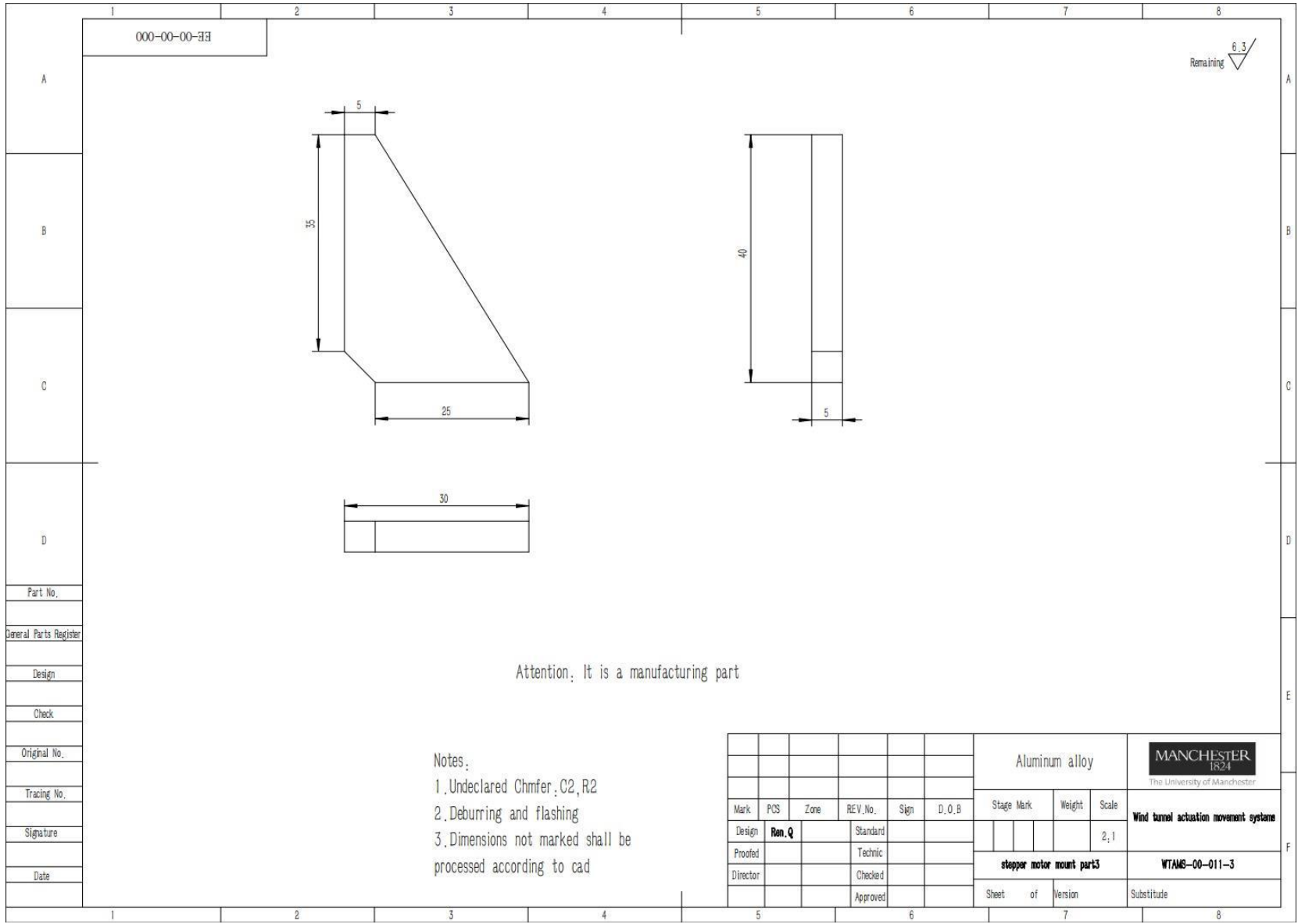

*Figure 95 Stepper motor mount part3*

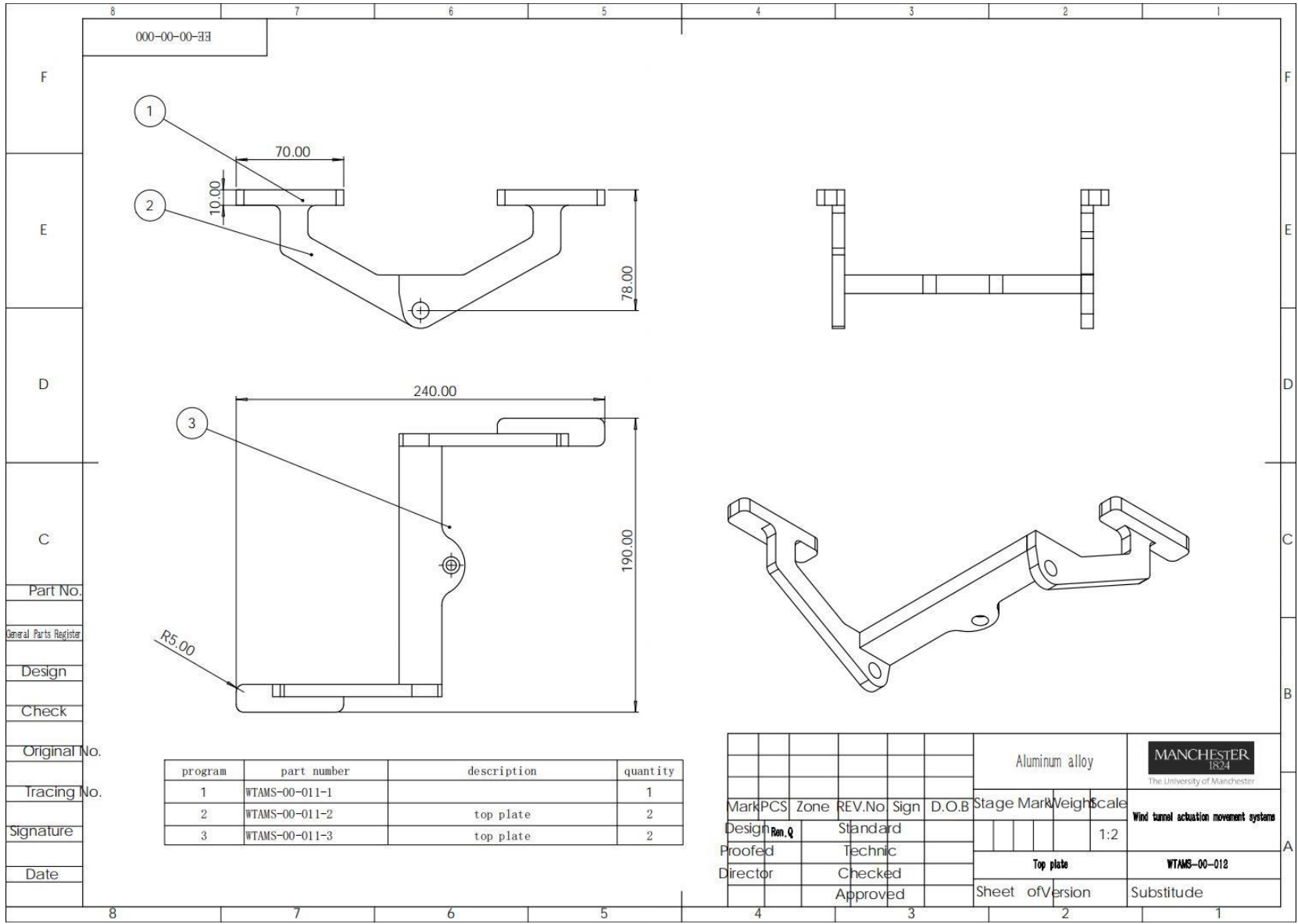

*Figure 96 Top plate assembly*

*Figure 97 Top plate part 1*

*Figure 98 Top plate part 2*

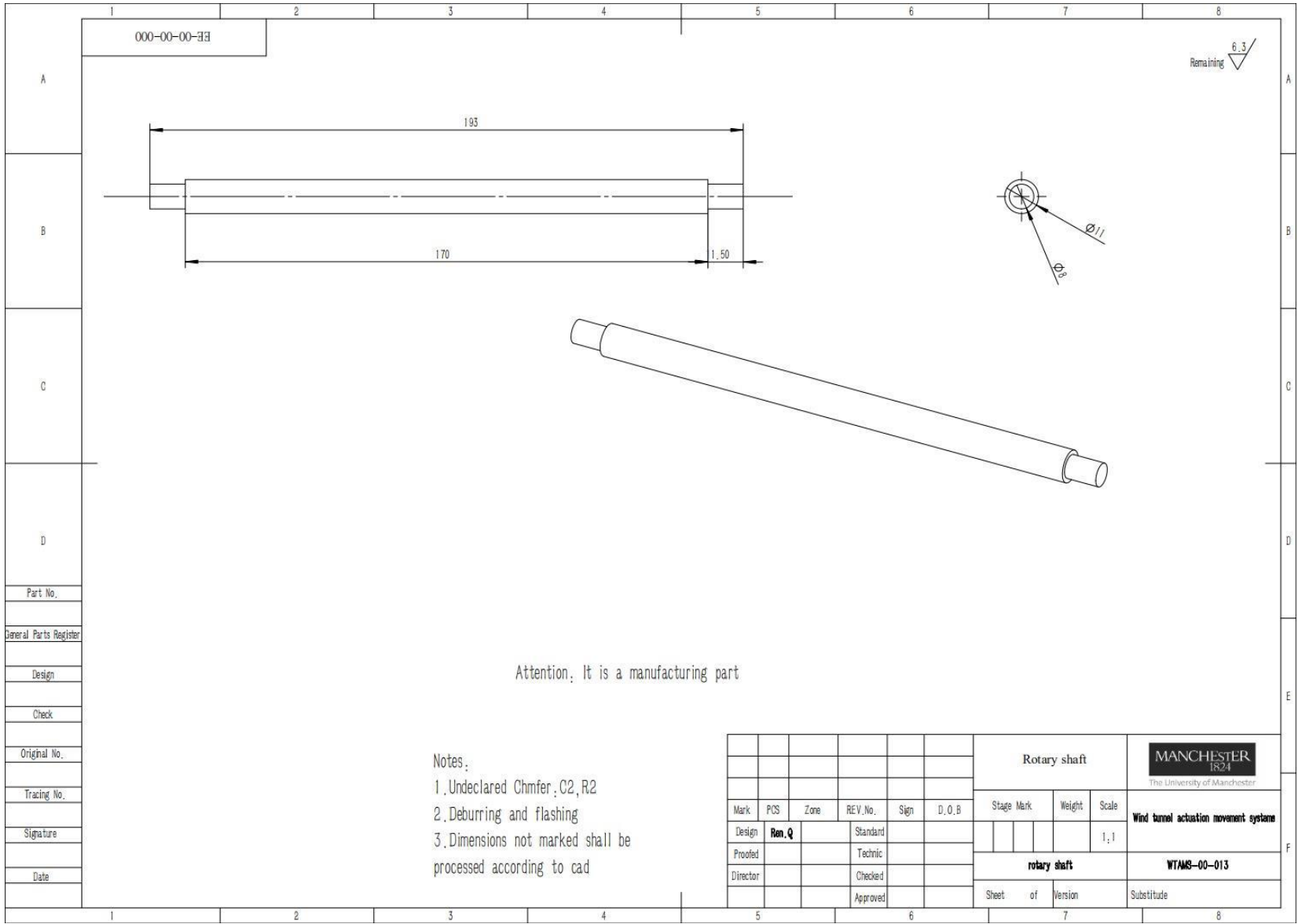

*Figure 99 Rotary shaft*

*Figure 100 Model supporting rod*

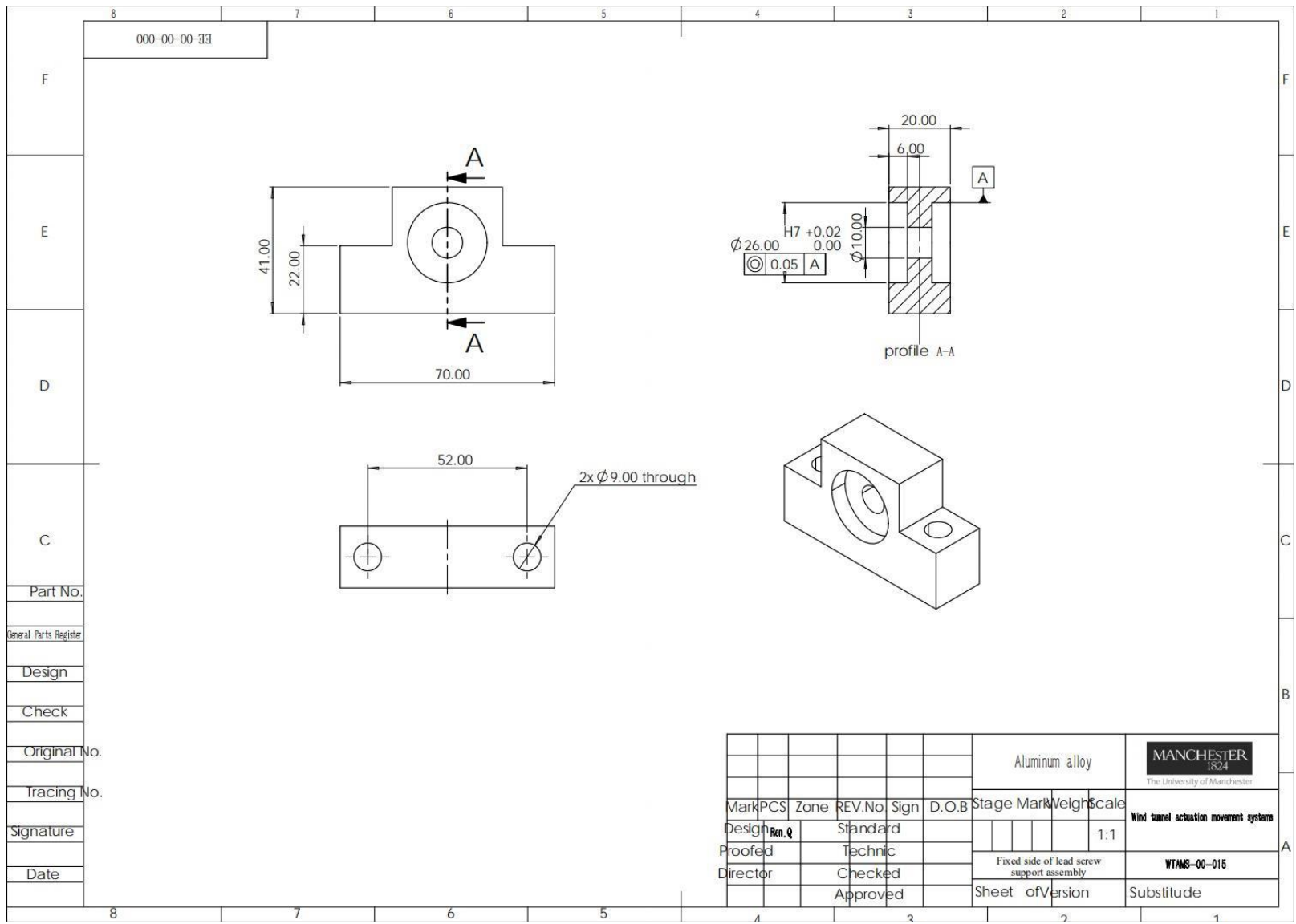

*Figure 101 Fixed side of the lead screw support assembly*

*Figure 102 Support side 1*

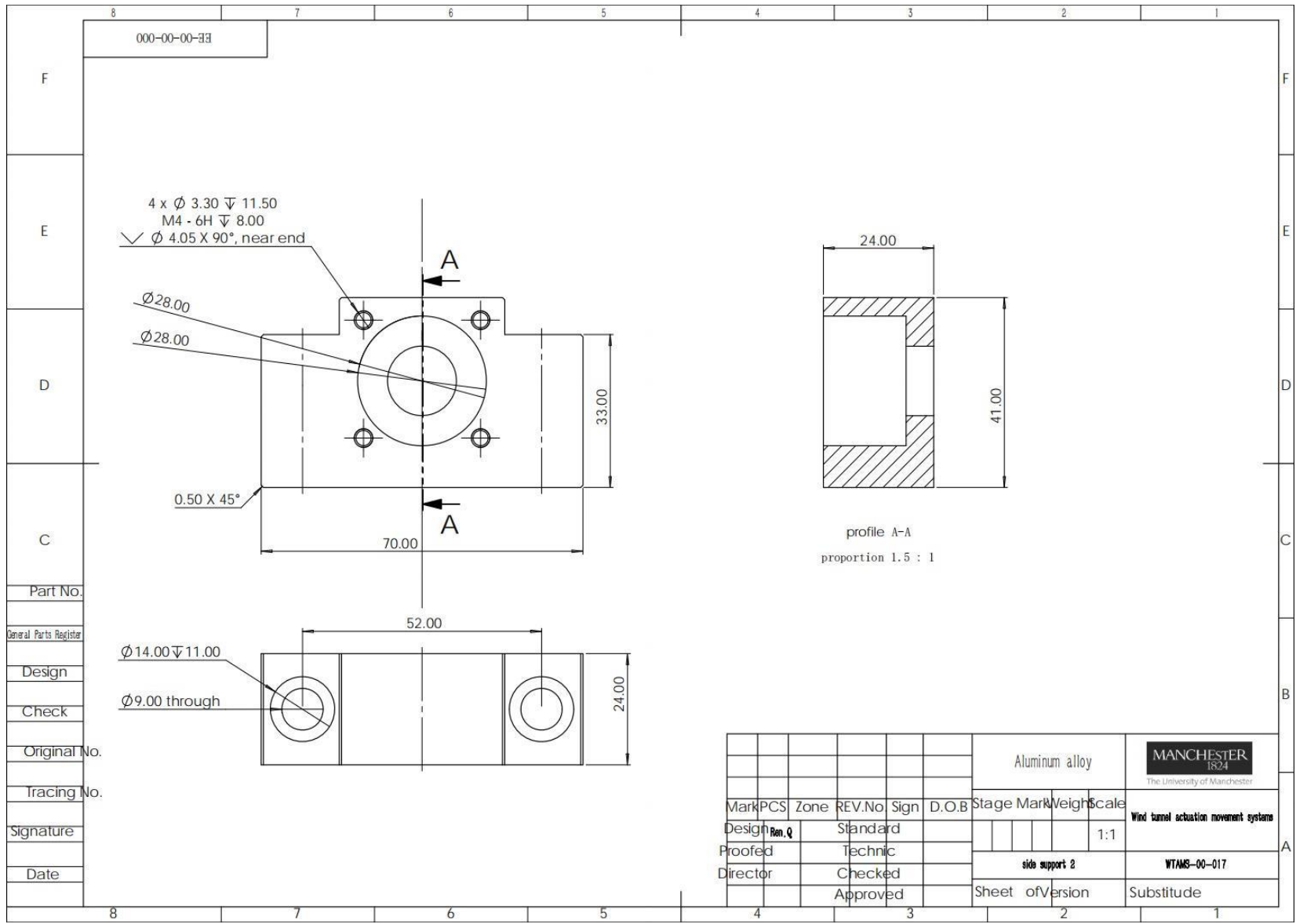

*Figure 103 Side support 2*

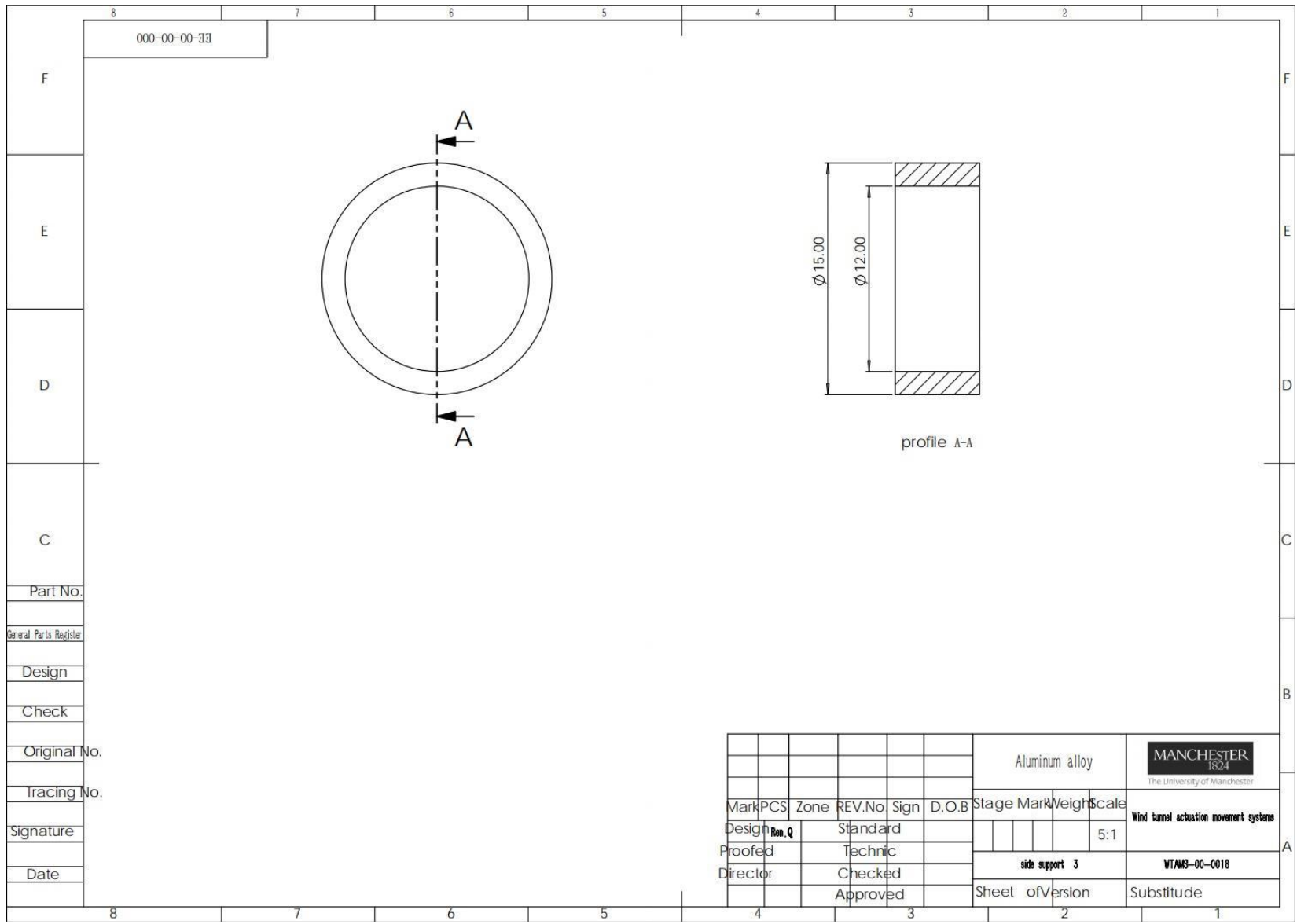

*Figure 104 Side support 3*

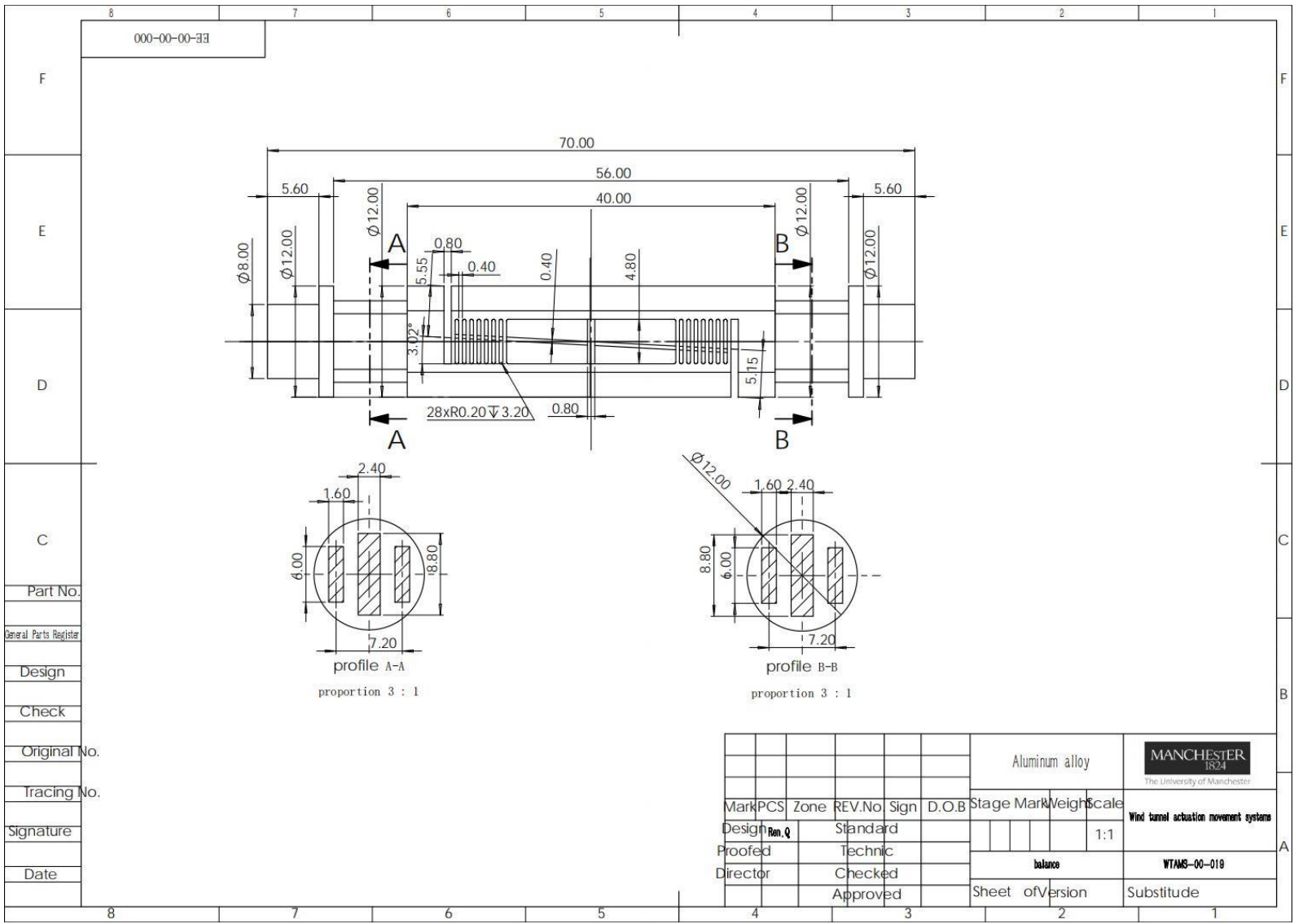

*Figure 105 Wind tunnel balance*

# LabVIEW code

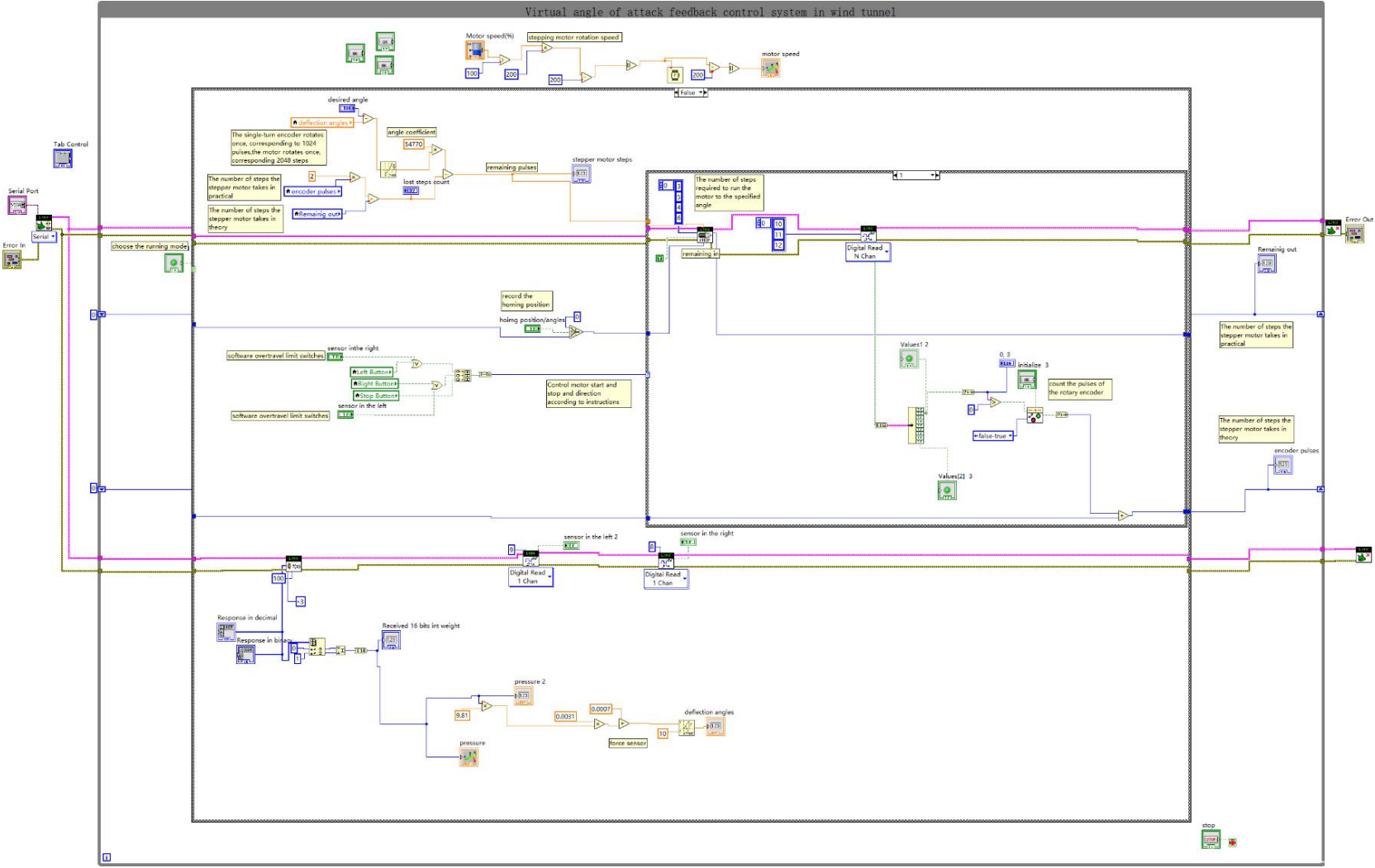

*Figure 106 Input angle mode-left direction*

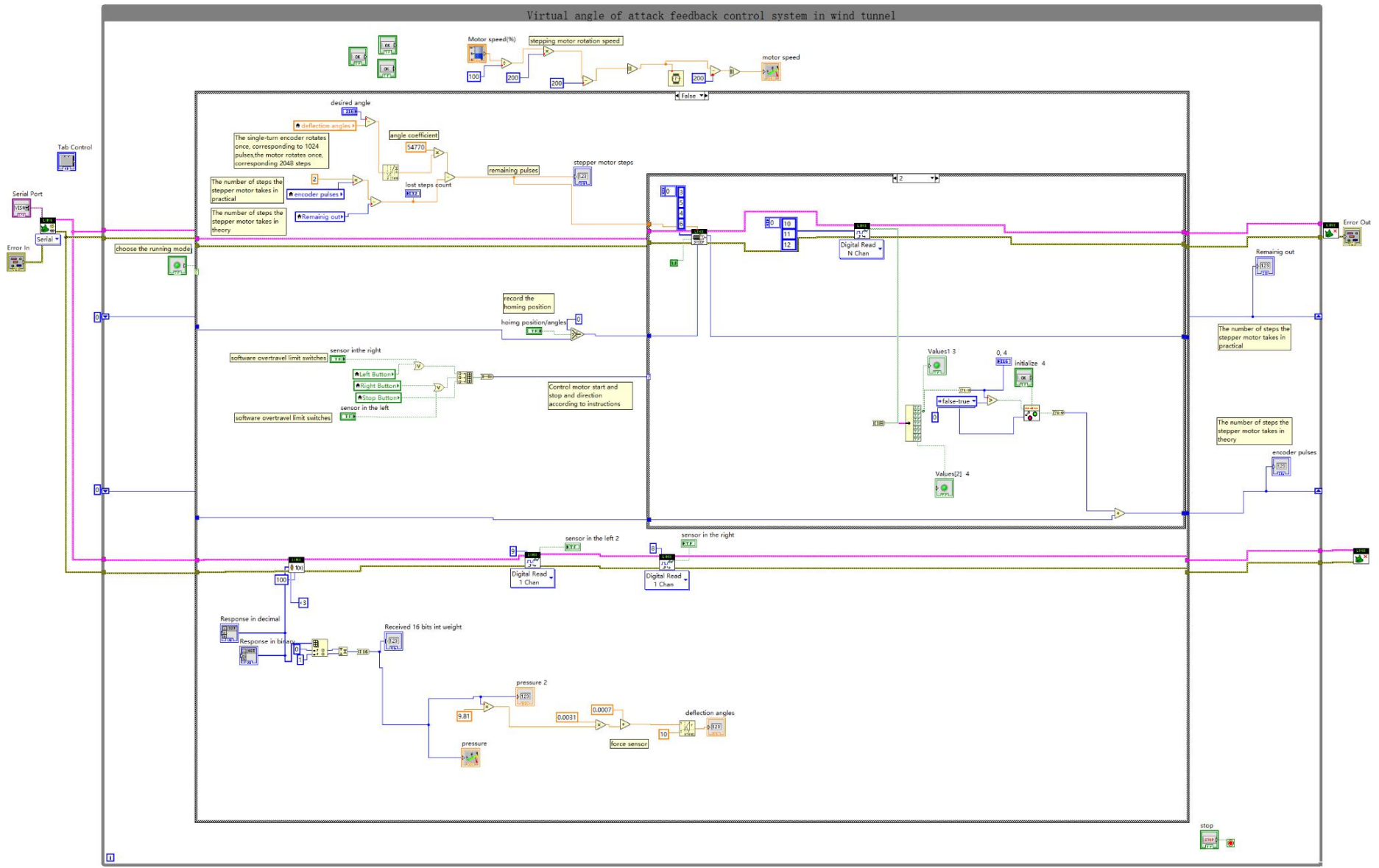

*Figure 107 Input angle mode-right direction*

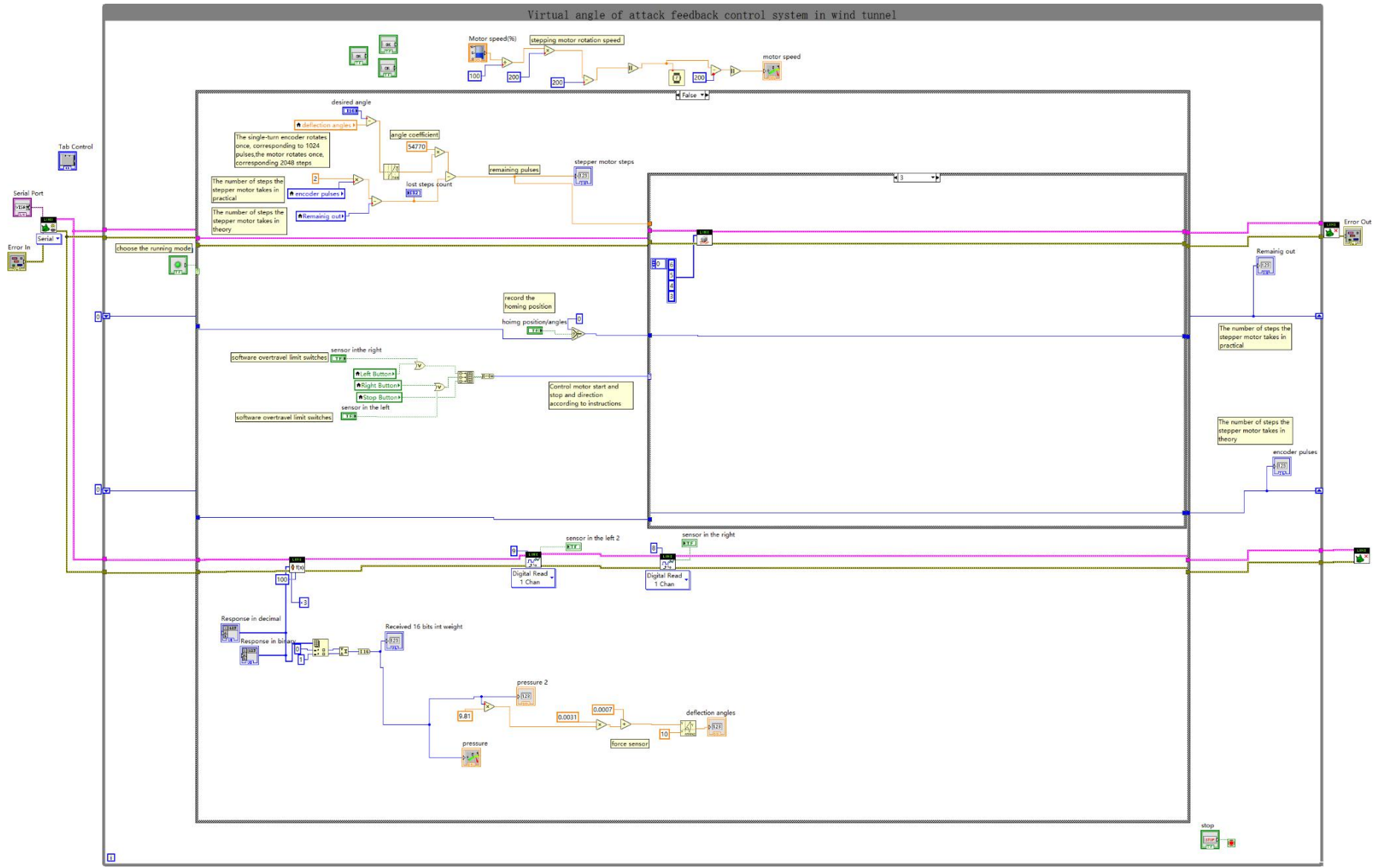

*Figure 108 Input angle mode-stop*

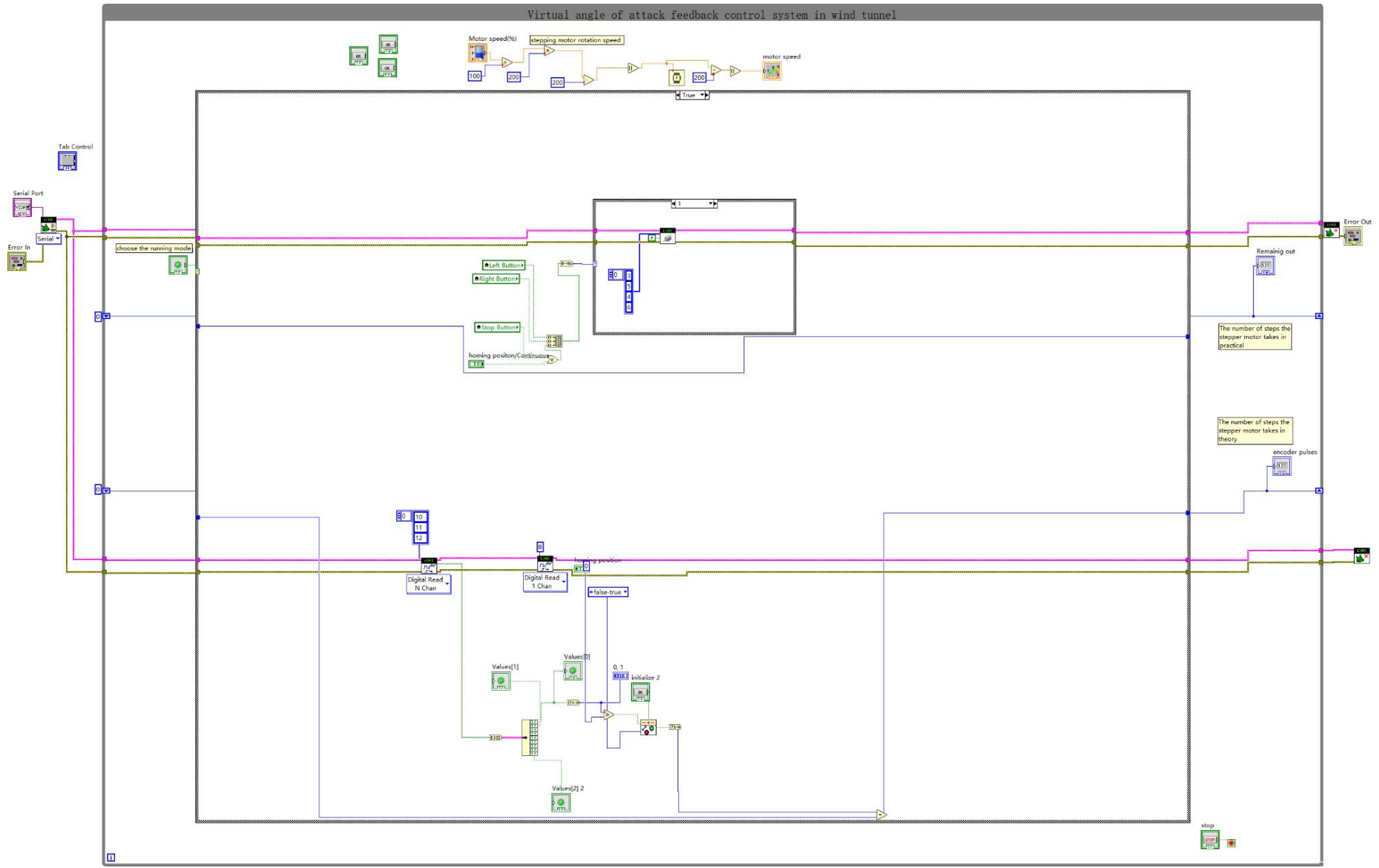

*Figure 109 Continous mode-left direction*

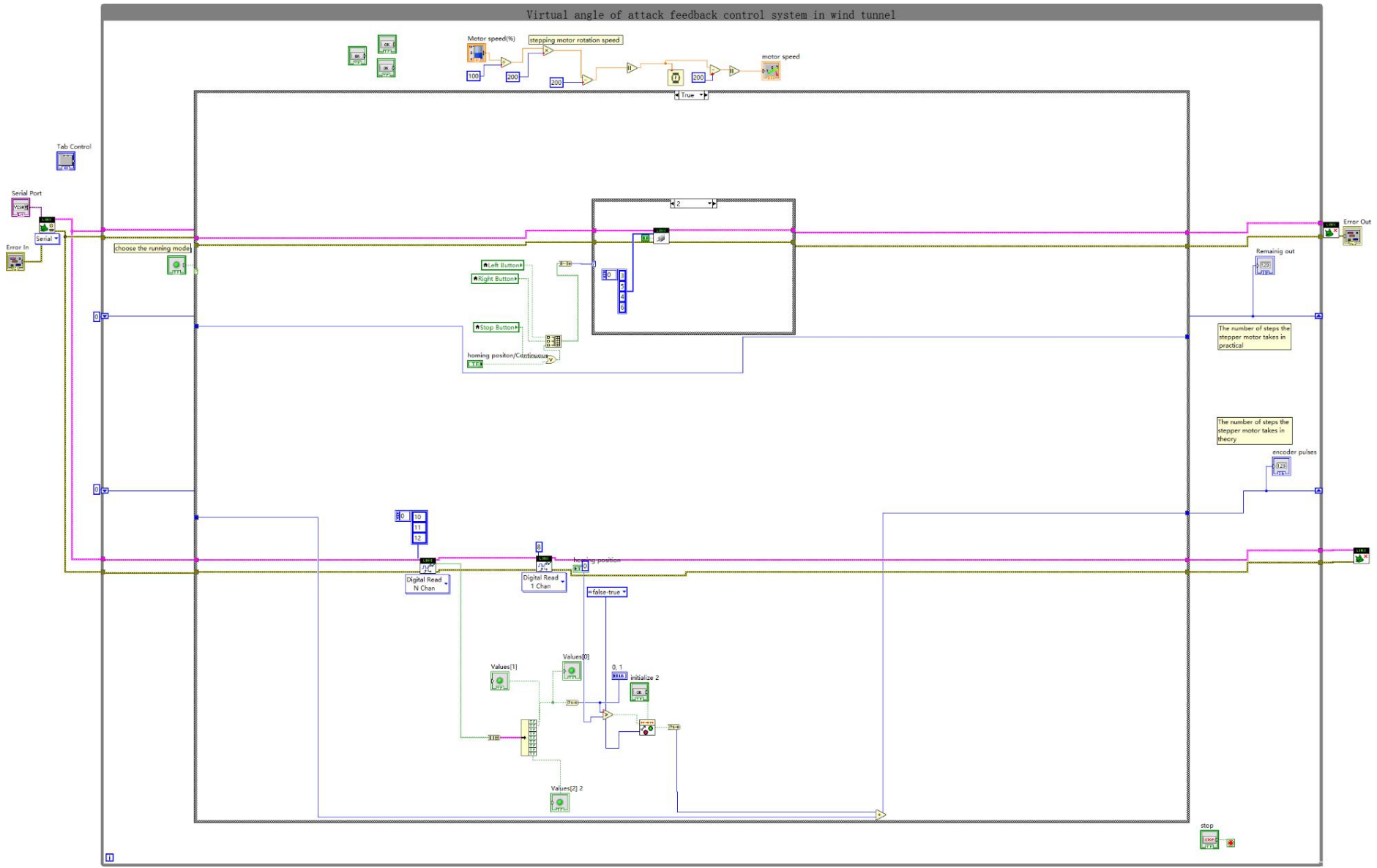

*Figure 110 Continous mode-right direction*

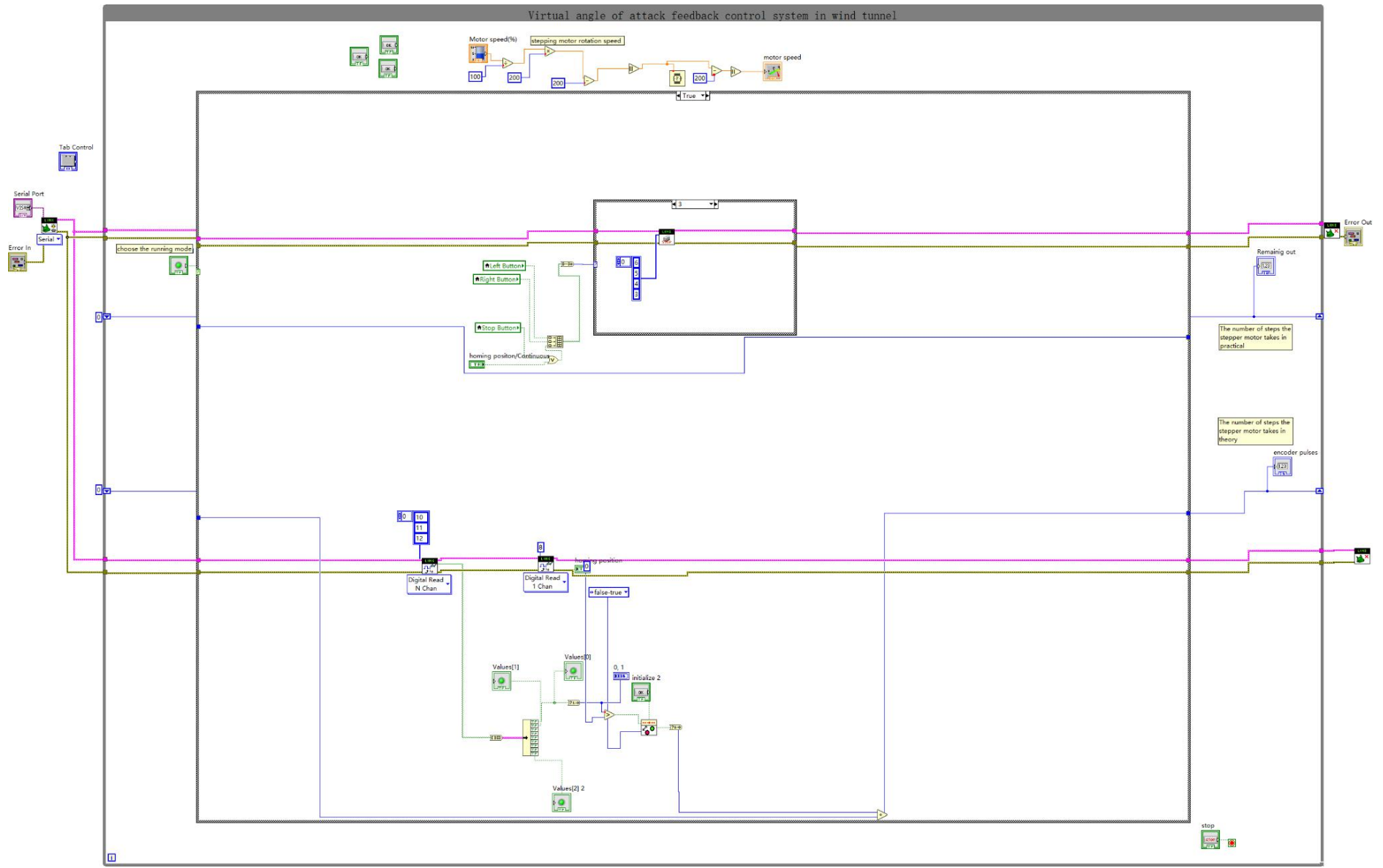

*Figure 111 Continuous mode-stop*

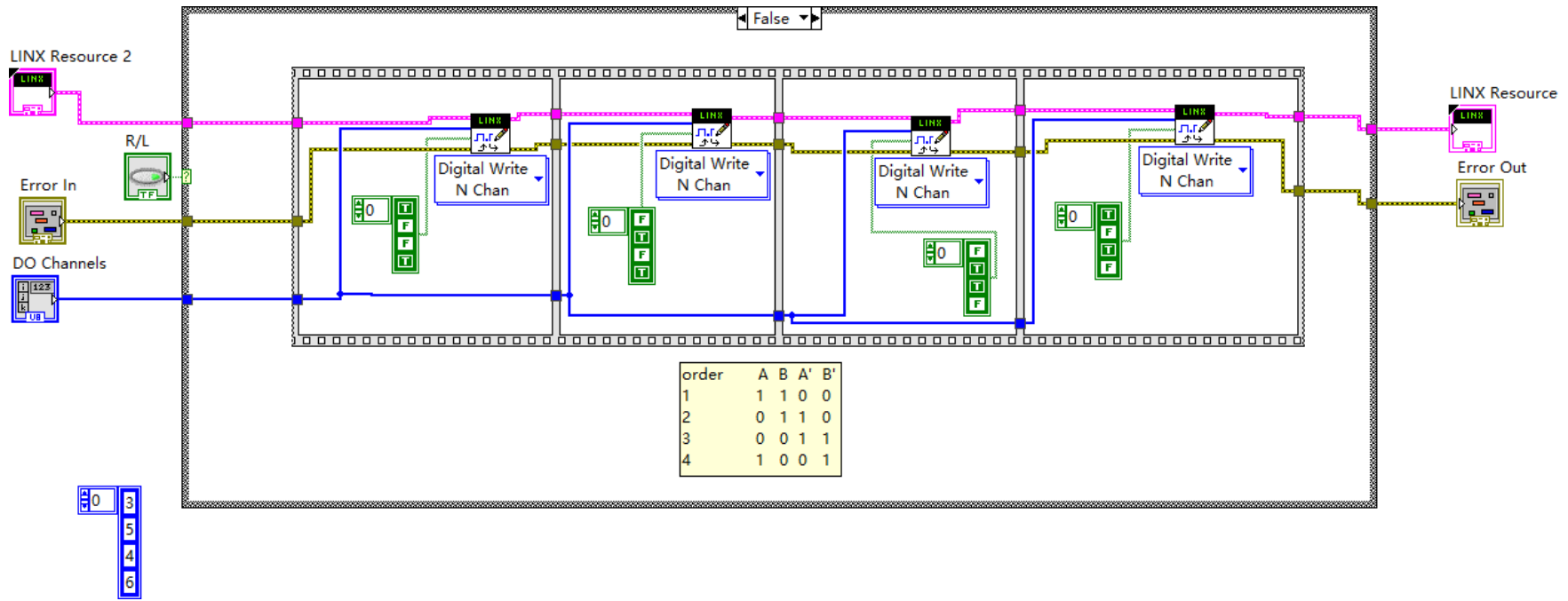

*Figure 112 SubVI of continuous mode*

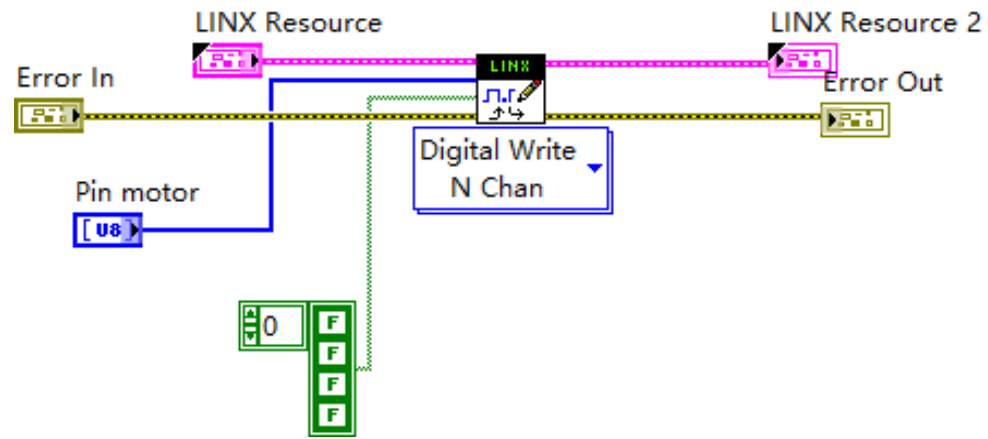

*Figure 113 SubVI of stop mode*

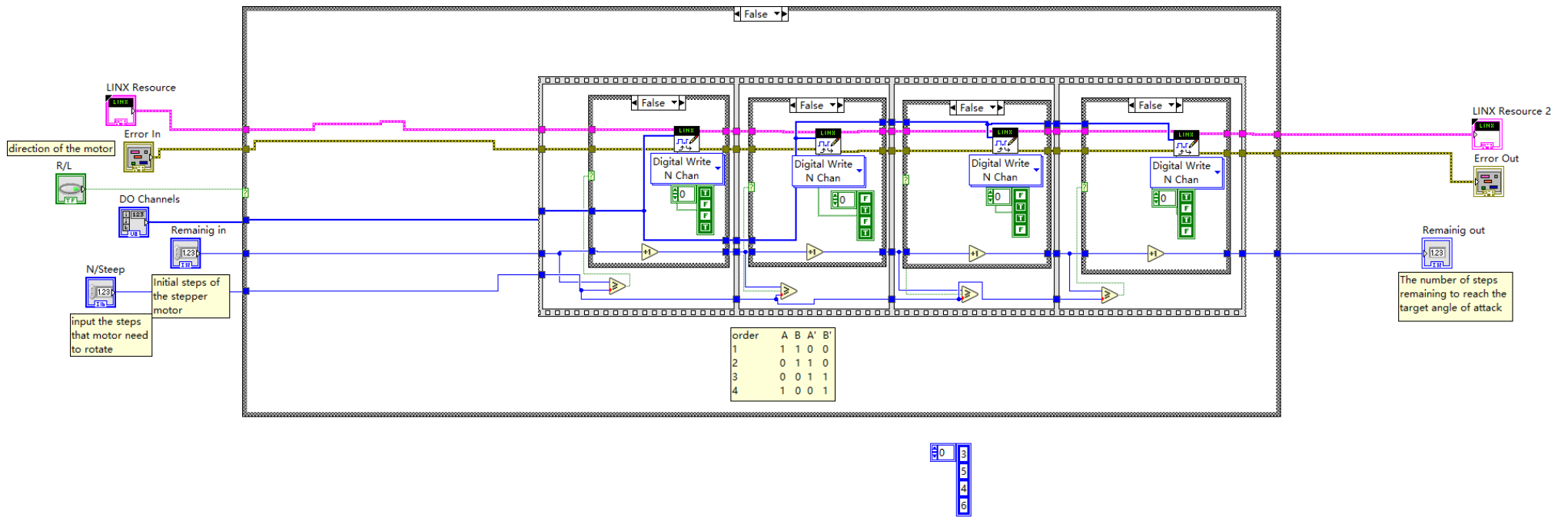

*Figure 114 SubVI of desired angles mode*

## Arduino code for custom command

```
1.  //Include All Peripheral Libraries Used By LINX
2.  #include <SPI.h>
3.  #include <Wire.h>
4.  #include <EEPROM.h>
5.  #include <Servo.h>
6.
7.  //Include Device Sepcific Header From Sketch>>Import Library (In This Case LinxArduinoUNO.h)
8.  //Also Include Desired LINX Listener From Sketch>>Import Library (In This Case LinxSerialListener.h)
9.  #include <LinxArduinoUno.h>
10. #include <LinxSerialListener.h>
11.
12. //Create A Pointer To The LINX Device Object We Instantiate In Setup()
13. LinxArduinoUno* LinxDevice;
14.
15. // Library inclusion, pin declaration and initialisation:
16. #include <Q2HX711.h>
17. const byte hx711_data_pin = 13; // HX711 data line connected to arduino D13
18. const byte hx711_clock_pin = 2; // HX711 clock line connected to arduino D2
19.
20. Q2HX711 hx711(hx711_data_pin, hx711_clock_pin);
21.
22. //Initialize LINX Device And Listener
23. void setup()
24. {
25.    //Instantiate The LINX Device
26.    LinxDevice = new LinxArduinoUno();
```

```
27.
28.   //The LINX Listener Is Pre Instantiated, Call Start And Pass A Pointer To The LINX Device And The UART Channel To Listen On
29.   LinxSerialConnection.Start(LinxDevice, 0);
30.
31. }
32.
33. void loop()
34. {
35.   //Listen For New Packets From LabVIEW
36.   LinxSerialConnection.CheckForCommands();
37.
38.   //Your Code Here, But It will Slow Down The Connection With LabVIEW
39.
40. LinxSerialConnection.AttachCustomCommand(0x3, readout_HX11);
41. // Custom commands should have number 0-15
42.
43. // here, custom command nr 9 was chosen ==> = hex '9'
44.
45. }
46. // %%%%%   Custom command definition      %%%%%%%%
47. int readout_HX11(unsigned char numInputBytes, unsigned char* input, unsigned char* numResponseBytes, unsigned char* response){
48. //
49. // Functionality:
50. // Converts a float into an int, then into 2 bytes and sends this to Labview,
51. // You can apply a scaling factor before, to choose the required resolution.
52. //
53. // Inputs (from Labview):
54. // * Res_inverse (default 100); // inverse of resolution. Default: 100. Range: 1-254
55. //      e.g. 8 means a resolution of 0.125, res_inverse = 100 a resolution up to to 0.01.
```

```
56. // Outputs:
57. //      2 bytes containing data of a float
58. //      1 byte containing the (inverse of) the requested resolution.
59. // Parameters:
60. // * float input_value. This is an example code with FIXED value that will be send to Labview.
61. //       You can easily extend this code by calling another (sub)function.
62.         // Range: 16 bits => number between -32767/res_inverse to +32767/res_inverse.
63.
64. //float input_value = -55.625; // °C Fictive example. Multiply by 8 to get an int with a resolution of 0.125
65. //unsigned int res_inverse = input[0]; // inverse of resolution. Default: 8. Range: 1-254
66. //signed int int_to_send = input_value* res_inverse ; // with input_value  =  -55.625 and res_inverse = 8, this is -445, or in bits: '1111 1110'  '0100 0011' => High byte: '254', low byte '67'
67. signed int rawsignal = int (float(hx711.read()/100.0));
68.
69. unsigned int value =rawsignal;     // Make the value an unsigned integer, to shift 0 in from the left, instead of ones.
70.
71. // send two bytes and res_inverse:
72. *numResponseBytes = 2;
73. //response[0] = (value & 0x0000) >> 16;    //mask all but MSB, shift 16 bits to the right; // to test: try B10101010 ; // default: MSB
74. response[0] = (value & 0xFF00) >> 8; //mask top bits and bottom bits, shift 8
75. response[1] = (value & 0x00FF) ; //mask top bits, no need to shift.
76. // response[2] = byte (res_inverse & 0x00FF);
77. return 0;
78. };
```

## Relationship between force and deflection

*Table 15 Deflection under different loads*

| 20 | 40 | 60 | 80 | 100 | 120 | 140 | 160 | 180 | 200 |
|---|---|---|---|---|---|---|---|---|---|
| 0.06mm | 0.13mm | 0.19mm | 0.25mm | 0.32mm | 0.38mm | 0.44mm | 0.50mm | 0.57mm | 0.63mm |

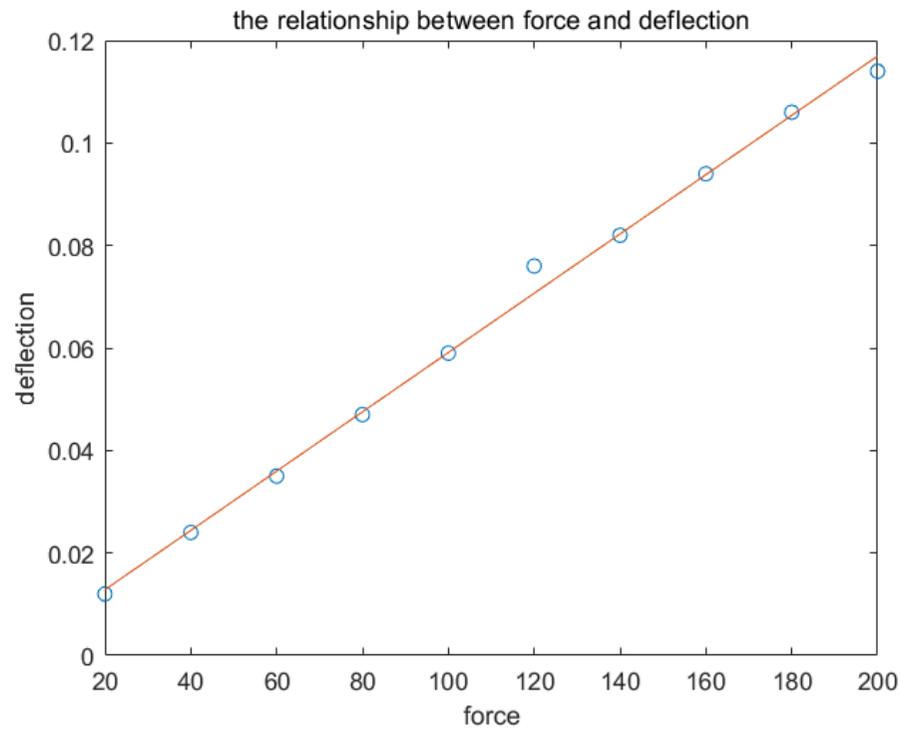

*Figure 115 The relationship between drag force and deflection*